%% file: XCOP_thermo.tex
\documentclass[traditabstract   ]{aa} %traditabstract

\usepackage{times,epsfig,amssymb,amsmath,natbib}

\newcommand{\xmm}{{\it XMM-Newton}}
\newcommand{\planck}{{\it Planck}}

\usepackage[labelfont=bf]{caption}

\usepackage{float}
\usepackage[toc,page]{appendix}
\usepackage{tabularx}
\usepackage{braket}
        
\usepackage{natbib,twoopt}
\usepackage{verbatim}
\usepackage{amsmath}

\usepackage[]{hyperref}
\hypersetup{pdftex,colorlinks=true,allcolors=blue}
\usepackage{hypcap}
\usepackage{txfonts}
\usepackage{color}

\begin{document}

\title{Universal thermodynamic properties of the intracluster medium over two decades in radius in the X-COP sample}
\titlerunning{X-COP: Thermodynamic properties}
\mail{vittorio.ghirardini2@unibo.it}

\author{V. Ghirardini\inst{1,2}\thanks{e-mail: \href{mailto:vittorio.ghirardini2@unibo.it}{\tt vittorio.ghirardini2@unibo.it}} \and D. Eckert\inst{3} \and S. Ettori\inst{2,4} \and E. Pointecouteau\inst{5}\and S. Molendi\inst{6}\and M. Gaspari\inst{7}\thanks{{\it Einstein} and {\it Spitzer} Fellow}\and M. Rossetti\inst{6}\and S. De Grandi\inst{8}\and M. Roncarelli\inst{1,2}\and H. Bourdin\inst{10,11}\and P. Mazzotta\inst{11}\and E. Rasia\inst{12}\and F. Vazza\inst{1,13}}
\authorrunning{V. Ghirardini et al.}

\institute{
 Dipartimento di Fisica e Astronomia Universit\`a di Bologna, Via Piero Gobetti, 93/2, 40129 Bologna, Italy 
 \and INAF, Osservatorio di Astrofisica e Scienza dello Spazio, via Pietro Gobetti 93/3, 40129 Bologna, Italy
 \and Max-Planck-Institut f\"ur Extraterrestrische Physik, Giessenbachstrasse 1, 85748 Garching, Germany
 \and INFN, Sezione di Bologna, viale Berti Pichat 6/2, I-40127 Bologna, Italy
 \and IRAP, Universit\'e de Toulouse, CNRS, CNES, UPS, (Toulouse), France
 \and INAF - IASF Milano, via Bassini 15, I-20133 Milano, Italy
\and Department of Astrophysical Sciences, Princeton University, 4 Ivy Lane, Princeton, NJ 08544-1001, USA
\and Dipartimento di Fisica, Universit\`a degli Studi di Milano, via Celoria 16, I-20133 Milano, Italy
\and INAF–Osservatorio Astronomico di Brera, via E. Bianchi 46, 23807 Merate, Italy
\and Harvard Smithsonian Center for Astrophysics, 60 Garden Street, Cambridge, MA 02138, USA
\and Dipartimento di Fisica, Universit\`a degli Studi di Roma Tor Vergata, via della Ricerca Scientifica, 1, I-00133 Roma, Italy
\and INAF, Osservatorio Astronomico di Trieste, via Tiepolo 11, I-34131, Trieste, Italy
\and Hamburger Sternwarte, Gojenbergsweg 112, 21029 Hamburg, Germany
}

%\offprints{V. Ghirardini}
\mail{vittorio.ghirardini2@unibo.it}
\abstract
{The hot plasma in a galaxy cluster is expected to be heated to high temperatures through shocks and adiabatic compression% at the boundary between the free-falling gas and the virialized intracluster medium
. The thermodynamical properties of the gas encode information on the processes leading to the thermalization of the gas in the cluster’s potential well and on non-gravitational processes such as gas cooling, AGN feedback, {shocks, turbulence, bulk motions, cosmic rays and magnetic field}.}
{In this work we present the radial profiles of the thermodynamic properties of the intracluster medium (ICM) out to the virial radius for a sample of 12 galaxy clusters selected from the \planck\ all-sky survey. We determine the universal profiles of gas density, temperature, pressure, and entropy over more than two decades in radius{, from 0.01$R_{500}$ to 2 $R_{500}$.}}
{We exploited  X-ray information from \xmm\ and Sunyaev-Zel'dovich constraints from \planck\ to recover thermodynamic properties out to $2 R_{500}$. We provide average functional forms for the radial dependence of the main quantities and quantify the slope and intrinsic scatter of the population as a function of radius.}
{We find that gas density and pressure profiles steepen steadily with radius, in excellent agreement with previous observational results. Entropy profiles beyond $R_{500}$ closely follow the predictions for the gravitational collapse of structures. The scatter in all thermodynamical quantities reaches a minimum in the range $[0.2-0.8]R_{500}$ and increases outward. Somewhat surprisingly, we find that pressure is substantially more scattered than temperature and density. }
{Our results indicate that once accreting substructures are properly excised, the properties of the ICM beyond the cooling region ($R>0.3R_{500}$) follow remarkably well the predictions of simple gravitational collapse and require few non-gravitational corrections.}


\maketitle % Insert title

%\href{mailto:vittorio.ghirardini2@unibo.it}{\tt vittorio.ghirardini2@unibo.it}
%----------------------------------------------------------------------------------------
%       ARTICLE CONTENTS
%----------------------------------------------------------------------------------------

\section{Introduction}

Galaxy clusters are the largest bound structures in the Universe. In the hierarchical structure formation scenario, they grow from  primordial density fluctuations to form the massive structures we observe today  \citep[e.g.,][]{kravtsov12}. Pristine gas falls into the dark matter potential and is progressively heated up to temperatures of $10^7 - 10^8$ K, such that the majority of the baryons end up in the form of a fully ionized plasma, the intracluster medium (ICM), which produces X-ray emission through { bremsstrahlung} radiation and line emission. In addition, photons of the cosmic microwave background (CMB) crossing galaxy clusters are subject to inverse Compton scattering off the hot ICM electrons. This produces a spectral shift in the CMB signal, the Sunyaev-Zel'dovich (SZ) effect \citep{SZ}, which is detectable at microwave wavelengths. X-ray emission and the SZ effect thus provide highly complementary diagnostics of the state of the ICM.

The thermodynamical properties of the ICM encode valuable information on the processes governing the formation and evolution of galaxy clusters. At first approximation, the state of the gas is determined by the properties of the gravitational potential and the merging history of the host halo alone. Gravitational collapse implies the existence of tight scaling laws between ICM properties and cluster mass \citep[the self-similar model,][]{kaiser86,bryan98} and non-radiative cosmological simulations predict that the scaled thermodynamical profiles of massive systems are nearly universal \citep[e.g.,][]{frenk99,borgani05,voit05rev}. Thus, the distribution of the thermodynamical properties across the ICM is a powerful tool for probing the formation of galaxy clusters. Deviations from gravitational collapse predictions can be used as a way to quantify the impact of non-gravitational physics such as gas cooling and feedback from supernovae and active galactic nuclei (AGN) \citep{tozzi01,kay02,borgani04,kravtsov05,gaspari14} and to probe hydrodynamical phenomena induced by structure formation such as shocks, turbulence and bulk motions \citep{dolag05,rasia+06,vazza09,burns10}, cosmic rays \citep{pfrommer+07}, and magnetic fields \citep{dolag+99}. In recent years, hydrodynamical simulations of increasing complexity have attempted to model simultaneously all known non-gravitational effects and to determine their impact on the structural properties of galaxy clusters \citep[e.g.,][]{lebrun14,hahn17,planelles17,barnes17,gaspari18}. These simulations highlight the role played by AGN feedback in maintaining the cooling/heating balance and regulating the star formation efficiency.%, pointing out the possible role of these processes in setting up the ICM properties up to large radii.

In the past decade, X-ray observations have provided constraints of increasing quality on the distribution of gas density \citep{croston+06,eckert12}, temperature \citep{degrandi2002,vikhlini+06,pratt07,lm08}, pressure \citep{arnaud+10}, and entropy \citep{cavagnolo+09,pratt+10} throughout the ICM. However, these observations sampled a radial range limited to less than half of the virial radius, and thus the majority of the cluster volume, and the entirety of the accretion region, remained unexplored. Thanks to its low instrumental background, the \emph{Suzaku} satellite allowed us to extend the accessible radial range to the virial radius in a dozen clusters \citep[e.g.,][]{reiprich13,akamatsu11,simi11,simi17,walker12a,walker12b,urban14}. Somewhat surprisingly, most studies find steeply decreasing temperature profiles and a flattening of the entropy at $R_{500}$ and beyond, at odds with the predictions of gravitational collapse. Possible explanations for these behaviors  include biases in X-ray measurements caused by gas clumping \citep{nagai11,simi11,vazza13,roncarelli+13}, non-equilibrium between electrons and ions \citep{hoshino10,avestruz15}, breakdown of hydrostatic equilibrium caused by turbulence and bulk motions \citep{lapi10,okabe14,avestruz15,khatri16}, or as-yet-unknown systematics in the subtraction of the \emph{Suzaku} background.  Clearly, accurate observational data out to the virial radius are required to understand the origin of these deviations.

{
Since the first X-ray observations of galaxy clusters it has become evident that clusters can be divided into two categories differing only in their central properties \citep{Jones+84}.
The origin of these two populations is still unclear;  however, several recent works \citep{rossetti17, lovisari17, andrade-santos17}  point out that there is a clear difference between X-ray and SZ selected samples, also generating possible biases  in the measured thermodynamic quantities.
}

In this paper, we present the universal thermodynamical properties of the galaxy clusters in the XMM Cluster Outskirts Project (X-COP), a sample of 12 massive, local galaxy clusters ($z < 0.1$) with deep \xmm\ and \planck\ data covering the entire azimuth out to the virial radius. Unlike previous studies that  exclusively utilized the X-ray signal, we take advantage of the high signal-to-noise ratio of our clusters in the \emph{Planck} survey \citep{psz1}, and combine X-ray and SZ data to increase the precision of our measurements while keeping a good control of systematic errors. This method was  applied to reconstruct the thermodynamical properties of a few clusters \citep{basu10,eckert13a,adam15,ruppin17}, and we demonstrated the ability of \xmm\ and \planck\ to measure accurately the state of the gas out to the virial radius in two pilot studies \citep{tchernin16, ghirardini18}. 

This paper is the first in a series presenting the results of the analysis of the full X-COP sample. Here we consider the global properties of the X-COP sample and present universal functional forms describing the structural properties of the ICM over more than two decades in radius ($[0.01-2]R_{500}$). We also determine the slope and intrinsic scatters of the various quantities as a function of radius. In Ettori et al. (2018) we present our reconstruction of the mass profiles of our clusters, whereas in Eckert et al. (2018) we estimate the amount of non-thermal pressure support affecting our sample.

The paper is organized as follows. In Sect. \ref{sec:data} we describe the available dataset and the analysis procedures. In Sect. \ref{sec:thermo} we present our results on the universal thermodynamic profiles, slopes, and intrinsic scatter. Our findings are discussed in Sect. \ref{sec:disc}. Throughout the paper, we assume a flat $\Lambda$CDM cosmology with $\Omega_m=0.3$, $\Omega_\Lambda=0.7$, and $H_{0}=70$ km s$^{-1}$ Mpc$^{-1}$. All our fitting is performed using the Bayesian nested sampling algorithm MultiNest \citep{multinest} unless otherwise stated. 

\section{Dataset and analysis procedures}
\label{sec:data}

\subsection{The X-COP project}

X-COP \citep{xcop} is a very large program (VLP) on \emph{XMM-Newton} whose  aim is to advance our understanding of the virialization region of galaxy clusters. The strategy adopted for the project is to target the most significant Sunyaev-Zel'dovich (SZ) sources in the \emph{Planck} survey in order to combine X-ray and SZ information out to the virial radius. This strategy was already applied to the cases of A2142 \citep{tchernin16} and A2319 \citep{ghirardini18} and it was shown to allow the detection and characterization of the ICM out to $R_{200}$ and even beyond. The sample was selected from the first \emph{Planck} SZ catalogue \citep[PSZ1,][]{psz1} according to the following criteria:
\begin{itemize}
 \item \textbf{PSZ1 S/N > 12 :} we select only the sources with the strongest SZ signal to ensure detection at high significance beyond $R_{500}$;
 \item $\mathbf{0.04<z<0.1}$ \textbf{:} this criterion allows us to select objects for which the region of interest can be covered with five \emph{XMM-Newton} pointings (one central and four offset);
 \item $\mathbf{\theta_{500}}$\footnote{Where $\theta_{500}$ refers to the apparent angular size of $R_{500}$}$\mathbf{>10}$ \textbf{arcmin :} given the 7 arcmin \emph{Planck} beam, this criterion ensures that the clusters are well resolved by \emph{Planck};
 \item $\mathbf{N_H}$\footnote{Where ${N_H}$ refers to the hydrogen column density along the line of sight}$\mathbf{<10^{21}\, cm^{-2}}$ \textbf{:}  since we use a soft energy band ([0.7-1.2] keV, see below) to extract surface brightness and gas density profiles, this cut avoids objects for which the X-ray flux is strongly suppressed below $\sim1$ keV.
 \end{itemize}
 
 These criteria lead to a set of 16 clusters from the PSZ1 catalogue. Of these systems, we excluded four: two  (A754 and A3667) because they exhibit strongly disturbed morphologies, rendering an azimuthally averaged analysis difficult; one (A2256)  because of its bad visibility for \emph{XMM-Newton}; and one (A3827) because its apparent size $\theta_{500}$ is very close to our cut and so may not be properly resolved by \emph{Planck}. 
 %To the resulting list of 12 clusters, we added Hydra A/A780 due to the availability of a high-quality \emph{XMM-Newton} mapping. However, in the cases where the sample selection is critical we exclude this system from our analysis, since a point source contaminates its SZ signal and therefore does not have \emph{Planck} data. 
 The final list of clusters is given in Table 1, together with some basic properties of these systems and classification into the cool-core and non-cool-core classes{, using the value of $K_0$ from \cite{cavagnolo+09} as discerning value (see also Sect.~\ref{sec:CC_NCC} for further details).} Like other \planck-selected samples \citep{rossetti17,andrade-santos17,lovisari17}, the X-COP sample is dominated by non-cool-core systems (8 out of 12).
 
The nominal \emph{XMM-Newton} exposure time was set to 25 ks per pointing, which allows us to reach a limiting surface brightness of $3\times10^{-16}$ ergs cm$^{-2}$ s$^{-1}$ arcmin$^{-2}$ in the [0.5-2.0] keV band. Including central and archival pointings, the total observing time is about 2 Ms. The observation log, observation IDs, and observing time, after applying flare filtering, are given in Table \ref{tab:log}.

\begin{table*}[t]
\caption{Basic properties of the X-COP sample.}
\begin{center}
\begin{tabular}{ c c c c c c c c c c}
\hline
Cluster & redshift & S/N & $M_{500}$ & $R_{500}$ & $M_{200}$ & $R_{200}$ & $K_0$ & R.A. & Dec \\
  &   & \emph{Placnk}  & $10^{14}$ $M_{\odot}$ & kpc & $10^{14}$ $M_{\odot}$ & kpc & keV cm$^2$  & deg & deg\\
\hline
       A1644 & 0.0473 & 13.2 & $3.48 \pm 0.20$ & $1054 \pm 20$ & $6.69 \pm 0.58$ & $1778 \pm 51$ & 19.0 (CC) & 194.3015 & -17.409729 \\
       A1795 & 0.0622 & 15.0 & $4.63 \pm 0.14$ & $1153 \pm 12$ & $6.53 \pm 0.23$ & $1755 \pm 21$ & 19.0 (CC) & 207.21957 & 26.589602 \\
       A2029 & 0.0766 & 19.3 & $8.65 \pm 0.29$ & $1414 \pm 16$ & $12.25 \pm 0.49$ & $2155 \pm 29$ & 10.5 (CC) & 227.73418 & 5.744432 \\
       A2142 & 0.0909 & 21.3 & $8.95 \pm 0.26$ & $1424 \pm 14$ & $13.64 \pm 0.50$ & $2224 \pm 27$ & 68.1 (NCC) & 239.58615 & 27.229434 \\
       A2255 & 0.0809 & 19.4 & $5.26 \pm 0.34$ & $1196 \pm 26$ & $10.33 \pm 1.23$ & $2033 \pm 81$ & 529.1 (NCC) & 258.21604 & 64.063058 \\
       A2319 & 0.0557 & 30.8 & $7.31 \pm 0.28$ & $1346 \pm 17$ & $10.18 \pm 0.52$ & $2040 \pm 35$ & 270.2 (NCC) & 290.30276 & 43.94501 \\
       A3158 & 0.0597 & 17.2 & $4.26 \pm 0.18$ & $1123 \pm 16$ & $6.63 \pm 0.39$ & $1766 \pm 35$ & 166.0 (NCC) & 55.717984 & -53.627728\\
       A3266 & 0.0589 & 27.0 & $8.80 \pm 0.57$ & $1430 \pm 31$ & $15.12 \pm 1.44$ & $2325 \pm 74$ & 72.5 (NCC) & 67.843372 & -61.429731 \\
        A644 & 0.0704 & 13.9 & $5.66 \pm 0.48$ & $1230 \pm 35$ & $7.67 \pm 0.73$ & $1847 \pm 59$ & 132.4 (NCC) & 124.35736 & -7.5086903 \\
         A85 & 0.0555 & 16.9 & $5.65 \pm 0.18$ & $1235 \pm 13$ & $8.50 \pm 0.36$ & $1921 \pm 27$ & 12.5 (CC) & 10.459403 & -9.3029207 \\
%      HydraA & 0.0538 & $-$ & $2.21 \pm 0.23$ & $904 \pm 32$ & $3.01 \pm 0.37$ & $1360 \pm 56$ & 13.3 (CC)  \\
     RXC1825 & 0.0650 & 13.4 & $4.08 \pm 0.13$ & $1105 \pm 12$ & $6.15 \pm 0.26$ & $1719 \pm 24$ & 217.9 (NCC) & 276.33547 & 30.436748 \\
      ZW1215 & 0.0766 & 12.8 & $7.66 \pm 0.52$ & $1358 \pm 31$ & $13.03 \pm 1.23$ & $2200 \pm 69$ & 163.2 (NCC) & 184.42191 & 3.6557217 \\
\hline 
\end{tabular}
\end{center}
\tablefoot{Cluster name, redshift, and signal-to-noise ratio from the PSZ1 catalogue \citep{psz1}. The mass information ($M_{500}$, $R_{500}$, $M_{200}$, $R_{200}$) is obtained from our own hydrostatic mass reconstruction (Ettori et al. 2018). The information on the central entropy was taken from the ACCEPT database \citep{cavagnolo+09}{, indicating  the cool-core (CC) clusters with $K_0 < 30$ keV cm$^2$, and the non-cool-core (NCC}) clusters. The last two columns indicate the center of the radial profiles in degrees.}
\label{table:data}
\end{table*}

\subsection{\emph{XMM-Newton} data analysis}

\begin{figure*}
\resizebox{\hsize}{!}{\includegraphics{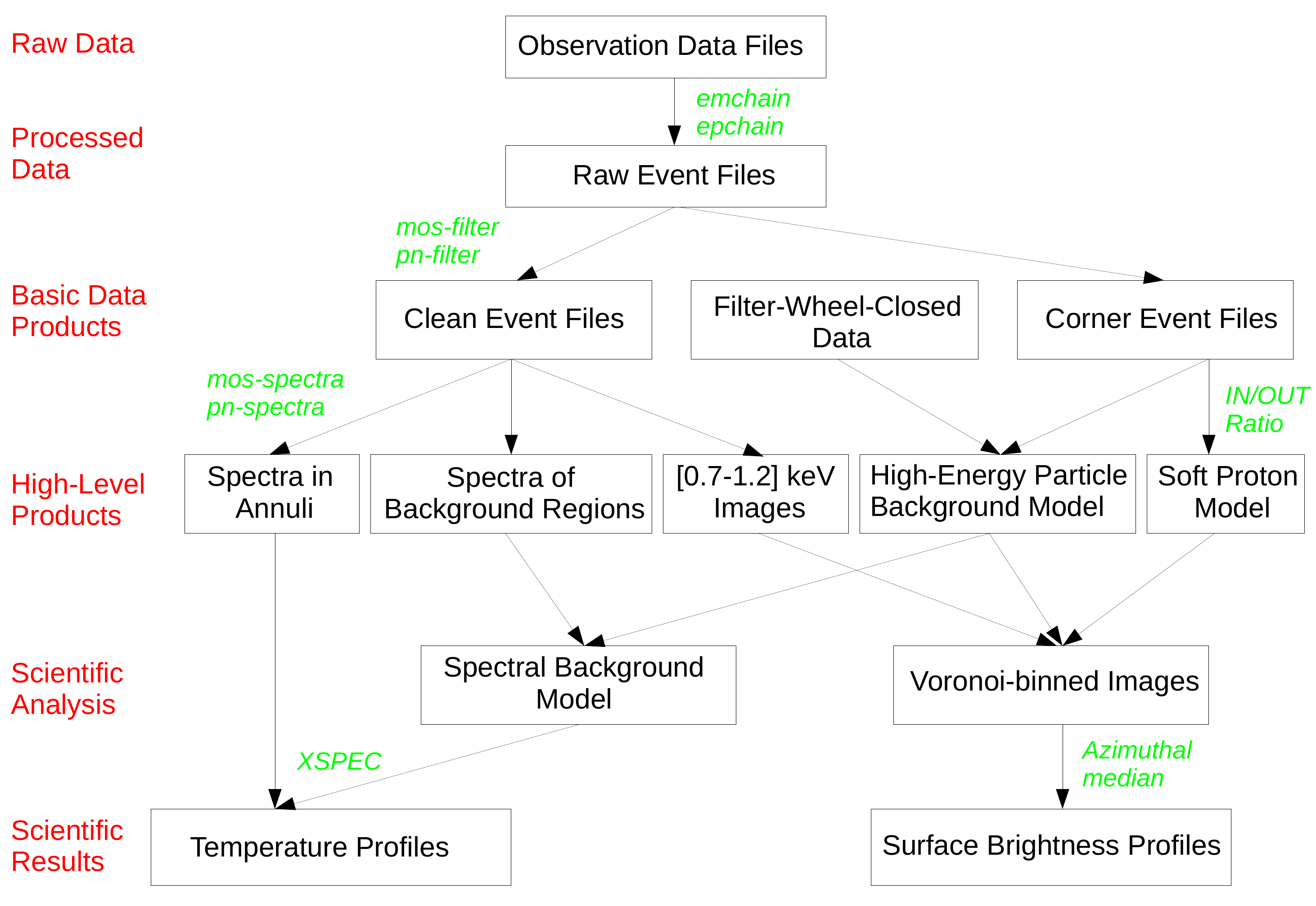}}
\caption{Flow chart of the \xmm\ data analysis pipeline. The steps of the analysis are shown in red, the main intermediate and final products are described in the black boxes, and the procedures are shown in green italics.}
\label{fig:XMM_flow_chart}
\end{figure*}

Here we describe in detail the data analysis pipeline that we set up for the analysis of the \xmm\ data. A flow chart describing the main steps of the analysis is presented in Fig. \ref{fig:XMM_flow_chart}.

\subsubsection{Data reduction}
We  reduced all the data using XMMSAS v13.5 and the Extended Source Analysis Software (ESAS) data reduction scheme \citep{snowden08}. To perform basic data reduction, we used the \texttt{emchain} and \texttt{epchain} pipelines to extract calibrated event files from the observations, and we reran \texttt{epchain} in out-of-time mode to create event files for pn out-of-time events. To filter out time periods affected by soft proton flares we ran the \texttt{mos-filter} and \texttt{pn-filter} executables, which extract the light curve of each observation in the hard band, and applied a sigma-clipping technique to exclude time intervals with enhanced background. We used the unexposed corners of the MOS detectors to monitor the particle background level during each observation, and measured the count rates in the high-energy band ([7.5-11.8] keV) of the MOS from the regions located inside and outside the field of view (FOV) of the telescope (hereafter IN and OUT). The comparison between IN and OUT count rates was then used to estimate the contamination of residual soft protons to the spectrum \citep{deluca04,lm08,salvetti17}.

\subsubsection{Image extraction and preparation}
\label{sec:red_ima}

We extracted photon count images from the three EPIC detectors in the [0.7-1.2] keV band. This narrow band maximizes the ratio between source and background emission and is thus best suited to minimize the systematics in the subtraction of the EPIC background \citep{ettori+10}. We used \texttt{eexpmap} to compute exposure maps taking vignetting effects into account for all three detectors independently. To create total EPIC images, we summed the count maps of the three detectors and repeated the same operation with the exposure maps, multiplying the pn exposure maps by a factor of 3.44 representing the ratio of pn to MOS effective areas in our band of interest.  
{
The high-energy particle background are modeled and subtracted by simple subtraction of the rescaled background images, taken from the unexposed corners of the detectors.
}

Even after cleaning the light curves from soft-proton flares, it is known that a fraction of residual soft proton contamination remains within the datasets. This component can introduce systematics at the level of $\sim20\%$ in the subtraction of the EPIC background in our band of choice \citep{tchernin16}. To model the contribution of residual soft protons, we follow the method outlined in \citet{ghirardini18}, which was calibrated using a large set of $\sim500$ blank-sky pointings. Namely, we measure the high-energy ([7.5-11.85] keV) MOS count rates in the exposed (IN) and unexposed (OUT) parts of the FOV, and we use the difference between IN and OUT count rates as an indicator of the contamination of each observation by residual soft protons \citep[see][for a detailed overview of this approach]{salvetti17}. We use our large blank field dataset to calibrate an empirical relation between the IN-OUT indicator and the required intensity of the soft proton component \citep[see Appendix A of][]{ghirardini18}, and we use this relation to create a 2D  soft proton model. This procedure was shown to bring the systematics in the subtraction of the EPIC background to an accuracy better than 5\%. 

\begin{figure*}[t]
\includegraphics[width=\textwidth]{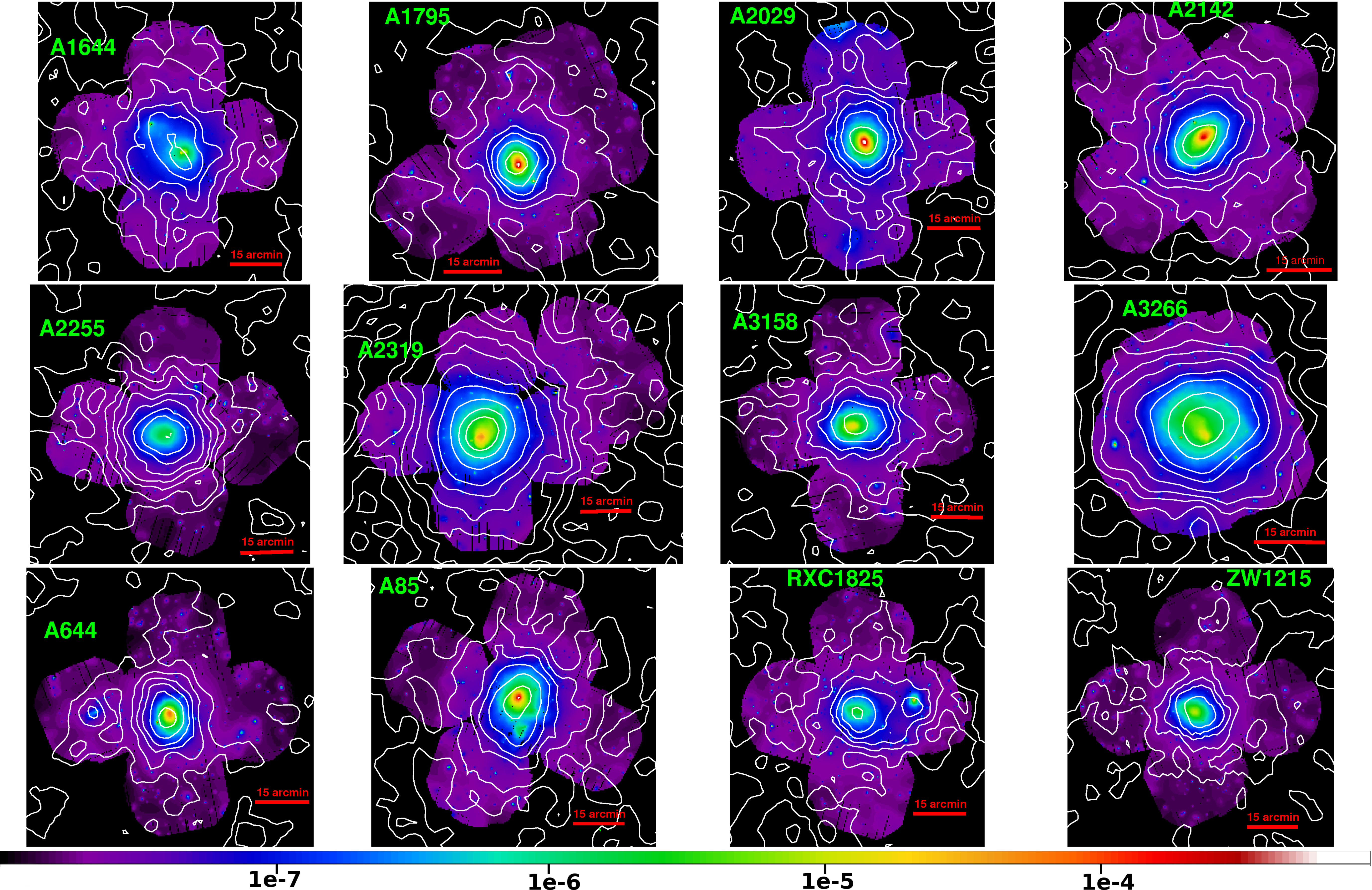}
\caption{\label{fig:XCOP_all}Adaptively smoothed and exposure corrected \xmm\ mosaic images in the [0.7-1.2] keV energy band for all X-COP clusters. The superimposed white contours represent the \emph{Planck} SZ S/N maps between 70 and 857 GHz. The contour levels correspond to 1, 3, 5, 7, 10, 15, 20, 30, 40, and 50$\sigma$. { The spatial scale is given by the red thick lines, which have a common length of 15 arcmin. The color scale is given below the images in units of cts s$^{-1}$ pixel$^{-1}$, and  is the same for all clusters.}}
\end{figure*}

For each cluster, we combine the resulting EPIC count maps in the [0.7-1.2] keV from each individual observation (central or offset) to create a mosaic image. We apply the same procedure to the combined EPIC exposure maps and to the background maps, summing  the non-X-ray background components (quiescent particle background and residual soft protons). We then obtain mosaicked photon maps, exposure maps, and non-X-ray background maps for each source. In Fig. \ref{fig:XCOP_all} we show the resulting background-subtracted and exposure-corrected mosaics, which we adaptively smoothed using the \texttt{asmooth} code \citep{ebeling06}.

\subsubsection{Point source subtraction}
\label{sec:ps}

To detect point sources present within the field, we extract photon count maps from the three detectors in  a soft ([0.5-2] keV) and a hard  ([2-7] keV) band, and we use the XMMSAS tool \texttt{ewavelet} with wavelet scales in the range 1--8 pixels and signal-to-noise ratio threshold of 5.0. We then cross-match the soft and hard band detections between the multiple (central and offset) observations of each cluster to create a global point source list per cluster. Since the vignetting and the point spread function of the \emph{XMM-Newton} telescopes depend on off-axis angle, the sensitivity threshold for source detection depends on the position of a source on the detector. At a fixed observing time, \emph{XMM-Newton} thus detects point sources down to lower fluxes near the aim point than close to the edge of the FOV, and the fraction of the cosmic X-ray background (CXB) that is resolved by the instrument is spatially dependent. To correct for this effect, we draw the distribution of measured count rates from the detected sources and we determine the count rate at which the distribution peaks. Since the logN-logS of distant sources contributing to the CXB is a monotonically decreasing function \citep[e.g.,][]{moretti03} the peak in the count rate distribution of our observation roughly corresponds to the threshold down to which our source detection is complete. We then excise only the sources with a measured count rate greater than our threshold and leave the fainter sources to enforce a constant flux threshold across the FOV and avoid biasing local measurements of the CXB intensity.

\subsubsection{Surface brightness profiles}

An important complication with the analysis of X-ray data of cluster outskirts lies in the presence of accreting structures and inhomogeneities in the gas distribution, which are expected to contribute substantially to the measured X-ray flux beyond $\sim R_{500}$ \citep[e.g.,][]{nagai11,vazza13,roncarelli+13}. Since the X-ray emissivity is proportional to the square of the gas density, overdense regions contribute predominantly to the measured X-ray flux and can bias high the recovered gas density. To avoid this problem we apply the azimuthal median method outlined in \citet{eckert15}. Numerical simulations show that the median surface brightness in concentric annuli is robust against the presence of outliers in the gas density distribution \citep{zhuravleva13}. We thus construct background-subtracted and exposure-corrected surface brightness maps using a Voronoi tessellation technique \citep{cappellari03} with a target number of 20 counts per bin. 

The intensity of the sky background is determined by averaging the surface brightness distribution in the regions with $R>2R_{500}$, where we assume the cluster emission to be negligible. A systematic uncertainty of 5\% of the background level (see Sect. \ref{sec:red_ima}) was added in quadrature to the error budget of the surface brightness profiles. This procedure was applied to all clusters except A3266, for which the current mosaic does not extend out to $2R_{500}$. In this case, we estimate the sky background intensity from the ROSAT all-sky survey background tool\footnote{\href{https://heasarc.gsfc.nasa.gov/cgi-bin/Tools/xraybg/xraybg.pl}{https://heasarc.gsfc.nasa.gov/cgi-bin/Tools/xraybg/xraybg.pl}} and included a systematic error of 30\% in quadrature to the full error budget.

To compute the surface brightness profiles, we draw the surface brightness distribution from the Voronoi-binned images in concentric radial bins starting from the X-ray peak{ and then choosing the annuli such that the emissivity in each bin is almost constant}. The errors on the azimuthal median are estimated from $10^4$ bootstrap resampling of the pixel distribution. Circular regions of 30 arcsec radius are excised around the positions of point sources selected through the procedure described in Sect. \ref{sec:ps}, corresponding to an encircled energy fraction of 90\% of the point source flux.

\subsection{Deprojection and gas density profiles}

To extract gas density profiles, we take advantage of the fact that the X-ray surface brightness in our energy band of choice is proportional to the squared gas density integrated along the line of sight. To convert surface brightness profiles into emission measure, we describe the emissivity of the source with a thin-plasma model absorbed by the Galactic $N_H$ and folded through the on-axis EPIC/MOS effective area. This approach allows us to calculate the conversion between the observed count rate in MOS units and the normalization of the APEC model, which is related to the plasma emission measure as 

\begin{equation}\mbox{Norm} = \frac{10^{-14}}{4\pi [d_A(1+z)]^2}\int_V n_en_H\, dV,\label{eq:apecnorm}\end{equation}

\noindent where $d_A$ is the angular diameter distance of the source, and $n_e$ and $n_H$ are the electron and ion number densities in units of cm$^{-3}$, with $n_e=1.17n_H$ in a fully ionized plasma \citep{ag89}. Since we are using a soft energy band, the conversion between count rate and emission measure shows little dependence on the temperature as long as the temperature exceeds $\sim1.5$ keV, which is the case in all X-COP systems. The resulting emission measure profiles can then be deprojected under the assumption of spherical symmetry by computing the projected volumes V of each spherical shell onto each 2D annulus. To recover the 3D emissivity and density profiles from the projected data, we apply two different deprojection methods that we briefly outline here.

\begin{itemize}
\item \textbf{L1 regularization:} This method builds on the non-parametric regularization approaches developed  by \citet{croston+06} and \citet{ameglio+07}, introducing a penalty term on the modulus of the second derivative of the 3D density profile to kill spurious small-scale fluctuations introduced by the random nature of the data \citep{diaz17}. Given an observed 2D emission measure profile  ${EM =(EM_1 \dots EM_n})$  and corresponding uncertainties { ${\sigma_{EM} =(\sigma_{EM,1} \dots \sigma_{EM,n}})$ }, the values of the 3D emissivity profile  { ${\epsilon =(\epsilon_1 \dots \epsilon_n})$ } are obtained by maximizing the following likelihood function
\begin{equation}
-2 \log \mathcal{L} = \chi^2 = \sum \frac{( V \# \epsilon - EM )^2}{\sigma_{EM}^2} + \lambda \sum \left| \frac{\partial^2 \log \epsilon }{\partial \log r^2} \right|
\label{eq:regularized}
,\end{equation}
{ where $V_{i,j}$ is the geometrical matrix volume of the j$\text{th}$ shell intercepted by the i$\text{th}$ annulus, $\#$ is the symbol for matrix product, and the sum is performed along all the annuli. Moreover the second derivative of the emissivity is computed as a numerical derivative of the $\epsilon(r)$ vector. }
The parameter $\lambda$ controls the degree of regularization of the profile. To maximize the likelihood function described in Eq. \ref{eq:regularized}, we use the Markov chain Monte Carlo (MCMC) tool \texttt{emcee} \citep{foreman-mackey13}, leaving the value of the 3D density profile at each radius as a free parameter and setting a logarithmic prior (i.e., uniform prior in logarithmic space) on each parameter value to enforce positivity of the resulting profile. The value of the parameter $\lambda$ is chosen such that the log-likelihood is about 1 per data point, to allow for typical statistical deviations of $1\sigma$. We note that $\lambda = 0$ is equivalent to using the onion-peeling technique directly \citep[see][]{kriss+83, ettori+02, ettori+10}.
\item \textbf{Multiscale fitting:} This method follows the technique developed in \citet{eckert16}, whereby the projected emission measure profile is decomposed into a sum of analytical multiscale functions which can be individually deprojected. Following \citet{eckert16} we write the observed 2D profile as a sum of $N$ King functions with fixed core radii and normalizations and slopes allowed to vary while fitting, choosing $N=N_{points}/4$; i.e., one model component is added for every set of four data points{, fixing a core radius to the mean radial value of these four data points}. Since the projection kernel is linear, each King function can be individually deprojected and the 3D profile can be analytically reconstructed from the fit to the projected data. As above, we use \texttt{emcee} to optimize for the parameters and reconstruct the error envelope around the best fitting curve.
\end{itemize}

In the top left panel of Fig. \ref{fig:density} we compare the density profiles reconstructed with the two methods and find a remarkable agreement between them, { with an average scatter} $<5\%$ at each radius. By construction, the profiles reconstructed with the L1 regularization method shows more pronounced features as the method imposes fewer constraints on the shape of the profile, whereas the profiles obtained with the multiscale method are smoother. Thus, we adopt the results of the L1 regularization when attempting to determine the exact shape of our profiles, whereas the multiscale technique is preferred when reconstructing hydrostatic mass profiles to provide better control over the gradient.

\subsection{Spectral analysis}
\label{sec:spec}

\subsubsection{Spectral extraction}
\label{sec:red_spec}

We extract spectra in concentric annuli around the X-ray peak covering approximately the radial range $[0-1]R_{500}${, removing the point sources which contaminates the spectra (see Sec.~\ref{sec:ps}). } 
{
In the spectral analysis, differently from the imaging case in Sect.~\ref{sec:red_ima}, we compute models of the high-energy particle background using the ESAS tools \texttt{mos-spectra} and \texttt{pn-spectra} in imaging mode. To this end, we select the filter-wheel-closed observations recorded at the nearest possible time to the observation, and we use the spectra of the unexposed corners of the detectors to rescale the filter-wheel-closed observations. This procedure is performed individually for all CCDs, and the CCDs operating in anomalous mode are discarded. We then extract an image from the rescaled filter-wheel-closed data in the [0.7-1.2] keV band using the \texttt{mos-back} and \texttt{pn-back} executables, which we use as our model for the high-energy particle background. In the case of the pn, we repeat the operation with the out-of-time event files and create a model for the intensity and spatial distribution of out-of-time events. }       

{We selected the binning such that the width of the bins increases exponentially, but choosing a minimum width of 0.5 arcmin for the innermost bins such that the instrumental PSF does not contribute much to the photon in each bin.}
We group the output spectra with a minimum of five counts per bin to ensure stable fitting results, and discard the data below 0.5 keV where the EPIC calibration is uncertain. We then use {\sc Xspec} { v12.9 and the C-statistic \citep{cash79}} to fit the spectra and determine the plasma parameters (see Sect. \ref{sec:spec}). 

\subsubsection{Spectral modeling}
\label{sec:spec}
To extract spectral diagnostics from the observed spectra (see Sect. \ref{sec:red_spec}), we proceeded using a full spectral modeling approach following the method described in detail in \citet{eckert14}. Here we describe our approach to model all the individual background components and the source spectra.

\begin{itemize}
\item \textbf{High-energy particle background :} We use the rescaled filter-wheel-closed spectra to determine the intensity and spectral shape of the particle background. We fit the filter-wheel-closed spectra using a diagonal response matrix and a phenomenological model including a broken power law and several Gaussians to reproduce the shape of the continuum and fluorescence lines. We then apply the fitted model to the source spectrum, leaving the normalization free to vary within $\pm10\%$ to account for possible systematics in the scaling of the filter-wheel-closed data.
\item \textbf{Sky background :} We model the X-ray background and foreground emission as the sum of three components: \emph{i}) an absorbed power law with a photon index fixed to 1.46 to describe the residual CXB \citep{deluca04}; \emph{ii}) an absorbed APEC thermal plasma model with a temperature allowed to vary in the range [0.15-0.6] keV to model the Galactic halo emission \citep{mccammon02}; and \emph{iii)} an unabsorbed APEC model with a temperature fixed to 0.11 keV to represent the local hot bubble. The Galactic hydrogen column density $N_H$ was fixed to the LAB value \citep{kalberla05}. Similarly to what was done for the imaging case, the parameters of the sky emission model are fitted to the spectra of background regions located at $R>2R_{500}$ from the cluster core. Again the exception to this procedure is A3266, for which we use the \emph{ROSAT} all-sky survey background tool to determine the sky background parameters. The best fit model is then applied to the source spectra, rescaling the intensity of the components according to the area of each region.
\item \textbf{Residual soft protons :} In cases where our IN-OUT indicator of soft proton contamination is found to be high (IN-OUT>0.1 counts/s), we include an additional model component to the particle background model to take soft protons into account. We model the soft proton component as a broken power law with fixed spectral shape \citep[slopes 0.4 and 0.8 and break energy 5 keV, ][]{lm08} and leave the normalization of this component free to vary in the overall fitting procedure.
\item \textbf{Source :} We model the source emission in each annulus as an absorbed single-temperature APEC model with temperature, emission measure, and metal abundance free to vary. In cases where multiple observations were available for the same regions, we fit all the available spectra jointly, tying the source parameters between the different spectra. The solar abundance table is set to \citet{ag89}. Since our objects are nearby and extended on scales much larger than the \emph{XMM-Newton} PSF, we neglect the potential cross-talk between the various annuli.
\end{itemize} 

All the spectra were fitted in the energy range [0.5-12] keV using {\sc Xspec} v12.9, ATOMDB v3.0.7, and the C-statistic \citep{cash79}.  When several observations of the same region were available, we extracted the spectra from each individual pointing and fit them jointly. We ignored the energy ranges [1.2-1.9] keV (MOS) and [1.2-1.7] keV, [7.0-9.2] keV (pn) where bright and time-variable fluorescence lines were present. We then constructed projected gas temperature profiles from the best fit results. We also deprojected our 2D temperature profiles using the projection matrix V and the emissivity in each annulus as weights, adopting the spectroscopic-like temperature scaling of \cite{mazzotta+04}.

\subsection{\emph{Planck} data analysis}
\label{s:plck}

The signal was recovered from the Planck survey  \citep{tauber+10,planckdr2015} making use of the six frequency maps provided by the High Frequency Instrument \citep[HFI][]{lamarre+10,planckHFI}. They were combined with a modified internal linear combination algorithm    \citep[MILCA,][]{hurier+13}  to produce  Comptonization parameter maps, i.e., $y$-maps tracing the intensity of the SZ effect. The maps used for the X-COP project are provided with a resolution of 7~arcmin FWHM.

The intra-cluster gas thermal pressure integrated is recovered from the $y$ parameter \citep{SZ} as 
\begin{equation}
y(\theta)=\frac{\sigma_T}{m_ec^2}\int P(\ell)d\ell,
\end{equation}
where the integral is computed along the line of sight, $\theta$ is the angular distance to the cluster core, $\sigma_T$ is the Thomson cross-section, $m_e$ is the mass of the electron, and $c$ is the speed of light. 

As illustrated for the cases of A2142 \citep{tchernin16} and A2319 \citep{ghirardini18}, we followed for the whole X-COP sample the methodology presented and used in \citet{planck+13}. 
%y-profile
Assuming azimuthal symmetry of the cluster, we computed the $y$ radial profile for each cluster over a regular grid scaled in units of $R_{500}$ with radial bins of size $\Delta\theta/\theta_{500}=0.2$ out to $10\times R_{500}$. The local background is assumed to be flat and constant and is computed from the area beyond $5\times \theta_{500}$. A covariance matrix is computed for each profile to account for the correlation between points due to the profile binning, and intrinsic noise correlation introduced from the $y$ map construction.

% Pressure profile
We corrected from the Planck beam redistribution through a real space deconvolution of the instrument PSF. Further assuming the spherical symmetry of the source, we reconstructed each pressure profile through a  geometrical deprojection. The two steps follow the method initially presented in \citet{croston+06}.  The associated covariance matrix for the pressure profile is obtained via a Monte Carlo procedure by randomizing  over the initial $y$ profile covariance matrix. In the following we ignore the innermost three \emph{Planck} data points because of the difficulty  deconvolving from the large \emph{Planck} beam. 

As an alternative method, we also extracted \emph{Planck} pressure profiles using the forward-modeling approach of \citet{bourdin+17}. In this case, a spectral model for the relevant components is constructed (CMB, dust, synchrotron, and thermal SZ). The model is folded through the \emph{Planck} response and fitted to the multi-frequency data points \citep[see][for details]{bourdin+17}. In Appendix \ref{app:bourdin} we compare the results obtained with this approach to the results of the  MILCA component separation method and show their consistency. For the remainder of the paper, we use the MILCA pressure profiles as our default choice.

\begin{figure*}[t]
\centering
\includegraphics[width=0.5\textwidth]{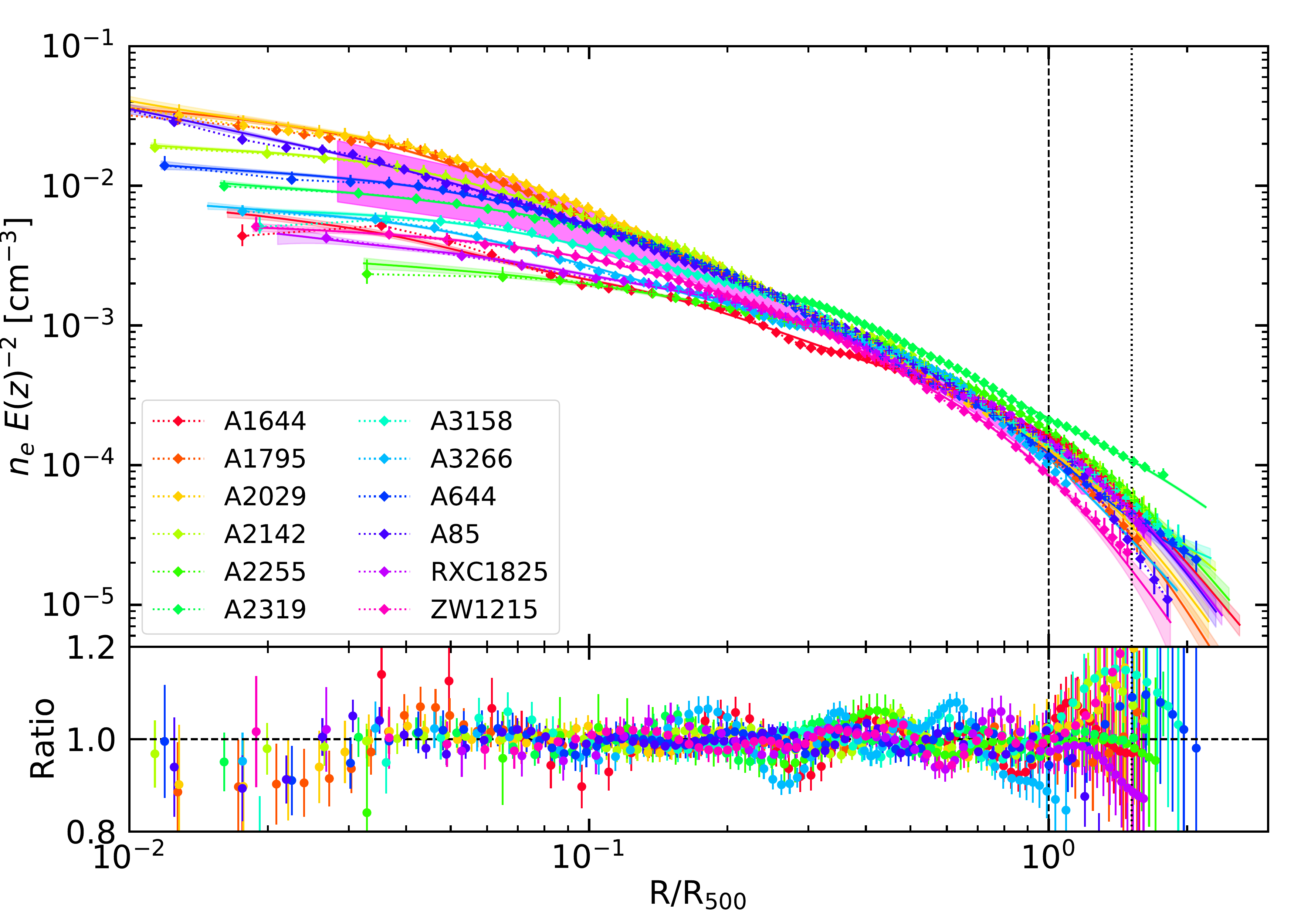}~
\includegraphics[width=0.5\textwidth]{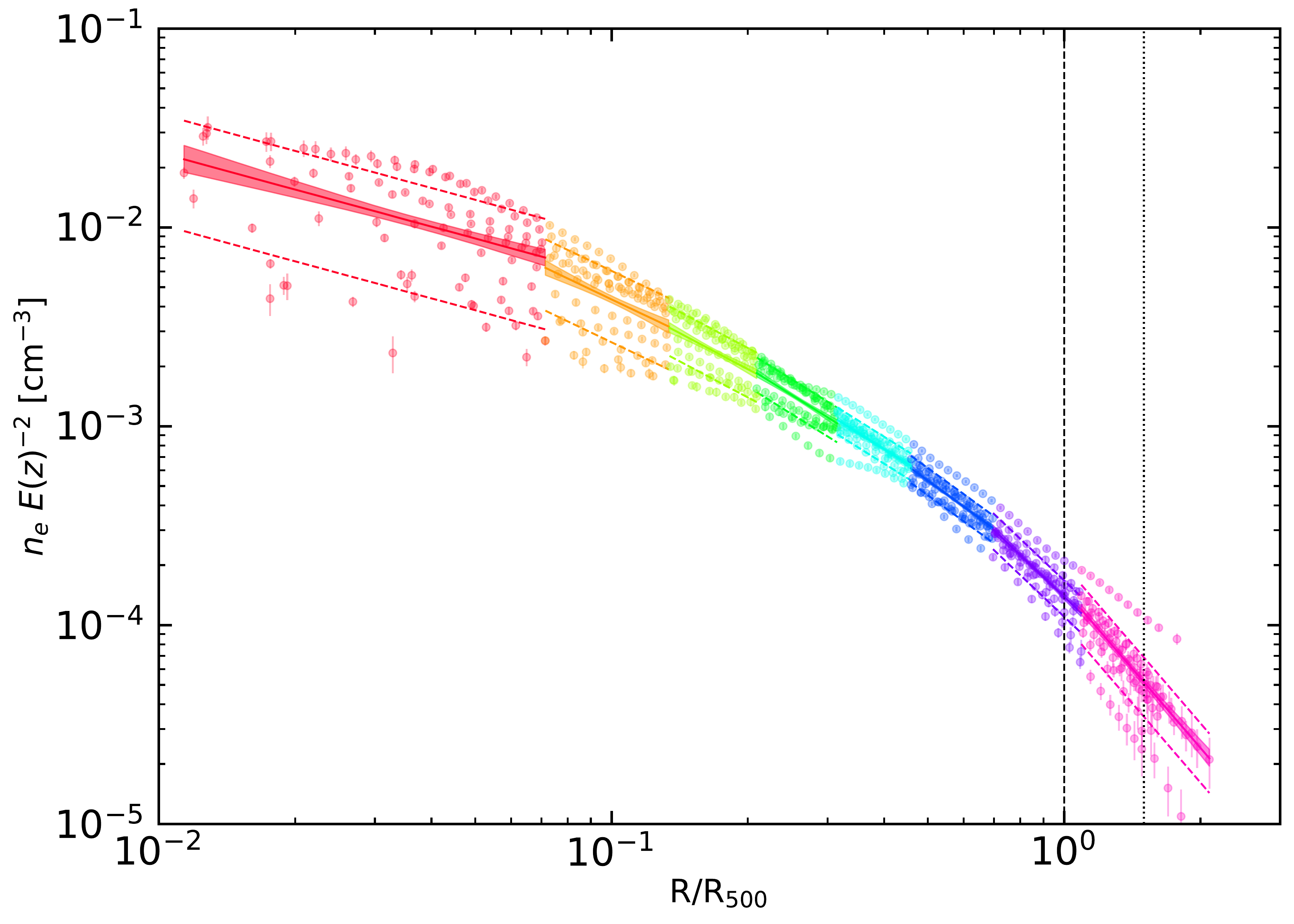}

\includegraphics[width=0.5\textwidth]{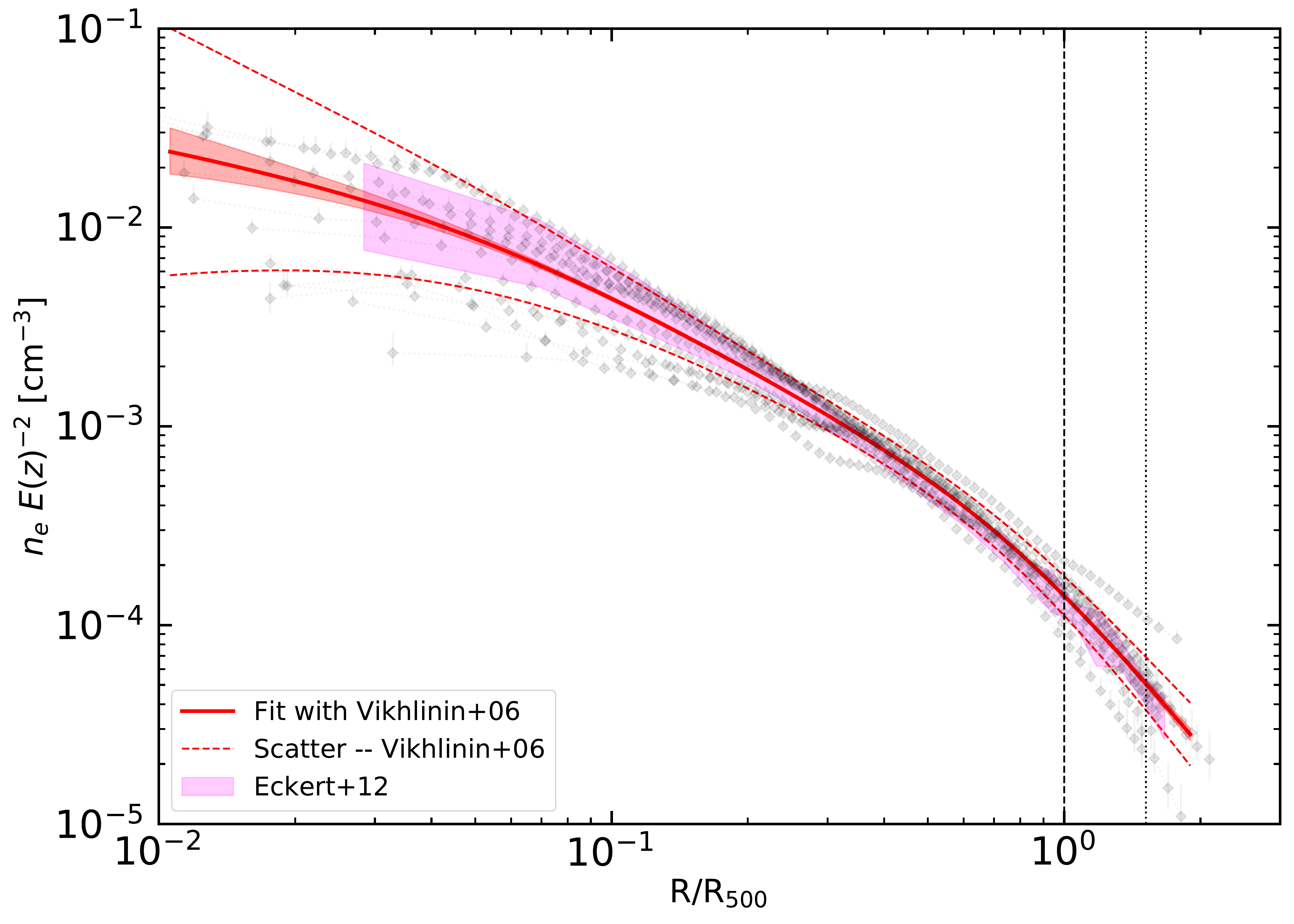}~
\includegraphics[width=0.5\textwidth]{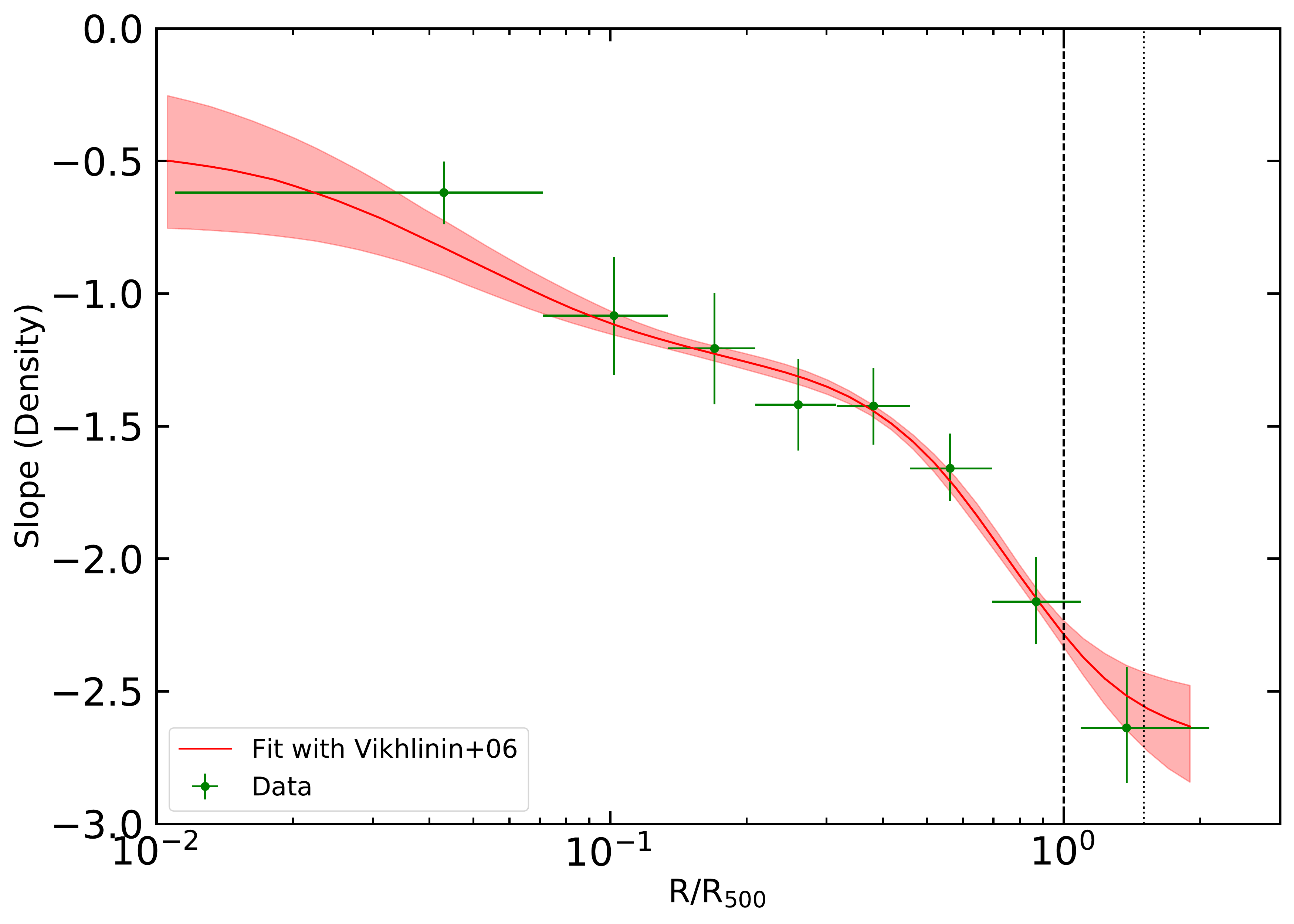}

\caption{{\it Top left panel}: Density profiles for all X-COP clusters obtained with two different deprojection methods: L1 regularization (data points) and multiscale fitting (solid lines). The  magenta shaded area represents the scatter of the median profile in \cite{eckert12}. The bottom panel represents the ratio of the two methods for each individual system. 
{\it Top right panel}: Joint fit to all the density profiles using piecewise power laws in several radial ranges (color-coded). The best fits and $1\sigma$ error envelope are shown by the solid lines, while the dashed lines represent the intrinsic scatter. {\it Bottom left panel}: Joint fit to the density profiles using the functional form introduced by \cite{vikhlini+06}, in red, with the shaded area indicating the 1$\sigma$ error envelope around the best fit. The dashed lines represent the intrinsic scatter in the functional form as a function of radius. {\it Bottom right panel}: Slope of the density profiles as a function of radius. The green data points show the results of the piecewise power law fits, whereas the red curve indicates the fit to the entire radial range using the \cite{vikhlini+06} functional form. In all panels, the vertical dashed and dotted lines represent the location of $R_{500}$ and $R_{200}$, respectively. }
\label{fig:density}
\end{figure*}

\begin{figure*}[t]
\centering
\includegraphics[width=0.5\textwidth]{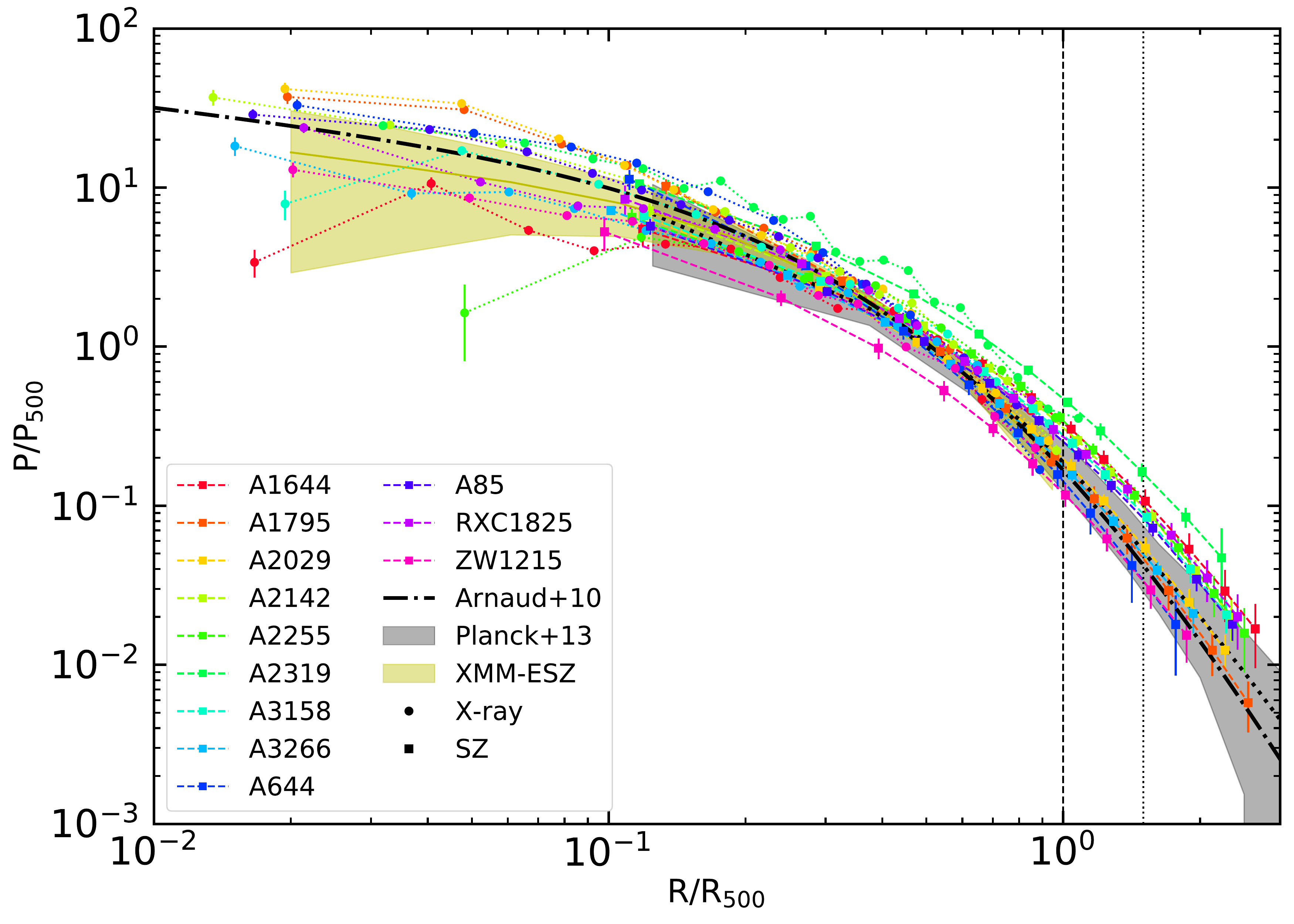}~
\includegraphics[width=0.5\textwidth]{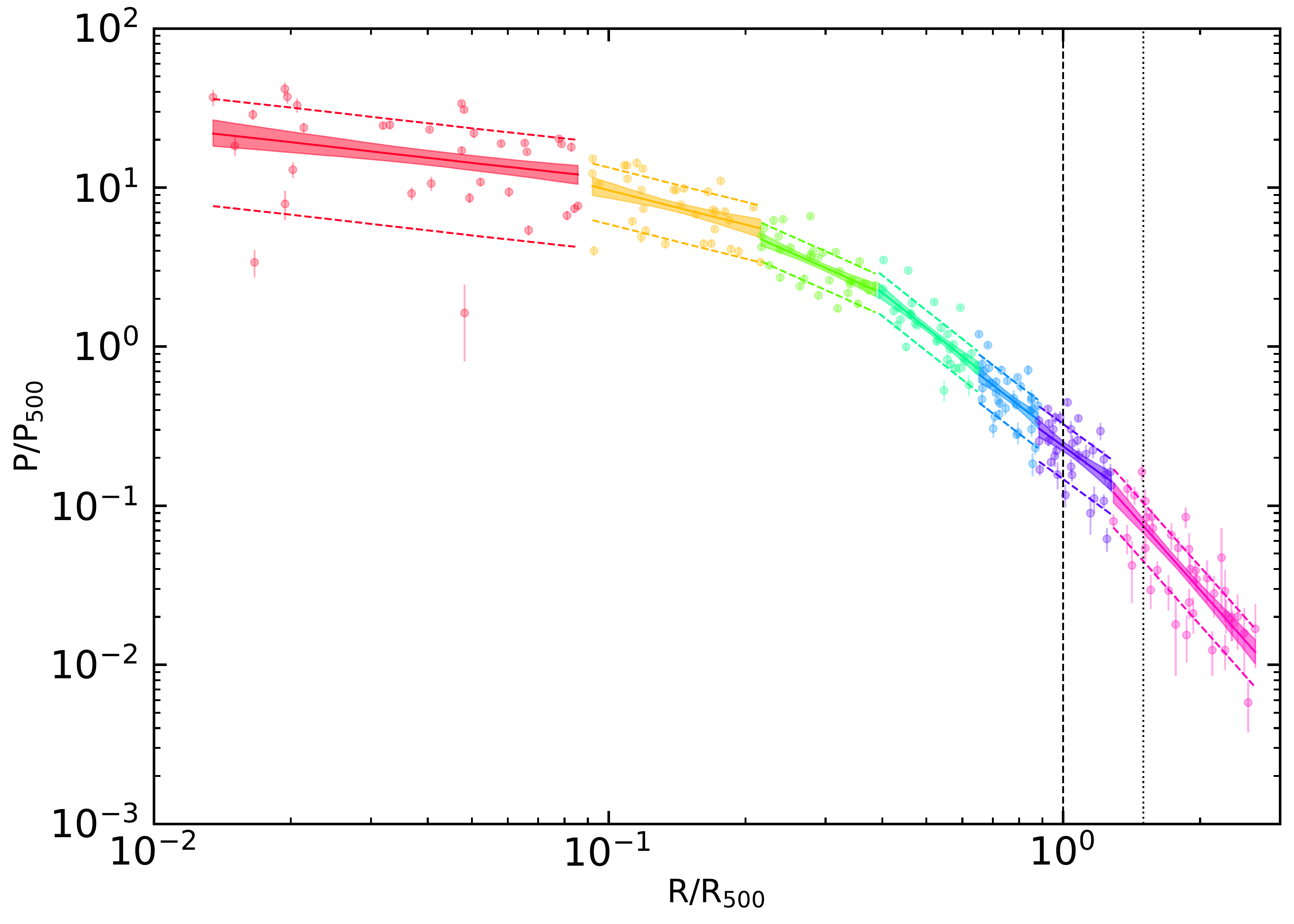}

\includegraphics[width=0.5\textwidth]{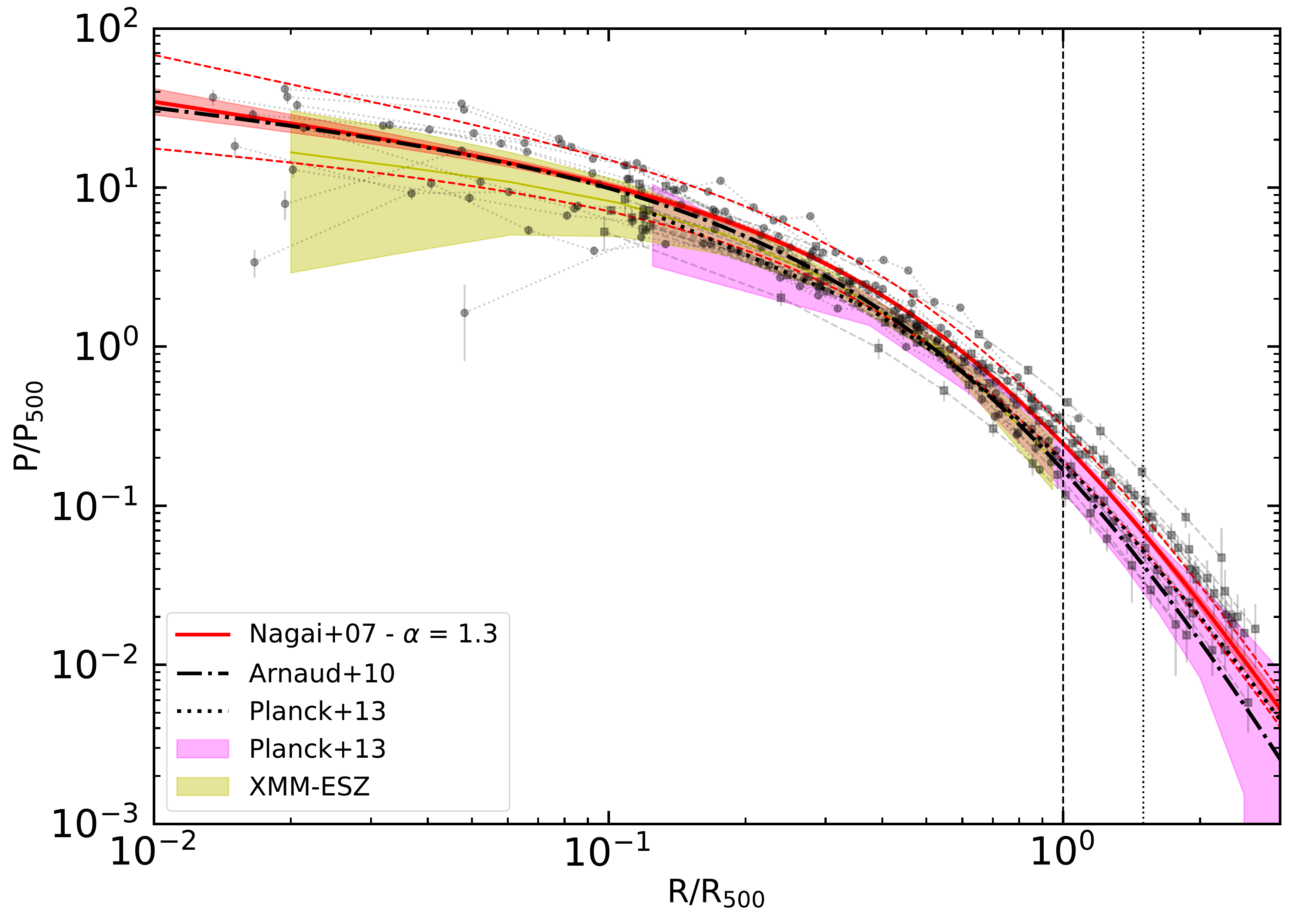}~
\includegraphics[width=0.5\textwidth]{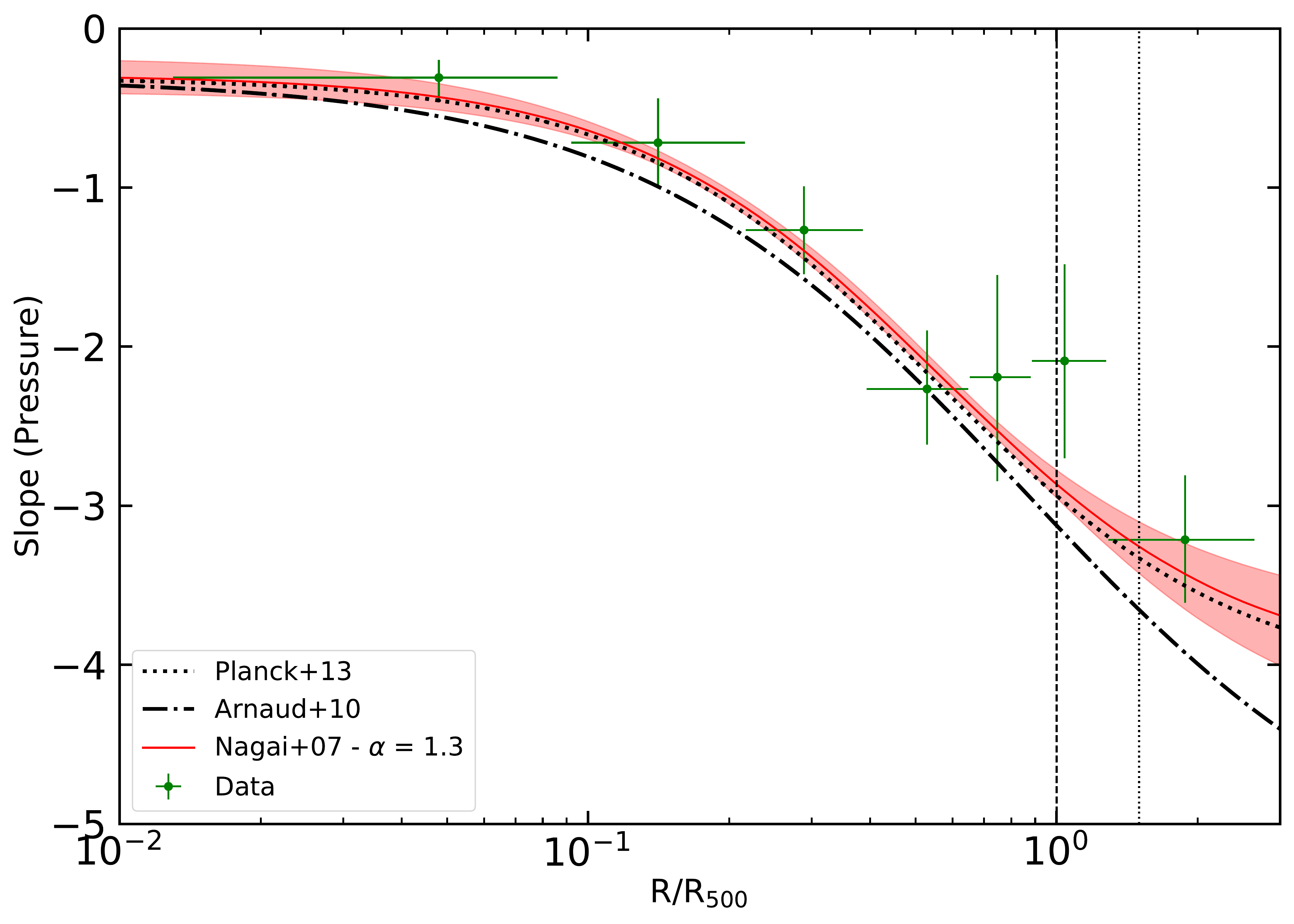}

\caption{Same as Fig. \ref{fig:density} for the pressure profiles rescaled by the self-similar quantity $P_{500}$ (Eq. \ref{eq:P500}). The squares indicate data points obtained from the deprojection of the SZ signal, while the filled circles are computed by combining the X-ray gas density profiles with the spectroscopic temperature. The solid red curve in the bottom panels shows the joint best fit to the data with the generalized NFW functional form \citep[see Eq. \ref{eq:nagai}]{nagai+07}. In all these plots the dotted and dashed-dotted lines represent the result of \cite{planck+13} and \cite{arnaud+10} respectively. The shadow areas represent the envelope obtained by \cite{planck+13} { and the Early release SZ sample\citep[XMM-ESZ,][]{PESZ}.}  }
\label{fig:pressure}
\end{figure*}

\begin{figure*}
\includegraphics[width=0.5\textwidth]{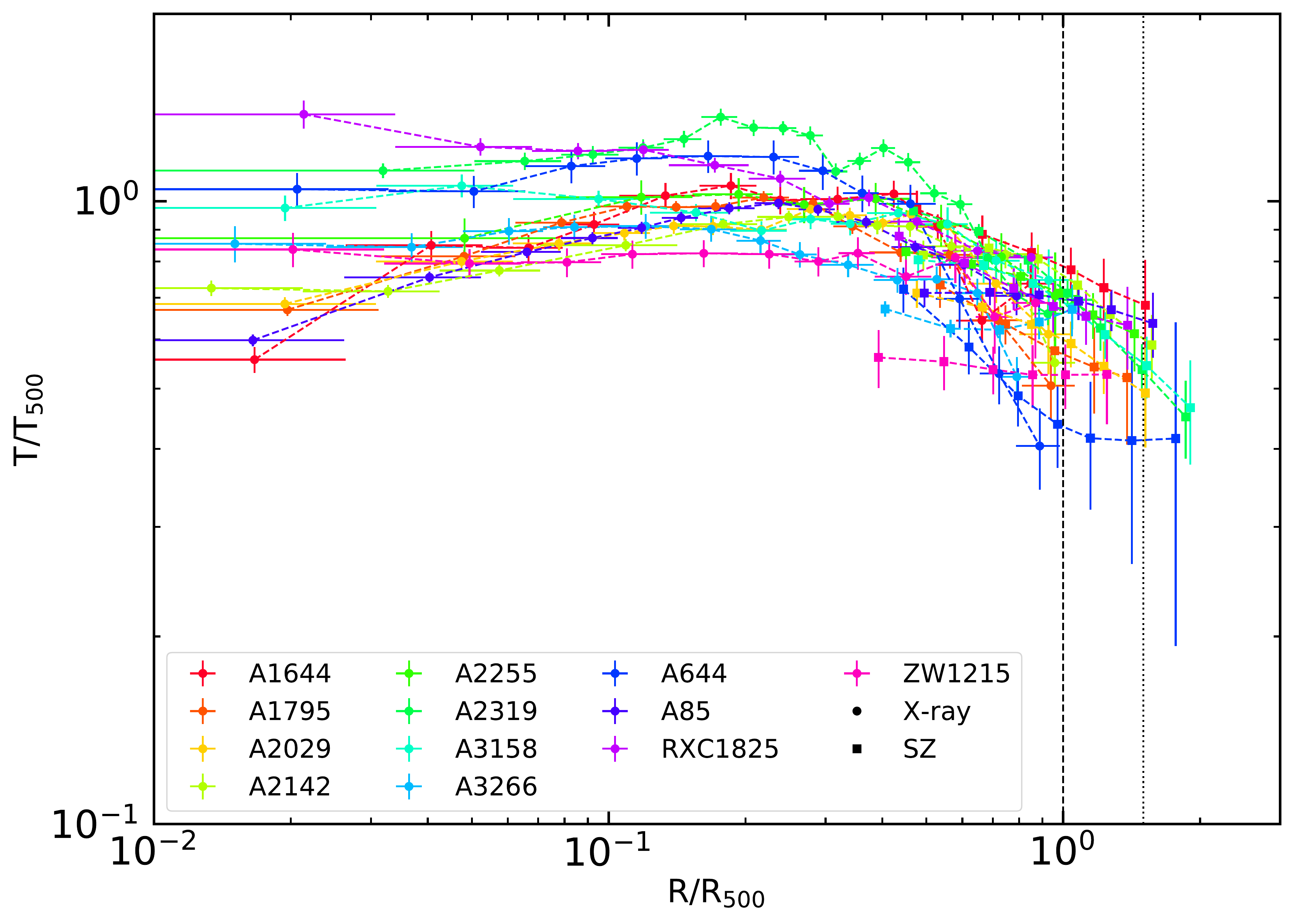}~
\includegraphics[width=0.5\textwidth]{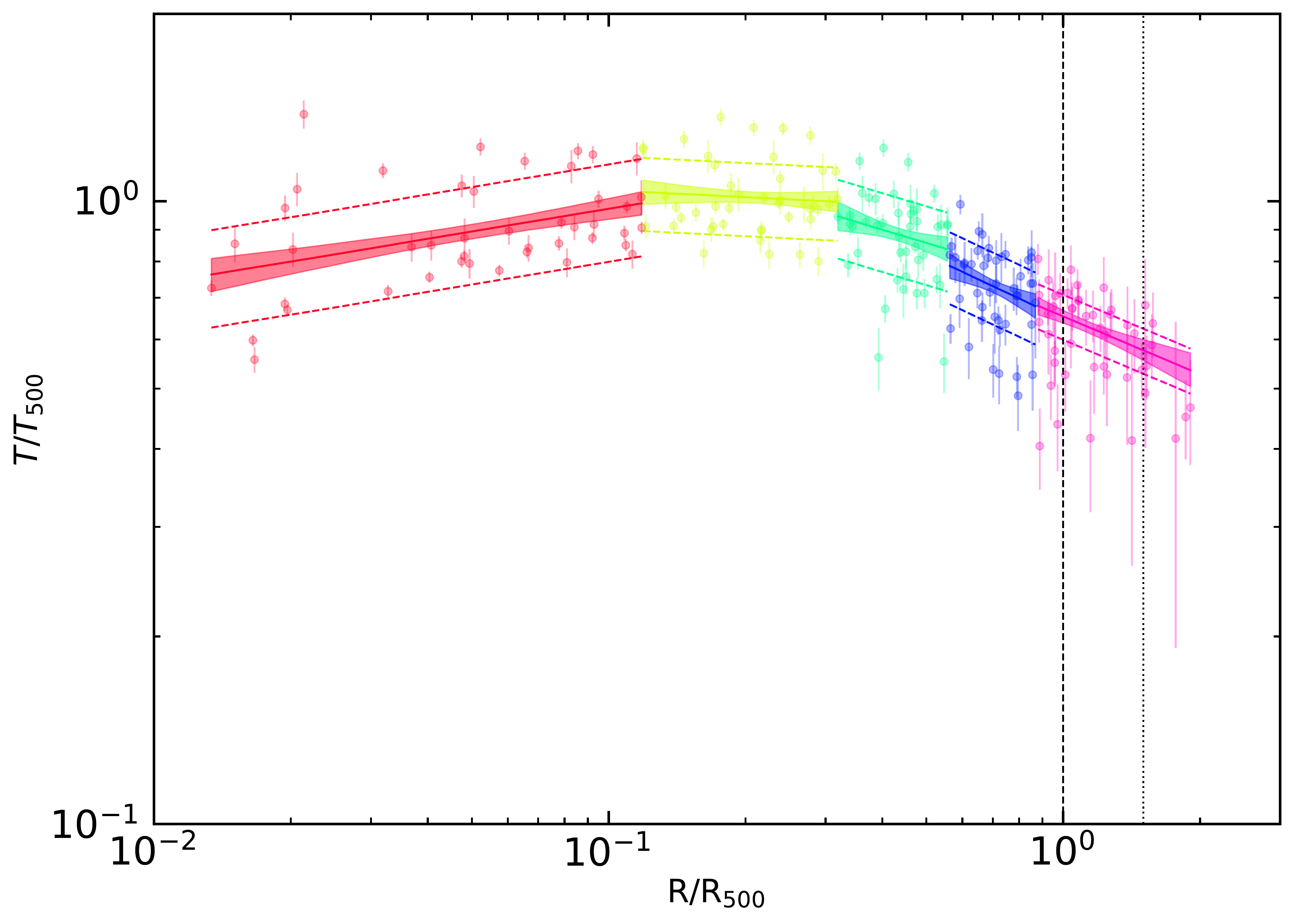}

\includegraphics[width=0.5\textwidth]{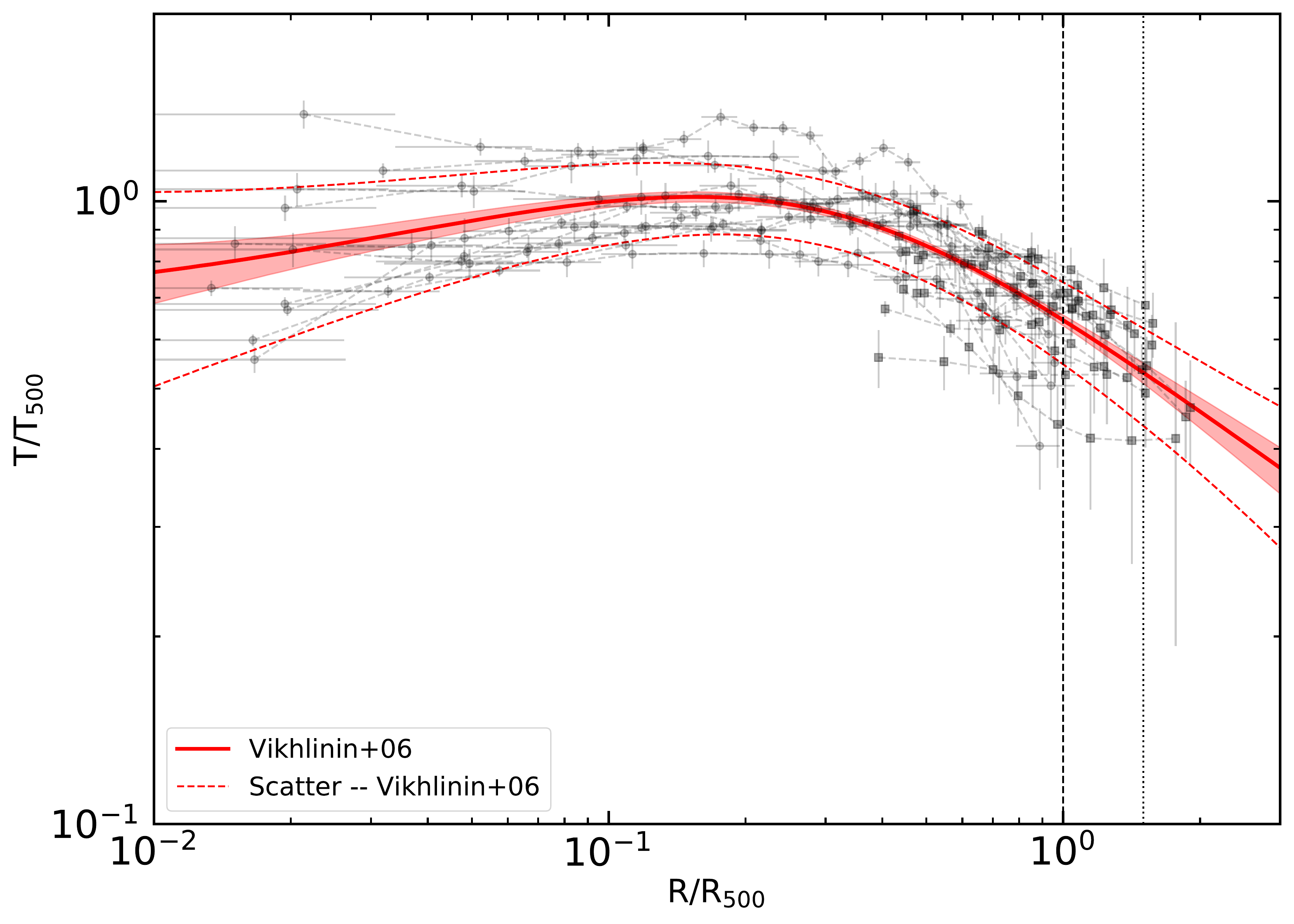}~
\includegraphics[width=0.5\textwidth]{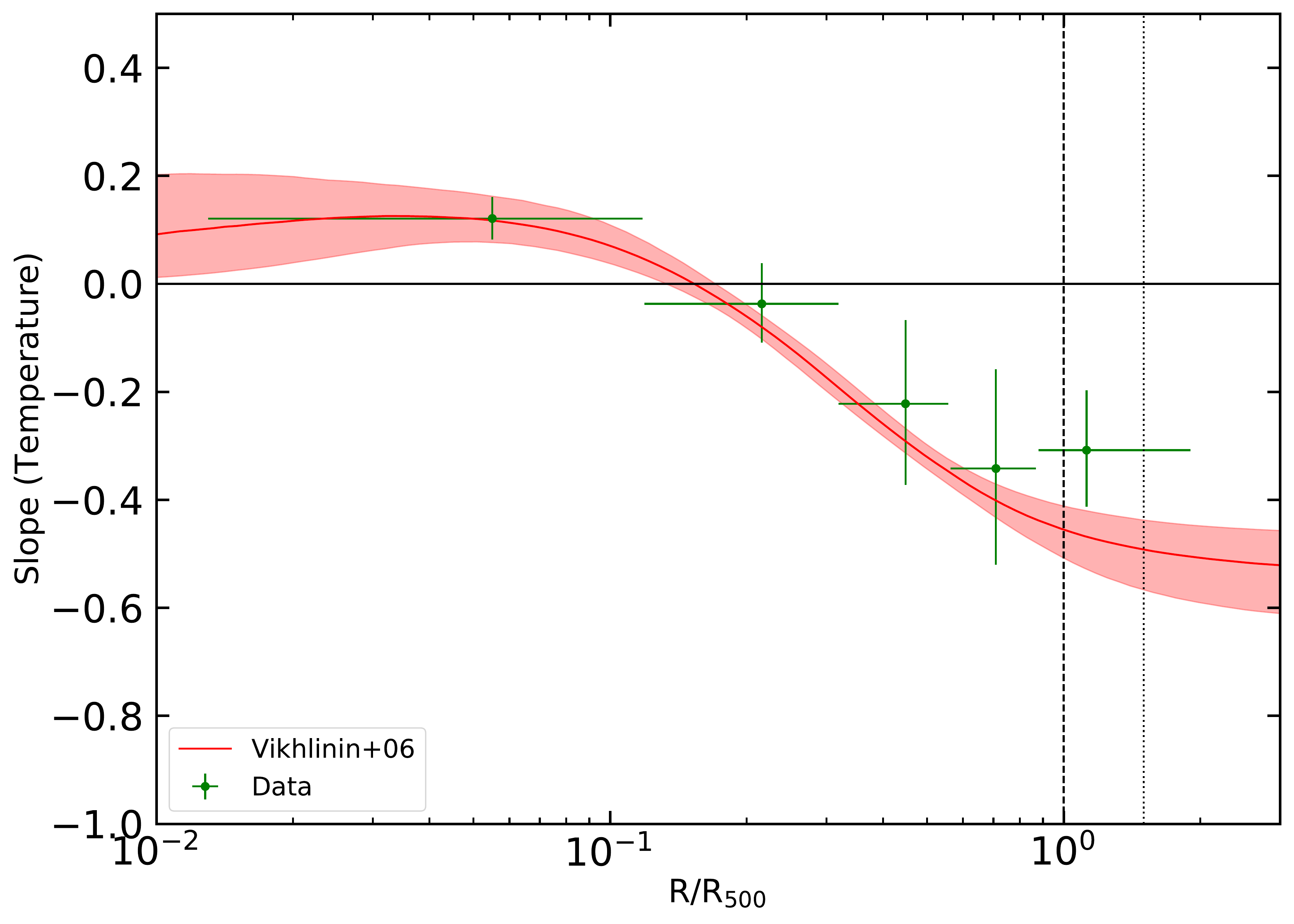}

\caption{Same as Fig. \ref{fig:density} for the projected temperature profiles rescaled by the self-similar quantity $T_{500}$ (Eq. \ref{eq:T500}). The filled circles show the measurements of the X-ray spectroscopic temperature (see Sect. \ref{sec:spec}), whereas the filled squares indicate the data points obtained by combining the SZ pressure with the gas density, projected along the line of sight assuming the spectroscopic-like scaling of \citet{mazzotta+04}. The solid red curves in the  bottom panels show the best fit to the joint dataset with the functional form introduced by \citet{vikhlini+06} (see Eq. \ref{eq:vikh_temperature}).}
\label{fig:temperature}
\end{figure*}

\begin{figure*}
\includegraphics[width=0.5\textwidth]{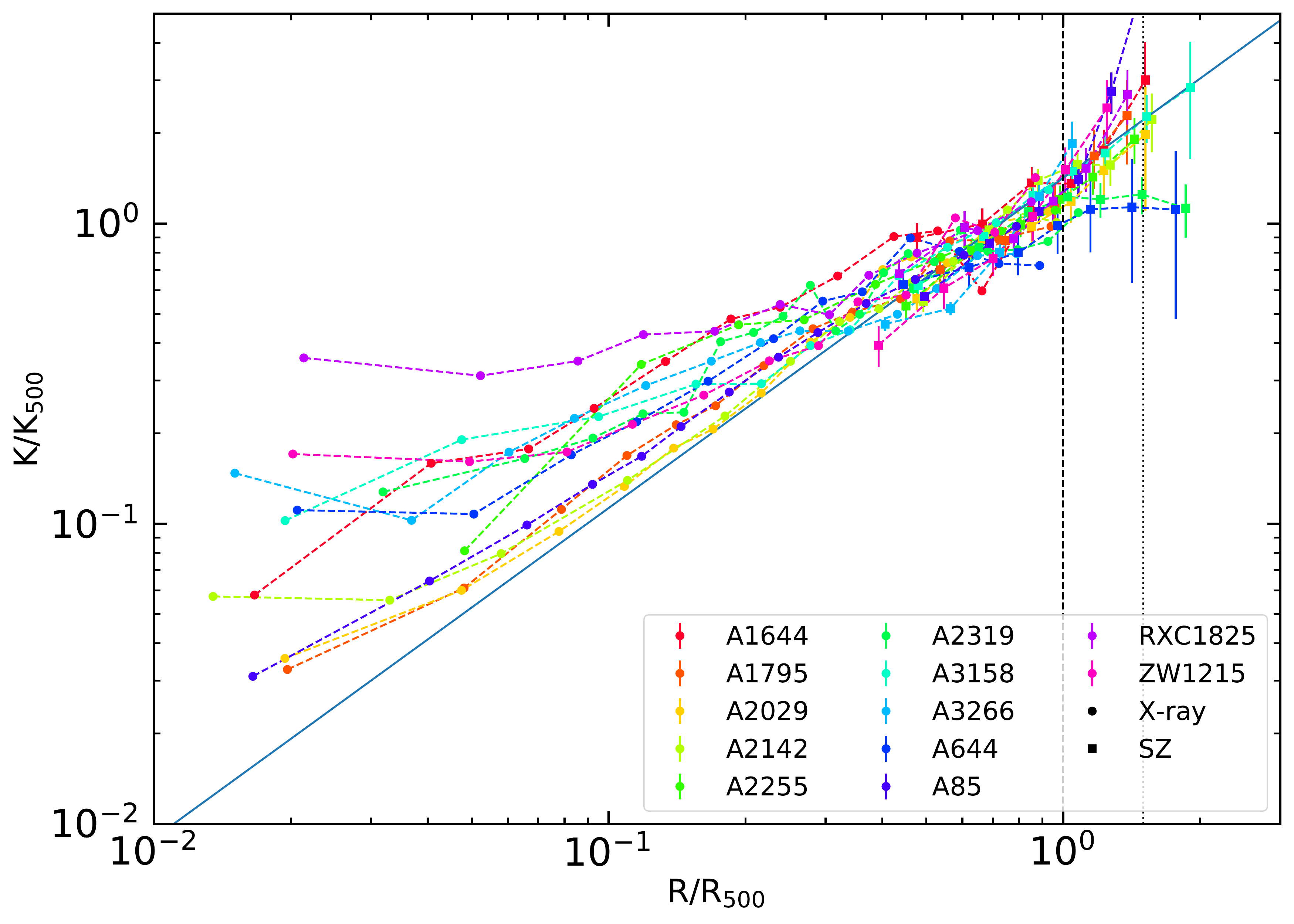}~
\includegraphics[width=0.5\textwidth]{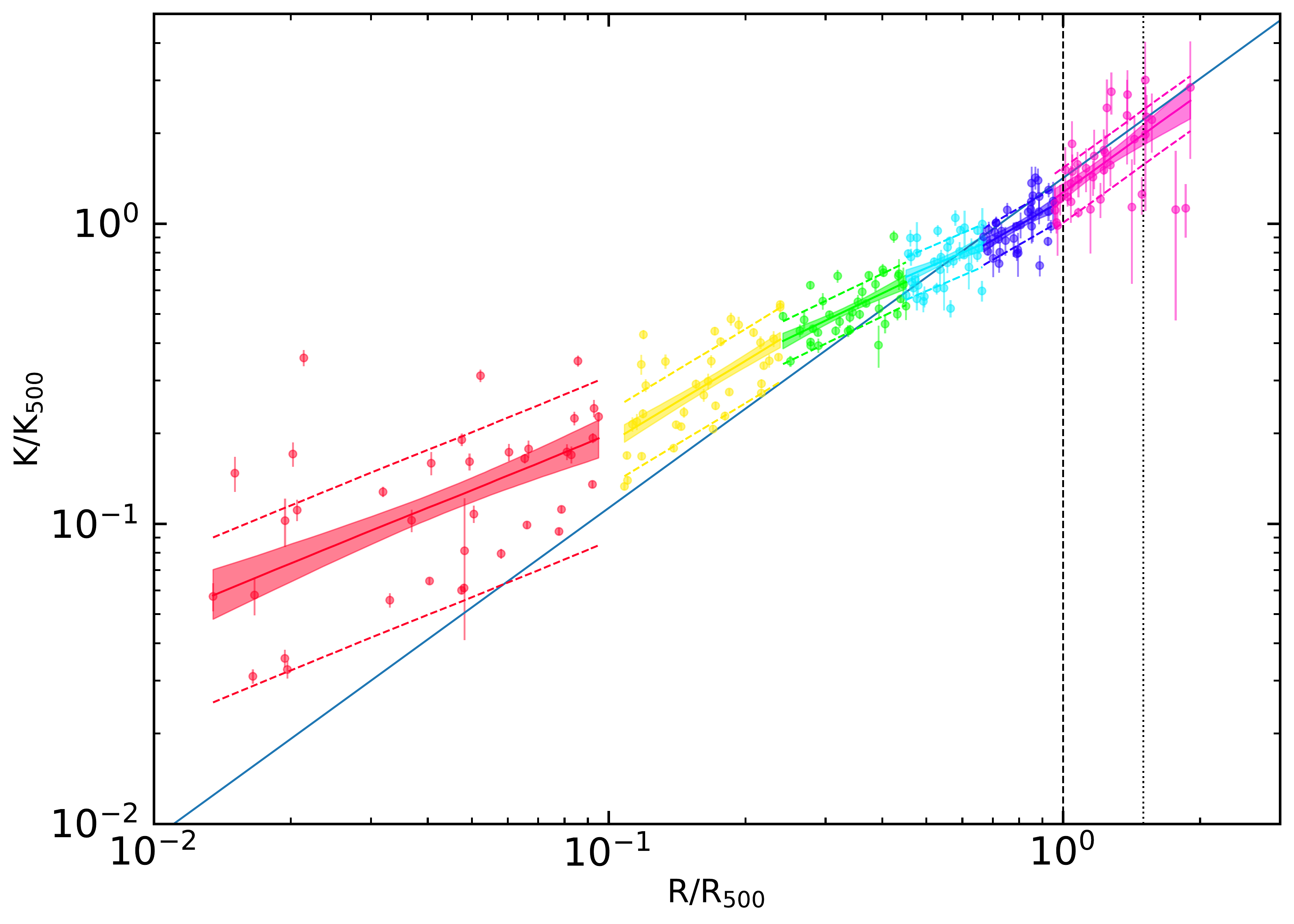}

\includegraphics[width=0.5\textwidth]{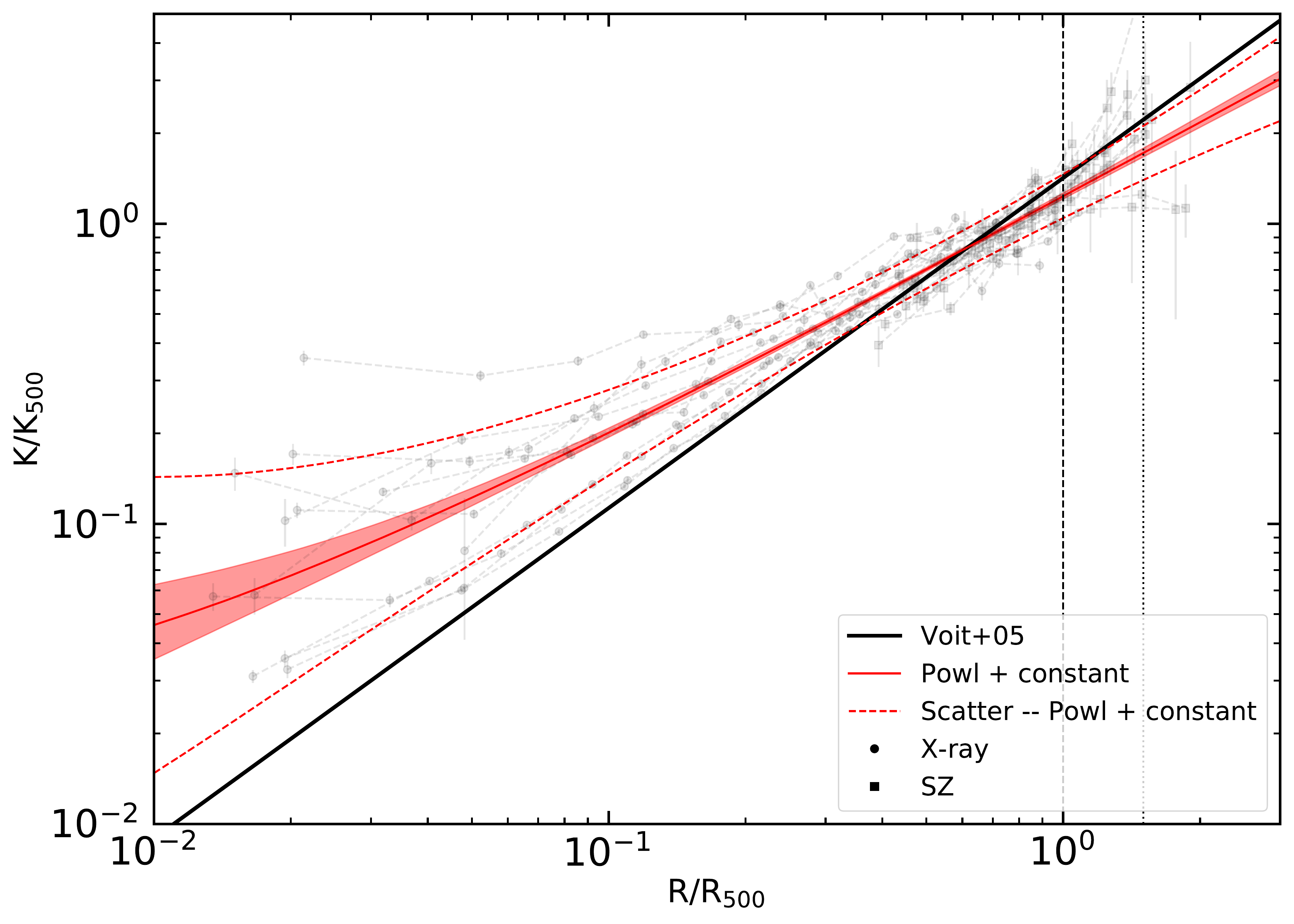}
\includegraphics[width=0.5\textwidth]{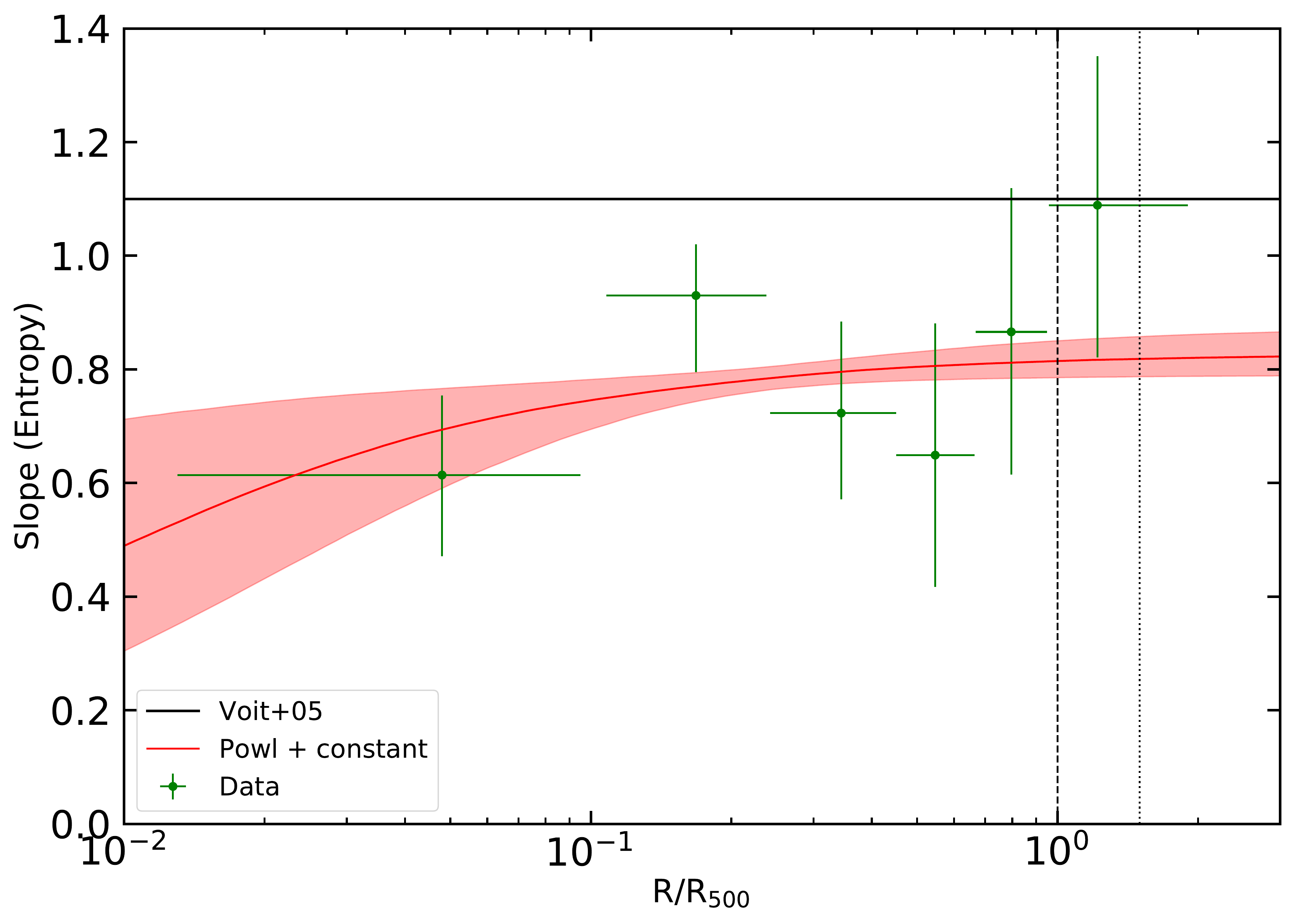}

\caption{Same as Fig. \ref{fig:density} for the entropy profiles rescaled by the self-similar quantity $K_{500}$ (Eq. \ref{eq:K500}). The filled circles show the measurements obtained from the X-ray spectroscopic temperature as $K=k_{B}T/n_{e}^{2/3}$, whereas the filled squares indicate the data points obtained by combining the SZ pressure with the gas density as $K=P_{e}/n_{e}^{5/3}$. The solid red curves in the  bottom panels indicate the best fit to the entire population with the functional form presented in Eq. \ref{eq:powlc_entropy}, whereas the solid black line shows the prediction of pure gravitational collapse \citep{voit+05}.}
\label{fig:entropy}
\end{figure*}

\subsection{Mass estimates}

To estimate scale radii and self-similar scaling quantities, we use the high-precision hydrostatic mass reconstructions presented in Ettori et al. (2018). The mass models were obtained by combining X-ray and SZ information for each individual system and solving the hydrostatic equilibrium equation. For the present work, we adopt as our reference mass model the \emph{backward NFW} results, which were obtained by assuming that the mass profile follows a Navarro-Frenk-White shape \citep{nfw96} with scale radius and concentration $c_{200}$ as free parameters. This method was shown to provide the best representation of the data (Ettori et al. 2018) and at $R_{500}$ it matches the results obtained without assuming a functional form for the mass profile with an accuracy of $\sim5\%$. Comparison of our mass reconstruction with weak lensing and SZ estimates (Ettori et al. 2018) and convergence toward the expected universal gas fraction (Eckert et al. 2018) show that our masses and the corresponding values of $R_{500}$ are accurate at the $10\%$ and $3\%$ level, respectively. For more details on the reconstruction of hydrostatic masses and estimates of systematic uncertainties we refer to Ettori et al. (2018) and Eckert et al. (2018).

\section{Thermodynamic properties}
\label{sec:thermo}

In the following section we describe how we derived the universal profiles, slopes, and intrinsic scatter of all our thermodynamic variables. We then present our main results and provide best fitting functional forms describing the X-COP cluster population.

\subsection{Fitting procedure}

We adopt two different approaches to fit the thermodynamic properties:

\begin{itemize}
\item \textbf{Piecewise power law fits: } In this case, we split our data into several radial ranges as a fraction of $R_{500}$ and we approximate the global behavior of the population in each range as a power law with free log-normal intrinsic scatter
\begin{equation}
\frac{Q(x)}{Q_{500}} = A \cdot  x^B \cdot \exp(\pm \sigma_{\rm int})
\label{eq:powl}
\end{equation}
with $x = R/R_{500}$; $Q/Q_{500}$ the rescaled thermodynamic quantity at an overdensity of 500; and $A$, $B$, and $\sigma_{\rm int}$ the normalization, slope, and intrinsic scatter in the radial range of interest. The values of $Q_{500}$ are computed adopting the virial theorem as in \cite{voit+05}, and are shown in Eq.~\eqref{eq:P500}, \eqref{eq:T500}, and \eqref{eq:K500}. The fitting procedure thus has three free parameters ($A$, $B$, and $\sigma_{\rm int}$) in each of the chosen radial ranges. We note that this procedure provides model independent measurements of the slope and intrinsic scatter at different radii. 

\item \textbf{Global functional forms:} In this case, we describe the radial dependence of the thermodynamic quantity of interest throughout the entire radial range with a parametric functional form found in the literature
\begin{equation}
\frac{Q(x)}{Q_{500}} = f(x) \cdot \exp\left[\pm \sigma_{\rm int} (x)\right]
\label{eq:general}
\end{equation}
with $Q/Q_{500}$ the rescaled thermodynamic quantity at
an overdensity of 500, as in Eq.~\eqref{eq:powl}, and $f(x)$ the chosen functional form for the thermodynamic quantity $Q$. In this case, since we model the whole radial range covered by our measurements, we allow the intrinsic scatter to vary with radius following a quadratic functional form to model the radial dependence of the intrinsic scatter
\begin{equation}
\sigma_{\rm int} (x) = \sigma_1  \log^2 \left( \frac{x}{x_0} \right) + \sigma_0
\label{eq:sigma_int_x}
\end{equation}
with $\sigma_1$ the width of the log-parabola, and $x_0$ and $\sigma_0$  respectively the location and the intercept of the minimum of the log-parabola. A total of $n+3$ parameters are allowed to vary during the fitting procedure, with $n$ the number of parameters of the adopted functional form $f(x)$. Optimizing jointly for the parameters of the intrinsic scatter profile allows us to determine the shape of $\sigma_{\rm int}(x)$.

\end{itemize}

In Figs. ~\ref{fig:density}--~\ref{fig:entropy} we show our rescaled thermodynamic quantities for the gas density, pressure, temperature, and entropy. The best fits with piecewise power laws and using functional forms are presented as well, together with the slopes for the parametric and non-parametric cases. 

\subsection{Density}
\label{sec:density}

We rescaled our density profiles by the self similar quantities, $E^2(z)=\Omega_m(1+z)^3+\Omega_\Lambda$ and $R_{500}$ for density and radius, respectively. In Fig. \ref{fig:density} we compare our scaled gas density profiles with the ``universal'' density profiles from \cite{eckert12} from a sample of 31 clusters with available \emph{ROSAT}/PSPC pointed data. We observe  excellent agreement with their results; A2319 is the only exception that deviates at large radii, as shown in \citet{ghirardini18}, because of its large non-thermal pressure support.

In the top right panel of Fig.~\ref{fig:density} we show our density profiles fitted with piecewise power laws in several radial ranges, { we show the best fitting parameters in Table~\ref{tab:piecewise}}. We parametrize the behavior of our density profiles over the whole radial range by adopting the \cite{vikhlini+06} functional form
\begin{equation}
f^2(x) = n_e^2(x)= n_0^2 \frac{(x/r_c)^{-\alpha}}{(1+x^2/r_c^2)^{3\beta-\alpha/2}} \frac{1}{(1+x^\gamma/r_s^\gamma)^{\epsilon/\gamma}}
\label{eq:vikh_density}
\end{equation}
with $x = R/R_{500}$ and $\gamma = 3$ fixed. The form thus has six free parameters ($n_0$, $r_c$, $\alpha$, $\beta$, $r_s$, and $\epsilon$) and is able to reproduce both the core and the outer parts of the density profile. We apply flat priors in logarithmic space to $n_0$, $r_c$, and $r_s$ and flat priors in linear space to the remaining parameters, constraining $\epsilon < 5$ (we specify the priors adopted in Table~\ref{tab:forms}). We show the posterior distributions of the parameters of this functional form in Fig.~\ref{fig:posterior_density_vikh} and the covariance between them.

The resulting profile is consistent at all radii with the universal envelope computed by \citet{eckert12}. We can see in all panels how the profiles become progressively less scattered going toward the outskirts. While the core is affected by a large scatter likely caused by cooling, AGN feedback and different merger states, the profiles show a high degree of self-similarity in the radial range $[0.3-1]R_{500}$. { Then in the outskirts the scatter increases again, likely caused by different accretion rates from one system to another.} We note from the plot in bottom right panel in Fig.~\ref{fig:density} that the slope of the density profiles steepens steadily with radius. The slopes computed from the piecewise power law fits and from the global fit with Eq. \ref{eq:vikh_density} agree within 1$\sigma$ at all points. Again, this result  agrees with the findings of \citet{eckert12} and \citet{morandi15}, but is at variance with the relatively flat slopes reported in several clusters observed by \emph{Suzaku}. For instance, several papers report density slopes flatter than $-2.0$, for example $-1.7$ in the outskirts of the Perseus cluster \citep{urban14} or even as low as $-1.2$ in A1689 \citep{kawa10} and Virgo \citep{simi17}. These measurements are clearly in tension with the slope of $-2.5$ at $R_{200}$ measured here for the X-COP cluster population. It must be noted, however, that thanks to  the azimuthal median technique, our gas density profiles are essentially free of the clumping effect, whereas the \emph{Suzaku} studies could not properly excise overdense regions because of the lower resolution of the instrument and/or observations performed along narrow arms.

\subsection{Pressure}
\label{sec:pressure}
Pressure in galaxy clusters is usually the smoothest thermodynamic quantity along the azimuth, if the cluster is not affected by an ongoing merger.%, as it is the most directly related to the gravitational potential of the underlying halo. 

We recover the gas pressure both through the combination of X-ray gas density and spectral temperature and through the direct deprojection of the SZ effect. In the former case, we deproject the spectral X-ray temperature \citep[e.g.,][see Sect. \ref{sec:spec}]{mazzotta+04} and combine the deprojected temperature with the gas density interpolated on the same grid to infer the pressure $P_X=k_BT_X\times n_e$. In the latter we recover the pressure directly from the \emph{Planck} data by deprojecting the measured $y$ profiles (see Sect.~\ref{s:plck}) from which we exclude the first three points from the analysis. We thus combine the higher resolution and precision of \xmm\ in the inner region with the high quality of the \planck\ data at $R_{500}$ and beyond, which allows us to constrain the shape and intrinsic scatter of the pressure profiles in the radial range $[0.01-2.5]R_{500}$.
%\begin{equation}
%y_{SZ}(r)=\frac{\sigma_T}{m_ec^2}\int P_e(\ell)d\ell 
%\label{eq:pressure_from_compton}
%\end{equation}
%For details on the method used to pass from comptonization parameter map to pressure profile we refer to \cite{hurier+13,planck+13,ghirardini18} and references therein.
We rescale our pressure profiles by the self-similar quantities at an overdensity of 500,
\begin{multline}
P_{500} = 3.426 \times 10^{-3} \text{\ keV} \text{\ cm}^{-3} \left(\frac{ M_{500} }{ h_{70}^{-1} 10^{15} M_\odot } \right)^{2/3} E(z)^{8/3} \cdot \\ \cdot \left(\frac{f_b}{0.16} \right) \left( \frac{\mu}{0.6} \right) \left( \frac{\mu_e}{1.14} \right)
\label{eq:P500},
\end{multline}
where $f_{b}$ is the Universal gas fraction, which we take to be $\Omega_b/\Omega_m = 0.16$ \citep[][rounding the number to 2 significant figures]{planck+16}, $\mu$ and $\mu_e$ are the mean molecular weight per particle and mean molecular weight per electron for which we adopt the values measured by \cite{ag89}. In Fig.~\ref{fig:pressure} we show the scaled pressure profiles of our 12 objects obtained through X-ray and SZ measurements of the ICM. We note that our profiles agree with the results obtained by the \textit{Planck} Collaboration for a sample of 62 clusters \citep{planck+13}, falling well within the two envelopes, with the exception, as in the case of the density, of A2319 \citep[see the discussion in][]{ghirardini18}. 

Similarly to the density, we fitted the profiles using piecewise power laws in several radial ranges, also obtaining  in this case a scatter that decreases with radius {out to $R_{500}$ and then (as for the density) that increases in the outskirts}; these profiles become progressively steeper with radius{, see Table~\ref{tab:piecewise}}. Our profiles in the outskirts are compatible with the results of the \textit{Planck} Collaboration, both for the central value and the slope. We also fitted  our data using the generalized NFW functional form introduced by \citet{nagai+07}
\begin{equation}
f(x) = \frac{P(x)}{P_{500}} = \frac{P_0}{(c_{500}x)^\gamma [1+(c_{500}x)^\alpha]^{\frac{\beta-\gamma}{\alpha}}},  
\label{eq:nagai}
\end{equation}
where $x = R/R_{500}$, with five free parameters $P_0$, $c_{500}$, and three slopes, $\gamma$, $\alpha$, and $\beta$ representing respectively the inner, intermediate, and outer slopes (we specify the priors adopted in Table~\ref{tab:forms}).
Since the parameters are strongly degenerate, it was advised  to fix at least one of the slopes \citep{arnaud+10}; therefore, we fixed the central slope $\alpha$ to the best fit value of 1.3 estimated by the \textit{Planck} Collaboration \citep{planck+13}. The resulting best fit, the intrinsic scatter around the median profile, and the slope computed from the fit are shown in Fig.~\ref{fig:pressure}. The posterior distributions of the parameters and the covariance between them are shown in Fig.~\ref{fig:posterior_pressure_nagai_fix1}. 

Similar to the case of the gas density, we find that the slope of the profiles steepens steadily with radius, as expected for a gas in hydrostatic equilibrium within a NFW potential. The best fit with the generalized NFW functional form does an excellent job of reproducing the slopes estimated from the piecewise power law fits. 

In the range where pressure measurements are available both from \xmm\ and \planck, we find  excellent agreement between the two (see Appendix~\ref{app:XvsSZ}) even though the pressure profiles were obtained using completely independent probes. This shows that X-ray and SZ observations provide a consistent picture of the state of the ICM and gives us confidence that systematics in our measurements are small and under control.

\subsection{Temperature}
\label{sec:temperature}

Temperature profiles in X-ray studies are usually obtained by performing spectral fitting in concentric annuli (see Sect. \ref{sec:spec}),  called the spectroscopic temperature. In addition, we also use our \planck\ SZ pressure profiles and combine them with the X-ray density profiles to obtain gas-mass-weighted temperatures $T_{gmw}=P_{SZ}/n_{e}$, which are then projected \citep[using the X-ray emissivity as weight, as in ][]{mazzotta+04} on the plane of the sky and overplotted  on the spectroscopic temperatures. Our X-ray and SZ measurements of pressure and temperature cover different radial ranges. X-ray spectroscopy probes the temperature of the gas within $R_{500}$, while SZ probes temperatures from $0.7 R_{500}$ to $2R_{500}$ (excluding the first three SZ data points), which highlights the complementarity of the two ICM diagnostics.
In Fig.~\ref{fig:temperature} we show our 2D spectral temperature profiles rescaled by $T_{500}$, defined as
\begin{equation}
T_{500} = 8.85 \text{\ keV} \left(\frac{ M_{500} }{ h_{70}^{-1} 10^{15} M_\odot } \right)^{2/3} E(z)^{2/3} \left( \frac{\mu}{0.6} \right)
\label{eq:T500}
\end{equation}

While density and pressure change by three to four orders of magnitude going from the center to the outskirts of the cluster, temperature shows much milder variations. In particular, it is almost constant out to $\sim0.5 R_{500}$, and then declines beyond this point.  

In Fig.~\ref{fig:temperature} we show the results of the piecewise power law fits in several radial ranges. We perform a global fit to the temperature profiles with the functional form described in \citet{vikhlini+06}, which is able to describe the temperature profiles of both the core and the outer parts of galaxy clusters
\begin{equation}
f(x) = \frac{T(x)}{T_{500}}  = T_0 \frac{\frac{T_{min}}{T_0} + \left( \frac{x}{r_{cool}} \right)^{a_{cool}}}{1+\left( \frac{x}{r_{cool}} \right)^{a_{cool}}}\frac{1}{\left(1+\left( \frac{x}{r_t} \right)^2 \right)^{\frac{c}{2}}}
\label{eq:vikh_temperature}
\end{equation}
with $x = R/R_{500}$, and six free parameters: $T_0$, $T_{min}$, $r_{cool}$, $a_{cool}$, $r_t$, and $c$ (we specify the priors adopted in Table~\ref{tab:forms}). The posterior distribution of these parameters and the covariances are shown in Fig.~\ref{fig:posterior_temperature_vikh}. This functional form provides an accurate description of the shape of the temperature profiles. The slopes estimated from the global fit follow the slopes measured from the piecewise power law fits at different radial ranges within $1\sigma$ at every radius. 

The average slope of the temperature profiles is slightly positive in the central regions because of effects due to cooling, especially in cool-core clusters. Beyond $\sim0.5R_{500}$ the slope remains relatively flat at a value of $-0.3$. This value is consistent with the slopes measured inside $R_{500}$ from \xmm\ and \emph{Chandra} data \citep{lm08,pratt07}, but it is flatter than the typical slopes measured in \emph{Suzaku} data. From a collection of a dozen clusters with published \emph{Suzaku} temperature profiles, \citet{reiprich13} report that the universal shape of the profiles can be well described by the form $T/\langle T\rangle=1.19-0.84(R/R_{200})$, i.e., the data are consistent with a slope of -1.0 in the outskirts, which is much steeper than the results presented here. Again, the low angular resolution of \emph{Suzaku} may have prevented the authors from removing cool, overdense regions that could bias low the measured spectroscopic temperature. On the other hand, SZ pressure profiles are much less sensitive to gas clumping \citep[e.g.,][]{khedekar13,roncarelli+13} and our density profiles were corrected for the statistical effect of gas clumping (see Sect. \ref{sec:red_ima}), thus our X/SZ temperatures are closer to gas-mass-weighted temperatures \citep[see the discussion in][]{adam17}.

\subsection{Entropy}
\label{sec:entropy}

%Entropy profiles trace the formation history of the ICM, as the radius at which an infalling gas particle will eventually reside in the potential well of the main halo is determined by its entropy.
Entropy  traces the thermal evolution of the ICM plasma, which can be altered via cooling/heating, mixing, and convection.
Simulations predict that in the presence of non-radiative processes only, entropy increases steadily with radius out to $\sim2\times R_{200}$, following a power law with a slope of 1.1 \citep{tozzi01,voit+05,borgani05,lau15}. The entropy profiles of the cluster population should scale self-similarly when rescaled by the quantity
\begin{multline}
K_{500} = 1667 \text{\ keV cm}^2 \left(\frac{ M_{500} }{ h_{70}^{-1} 10^{15} M_\odot } \right)^{2/3} E(z)^{-2/3} \cdot \\ \cdot \left(\frac{f_b}{0.16} \right)^{-2/3} \left( \frac{\mu}{0.6} \right) \left( \frac{\mu_e}{1.14} \right)^{2/3}
\label{eq:K500}
\end{multline}

Similarly to pressure and temperature, entropy can be recovered from X-ray-only quantities as $K=k_BT_X\times n_{e}^{-2/3}$ { (using the deprojected temperature; see Sect.~\ref{sec:spec})} or by combining SZ pressure with X-ray density as $K=P_{SZ}\times n_{e}^{-5/3}$ (ignoring the first three \emph{Planck} points). We show our scaled entropy profiles in Fig.~\ref{fig:entropy}. Our profiles very closely match  the predicted power law model beyond $0.3 R_{500}$, with just A2319 showing a significant flattening not compatible within the error bars. In the central regions our profiles flatten, with non-cool-core clusters flattening more than cool-core clusters, as already noted in numerous studies \citep[e.g.,][and references therein]{pratt+10,cavagnolo+09}.

By fitting the profile using piecewise power laws we observe a gradual steepening of the entropy slope, which becomes consistent with the predictions of gravitational collapse \citep{voit+05} beyond $\sim0.5R_{500}$, i.e.,
from a slope of $\sim 0.6$ in the core to $\sim 1.1$ in the outskirts.
As for the previous cases, we fitted our profiles throughout the entire radial range with the functional form introduced by \citet{cavagnolo+09}, which consists of a power law with a constant entropy floor
\begin{equation}
f(x) = \frac{K(x)}{K_{500}} = K_{0} + K_{1} \cdot x^\alpha
\label{eq:powlc_entropy}
\end{equation}
with $x = R/R_{500}$, and three free parameters $K_0, K_1$, and $\alpha$ (we specify the priors adopted in Table~\ref{tab:forms}). The posterior distributions of the parameters and the covariances are shown in Fig.~\ref{fig:posterior_entropy_powlc}. We note that this functional form does not provide an accurate description of the data in the outer parts of the profiles where it is not able to follow the observed gradual change in slope throughout the radial range covered. At large radii, the best fitting slope using Eq. \ref{eq:powlc_entropy} reads $\alpha=0.84\pm0.04$ (see Table \ref{tab:forms}), whereas the data prefer a slope consistent with the self-similar prediction of 1.1 beyond $0.6R_{500}$ (see Table \ref{tab:piecewise}).

\begin{table}
\caption{\label{tab:piecewise}Results of the piecewise power law fits (normalizations, slopes, and intrinsic scatter; see Eq. \ref{eq:powl}) for the various thermodynamic quantities in several radial ranges{, marked by the inner and outer rescaled radii $x_{in}$ and $x_{out}$.  $\rho_{A,B}$ is the correlation coefficient between A and B.}}
\begin{center}
{\bf Density}
\resizebox{\columnwidth}{!}{%

\begin{tabular}{ c c c c c c }
$x_{in}$ & $x_{out}$ & A ($10^{-4}\rm{cm}^{-3}$) & B (slope) & $\sigma_{int}$ & $\rho_{A,B}$\\
\hline
0.01 & 0.07 & $13.00 \pm 3.83$ & $-0.48 \pm 0.10$ & $0.38 \pm 0.04$ & 0.9884\\
0.07 & 0.13 & $3.44 \pm 1.73$ & $-1.04 \pm 0.24$ & $0.28 \pm 0.03$ & 0.9969\\
0.13 & 0.21 & $2.60 \pm 0.79$ & $-1.21 \pm 0.19$ & $0.18 \pm 0.02$ & 0.9974\\
0.21 & 0.31 & $2.85 \pm 0.72$ & $-1.16 \pm 0.21$ & $0.15 \pm 0.01$ & 0.9963\\
0.31 & 0.46 & $1.84 \pm 0.31$ & $-1.60 \pm 0.20$ & $0.15 \pm 0.01$ & 0.9938\\
0.46 & 0.72 & $1.63 \pm 0.17$ & $-1.80 \pm 0.21$ & $0.18 \pm 0.02$ & 0.9767\\
0.72 & 1.14 & $1.42 \pm 0.06$ & $-2.38 \pm 0.29$ & $0.27 \pm 0.03$ & 0.7102\\
1.15 & 2.00 & $1.53 \pm 0.19$ & $-2.47 \pm 0.31$ & $0.37 \pm 0.04$ & -0.8783\\
\end{tabular}
}

{\bf Pressure}
\resizebox{\columnwidth}{!}{%

\begin{tabular}{ c c c c c c }
$x_{in}$ & $x_{out}$ & A & B (slope) & $\sigma_{int}$ & $\rho_{A,B}$\\
\hline
0.01 & 0.09 & $5.75 \pm 2.39$ & $-0.31 \pm 0.13$ & $0.65 \pm 0.09$ & 0.9353\\
0.09 & 0.22 & $1.84 \pm 1.06$ & $-0.72 \pm 0.28$ & $0.39 \pm 0.05$ & 0.9321
\\
0.22 & 0.39 & $0.68 \pm 0.24$ & $-1.27 \pm 0.28$ & $0.27 \pm 0.04$ & 0.9550
\\
0.39 & 0.65 & $0.27 \pm 0.07$ & $-2.27 \pm 0.36$ & $0.29 \pm 0.04$ & 0.9636
\\
0.65 & 0.88 & $0.26 \pm 0.05$ & $-2.19 \pm 0.65$ & $0.34 \pm 0.05$ & 0.9371
\\
0.89 & 1.28 & $0.24 \pm 0.02$ & $-2.09 \pm 0.61$ & $0.38 \pm 0.06$ & -0.3866
\\
1.29 & 2.65 & $0.27 \pm 0.07$ & $-3.21 \pm 0.40$ & $0.40 \pm 0.07$ & -0.9363
\\
\end{tabular}
}

{\bf Temperature}
\resizebox{\columnwidth}{!}{%

\begin{tabular}{ c c c c c c }
$x_{in}$ & $x_{out}$ & A & B (slope) & $\sigma_{int}$ & $\rho_{A,B}$\\
\hline
0.01 & 0.12 & $1.29 \pm 0.16$ & $0.12 \pm 0.04$ & $0.18 \pm 0.02$ & 0.9713\\
0.12 & 0.32 & $0.96 \pm 0.11$ & $-0.04 \pm 0.07$ & $0.14 \pm 0.02$ & 0.9804
\\
0.32 & 0.56 & $0.74 \pm 0.10$ & $-0.22 \pm 0.15$ & $0.14 \pm 0.02$ & 0.9789
\\
0.56 & 0.87 & $0.65 \pm 0.04$ & $-0.34 \pm 0.18$ & $0.13 \pm 0.02$ & 0.9388
\\
0.88 & 1.90 & $0.65 \pm 0.02$ & $-0.31 \pm 0.11$ & $0.08 \pm 0.02$ & -0.4798
\\
\end{tabular}
}

{\bf Entropy}
\resizebox{\columnwidth}{!}{%

\begin{tabular}{ c c c c c c }
$x_{in}$ & $x_{out}$ & A & B (slope) & $\sigma_{int}$ & $\rho_{A,B}$\\
\hline
0.01 & 0.10 & $0.82 \pm 0.38$ & $0.61 \pm 0.14$ & $0.56 \pm 0.07$ & 0.9394\\
0.11 & 0.24 & $1.57 \pm 0.32$ & $0.93 \pm 0.11$ & $0.28 \pm 0.03$ & 0.9665
\\
0.24 & 0.45 & $1.14 \pm 0.20$ & $0.72 \pm 0.16$ & $0.16 \pm 0.02$ & 0.9783
\\
0.45 & 0.66 & $1.11 \pm 0.16$ & $0.65 \pm 0.23$ & $0.16 \pm 0.02$ & 0.9748
\\
0.67 & 0.95 & $1.20 \pm 0.08$ & $0.87 \pm 0.25$ & $0.14 \pm 0.02$ & 0.9057
\\
0.96 & 1.90 & $1.27 \pm 0.08$ & $1.09 \pm 0.27$ & $0.21 \pm 0.05$ & -0.6778
\\
\end{tabular}
}

\end{center}
\end{table}

\begin{figure}[t]
\centering
\includegraphics[width=0.5\textwidth]{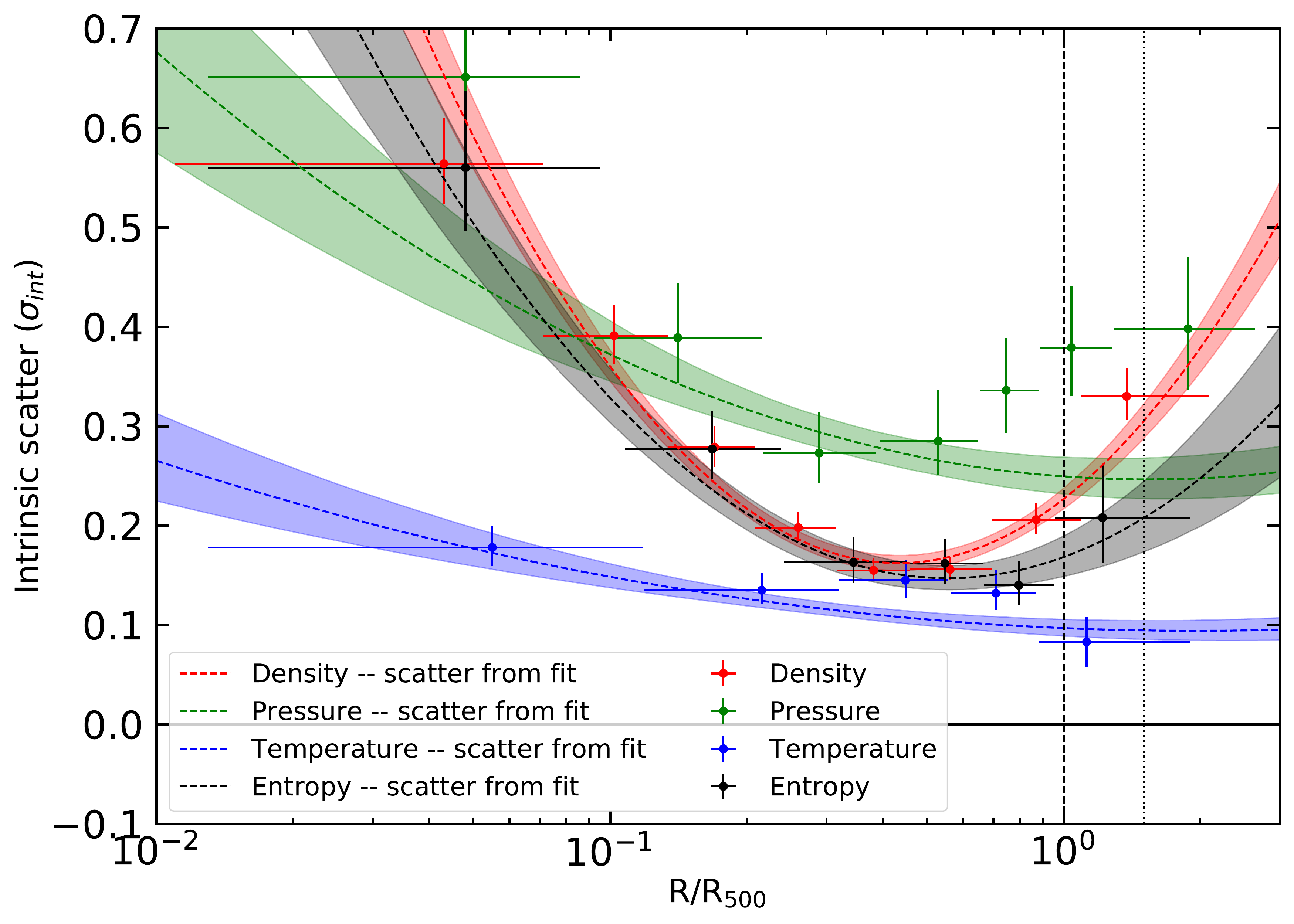}
\caption{Measured intrinsic scatter of all our thermodynamic quantities, density (red), pressure (green), temperature (blue), and entropy (black). The data points indicate the results of piecewise power law fits on several radial ranges, whereas the dashed lines and shaded areas show the intrinsic scatter described as a log-parabola (Eq. \ref{eq:sigma_int_x}) around the best fitting functional forms. }
\label{fig:scatter_all}
\end{figure}

\subsection{Scatter}
\label{sec:scatter}
The high data quality of X-COP allows us to probe the intrinsic scatter of our profiles as a function of radius. The piecewise fit using power laws allows us to measure the scatter in a nearly model independent way, whereas the global fit with functional forms and scatter described as a log-parabola provides a consistent description of both the profile shape and the intrinsic scatter throughout the entire radial range.
In Fig.~\ref{fig:scatter_all} we show the scatter of all our thermodynamic quantities obtained in both cases.
We recall that our definition of the intrinsic scatter is relative: a value of 0.1 on the $y$-axis indicates that the considered quantity is intrinsically scattered by 10\% of its value.

We notice that our thermodynamic profiles generally exhibit a high scatter in the central parts of the profile. The scatter decreases toward the outskirts, reaching a minimum in the range $[0.2 - 0.8] R_{500}$, and increases slightly beyond this point. We find that temperature is the least scattered thermodynamic quantity, with intrinsic scatters ranging from 10\% to 20\%. On the contrary, and surprisingly, pressure is the most scattered quantity at all radii (looking at the scatter reconstructed from the piecewise power law fits), ranging from 25\% to 60\%. 
%This result is at odds with the predictions of several numerical simulations \citep[e.g.][which however did not have AGN feedback in its simulations]{kravtsov06}, which predict an anti-correlation between the scatter in density and temperature, leading to a scatter of $\sim10\%$ in the scaled pressure profiles \citep{arnaud+10}. Such an anti-correlation would be expected if the scatter in the relation is caused by the presence of cool, overdense accreting clumps, which are roughly in pressure equilibrium with their environment. {\bf go through Gupta+17} However, we observe a higher intrinsic scatter on the pressure than on either density or temperature, suggesting conversely that the scatters in temperature and density are positively correlated. This conclusion is corroborated by the relatively low scatter in our entropy profiles beyond the core. Indeed, in case temperature and density were anti-correlated one would expect the scatter in entropy $K\propto T_X\times n_e^{-2/3}$ to be substantially larger than the scatter in temperature and density, opposite to the trend observed here. 

In all cases, we note that our profiles present a high degree of self-similarity in the radial range $[0.2-0.8]R_{500}$, with a typical intrinsic scatter less than 0.3 ($\sim$ 0.1 dex) in all the measured quantities. This radial range corresponds to the region where gravity dominates and baryonic physics (gas cooling, AGN, and supernova feedback) is relatively unimportant, whereas gas accretion still plays a subdominant role. 
This is consistent with tightly self-regulated mechanical AGN feedback \citep[e.g., via chaotic cold accretion][]{gaspari12}, which can only affect the region $< 0.1\,R_{500}$, with predicted moderate scatter in T/$n_e$ as similarly retrieved here.

We note that the intrinsic scatter profiles shown in Fig. \ref{fig:scatter_all} include the scatter that is induced by uncertainties on the cluster mass, hence on the self-similar scaling quantities. In Appendix \ref{app:sigma_mass} we estimate numerically the residual scatter coming from uncertainties in the self-similar scaling on the various thermodynamic quantities. We found that the scaled pressure is the quantity that is most strongly affected by mass uncertainties, which introduce a scatter of $\sim11\%$ at $R_{500}$, compared to $6\%$ for the temperature, $5\%$ for the density, and $3\%$ for the entropy. However, this effect appears insufficient to fully explain the difference in scatter between  density and pressure at $0.5R_{500}$. We also checked whether the higher scatter in pressure could be explained by intrinsic differences between X-ray and SZ pressure profiles (see Appendix \ref{app:XvsSZ}). However, we find no statistically significant differences between the pressure profiles measured with the two methods, and the scatter in pressure remains the same when considering X-ray and SZ data separately. 

%\subsubsection{Mass induced scatter}
%{\bf Because errors on the mass measurement induced by either statistical errors and by any hydrostatic bias can influence the intrinsic scatter measured, we investigate how important is its role. We started from scatter-free thermodynamic profiles, and we draw a distribution of possible hydrostatic bias between 0\% and 15\% ($b$ = Uniform[0,15]), similarly to values find in Paper 3. We draw new perturbed mass estimates starting from our measured masses, correcting by hydrostatic bias and considering errors on mass measurements ($M_{new}$ = Gaussian[$\mu = M_{500} \cdot (1-b)$, $\sigma = \sigma_{M_{500}}$]). Then we perturbed the scatter-free thermodynamic profiles of each object by scaling with respect to these perturbed mass, therefore changing both the scaling on the $x$-axis and $y$-axis (with the exception of density which does not have a scaling dependent on mass). Finally we computed the scatter in the resulting cluster population at fixed radius, since the scatter generated this way is independent of radius, by taking the standard deviation of the relative variation of rescaled thermodynamic quantities.
%This effect induces an intrinsic scatter of 6\% on density, 12\% on pressure, 12\% in temperature, and 3\% in entropy.
%}

\begin{figure*}
\includegraphics[width=0.5\textwidth]{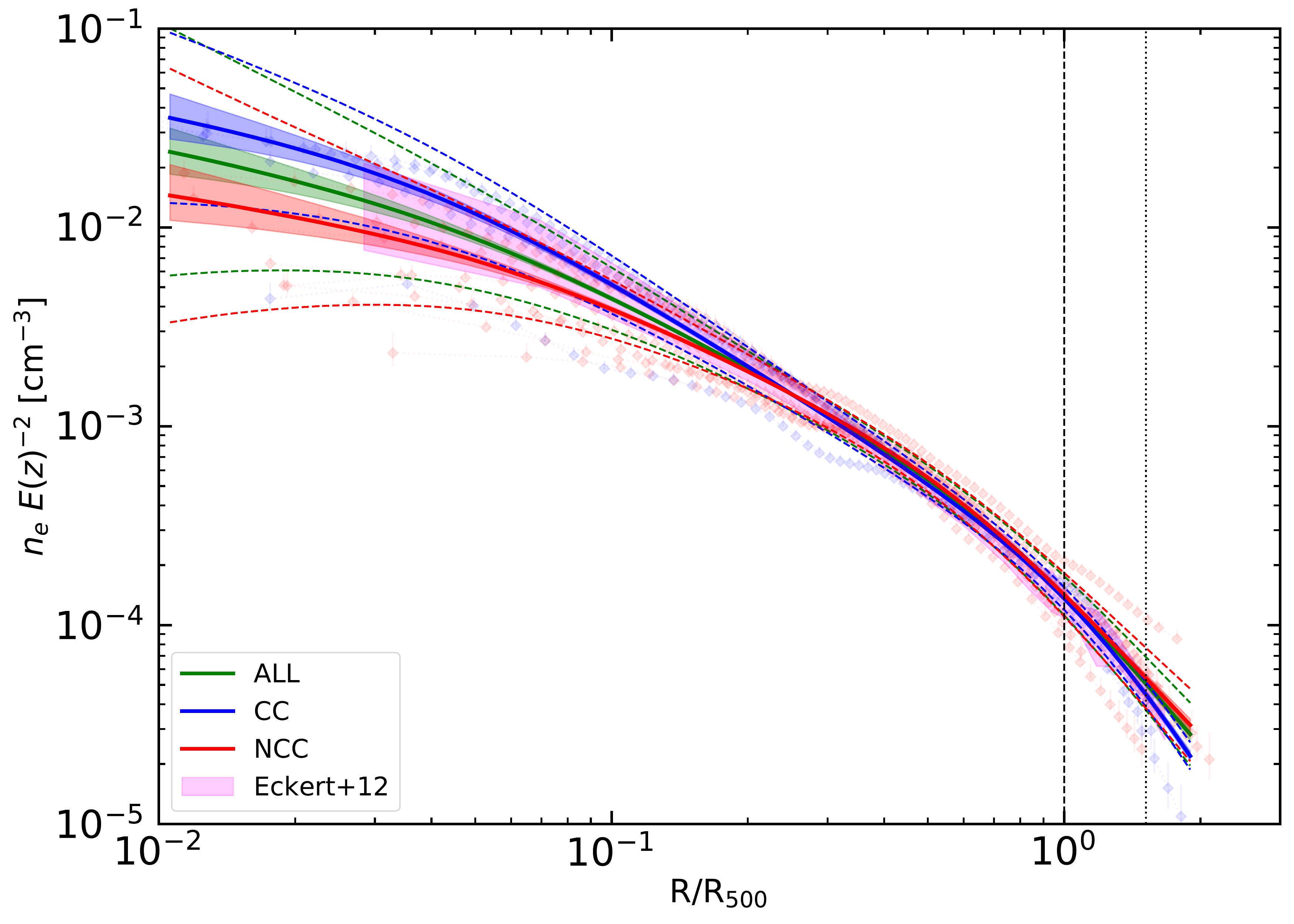}~
\includegraphics[width=0.5\textwidth]{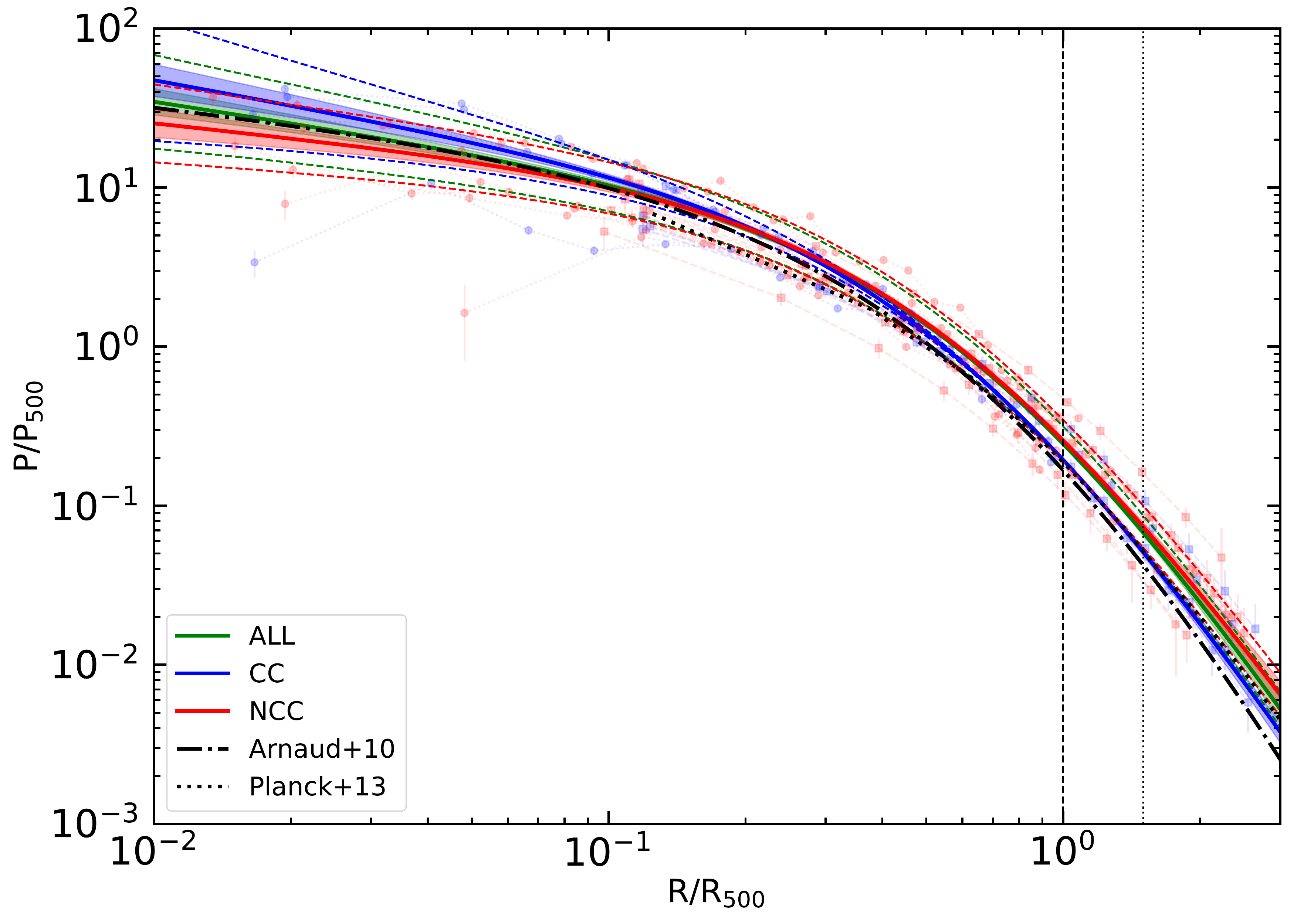}

\includegraphics[width=0.5\textwidth]{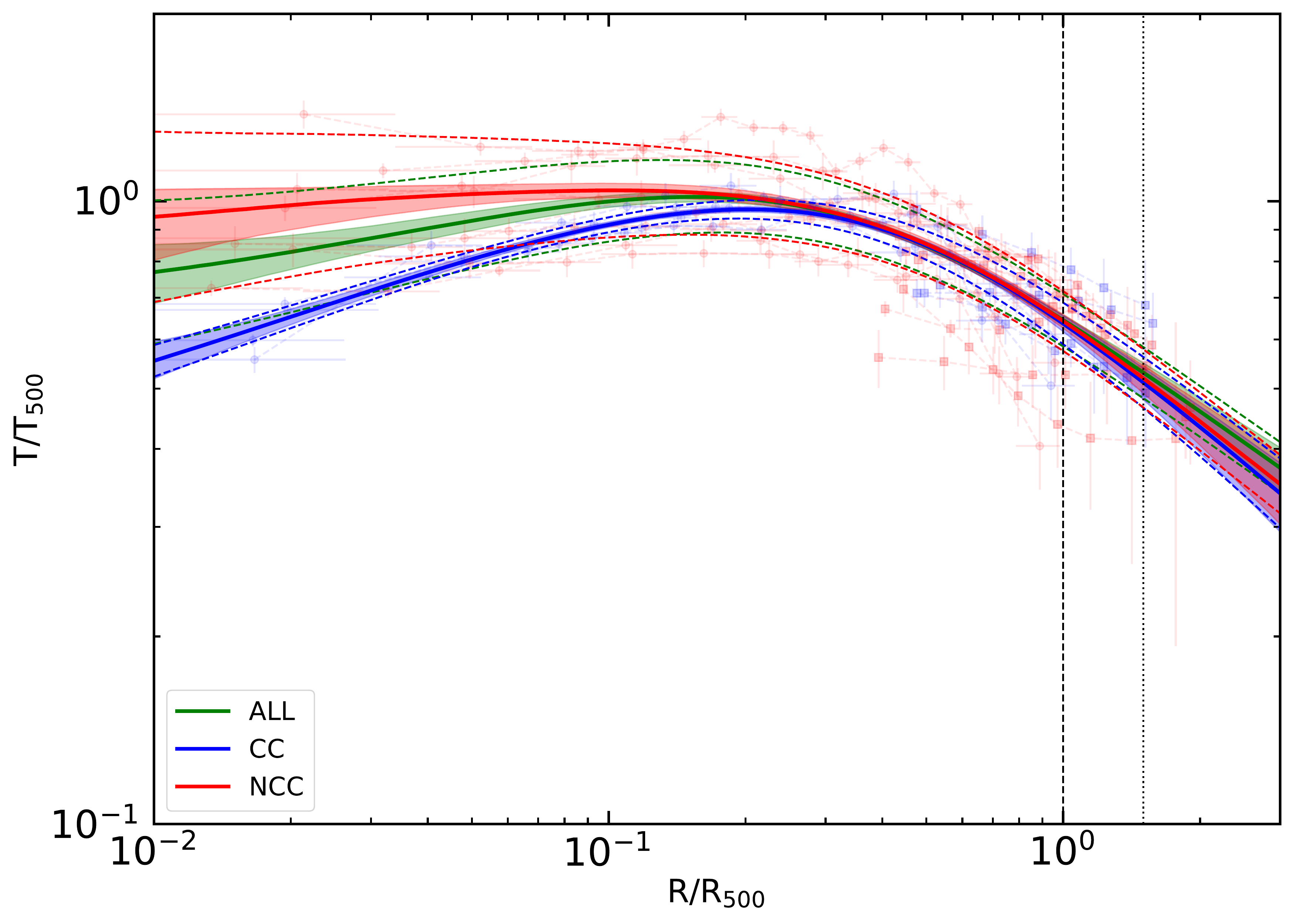}~
\includegraphics[width=0.5\textwidth]{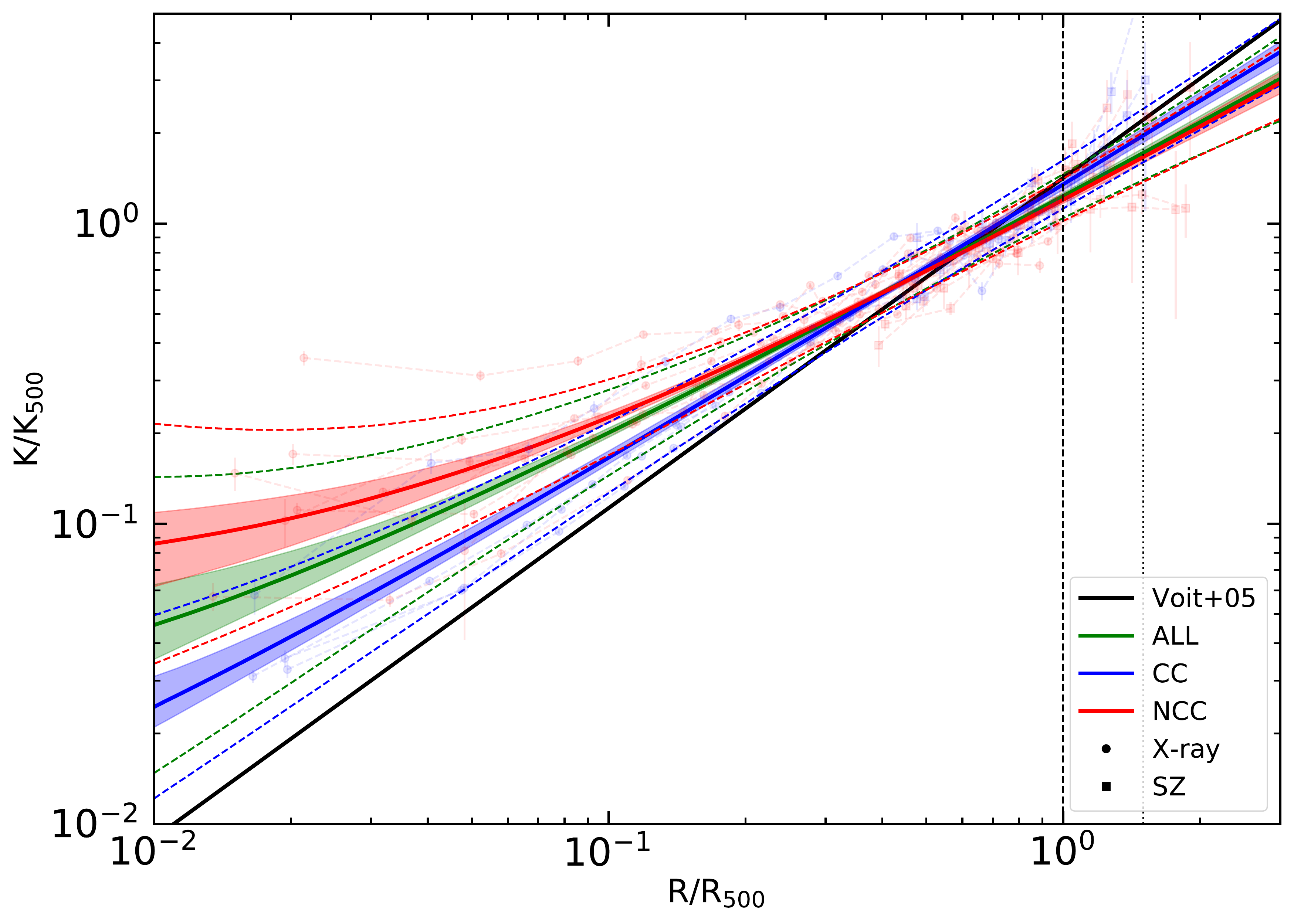}

\caption{Thermodynamic quantities of the X-COP clusters dividing the clusters into cool-core  (CC, in blue) and non-cool-core  (NCC, in red) populations, compared with the entire population (ALL, in green). \emph{Top left:} Density profiles fitted using the functional form presented in Eq.~\eqref{eq:vikh_density}, overplotted on the data and on the ``universal'' density profile of \citet[pink shaded area]{eckert12}. \emph{Top right:} Pressure profiles fitted using the functional form presented in Eq.~\eqref{eq:nagai}, overplotted on the data and compared to the \textit{Planck} results \citep[dotted black line]{planck+13} and the universal pressure profile of \citet[dash-dotted black line]{arnaud+10}.
\emph{Bottom left:} Temperature profiles fitted using the functional form presented in Eq.~\eqref{eq:vikh_density} and overplotted on the data.
\emph{Bottom right:} Entropy profiles fitted using the functional form presented in Eq.~\eqref{eq:powlc_entropy}, overplotted on the data and on the gravitational collapse predictions \citep[solid black line]{voit+05}. }
\label{fig:CCvsNCC}
\end{figure*}

\begin{table*}
\caption{\label{tab:forms}Best fit parameters of the functional forms describing the universal thermodynamic quantities. In all cases, we provide the results of the fit to the entire population (ALL) and to the  cool-core (CC) and non-cool-core (NCC) populations separately.  $\sigma_1$, $\sigma_0$, and $x_0$ are the parameters of the log-parabola describing the behavior of the intrinsic scatter.
We indicate the priors adopted on the parameters, indicating uniform priors between \emph{a} and \emph{b} with $U(a, b)$.
}

\begin{center}

{\bf Density}: Eq.~\eqref{eq:vikh_density}

\begin{tabular}{ c | c c c c c c | c c c }
Data & $\log(n_0)$ & $\log(r_c)$ &  $\log(r_s)$ & $\alpha$ & $\beta$ & $\epsilon$ & $\sigma_1$ & $x_0$ & $\sigma_0$ \\ 
\hline
Priors & \emph{U}(-7, -2) & \emph{U}(-7, -2) & \emph{U}(-2.5, 2.5) & \emph{U}(0, 5)  & \emph{U}(0, 5) & \emph{U}(0, 5) & \emph{U}(0, 0.5) & \emph{U}(0, 1) & \emph{U}(0, 0.5) \\

ALL & $-4.4 \pm 0.5$ &$-3.0 \pm 0.5$ &$-0.29 \pm 0.15$ &$0.89 \pm 0.59$ &$0.43 \pm 0.02$ &$2.86 \pm 0.38$ &$0.09 \pm 0.01$ &$0.43 \pm 0.02$ &$0.16 \pm 0.01$ \\
CC & $-3.9 \pm 0.4$ &$-3.2 \pm 0.3$ &$0.17 \pm 0.07$ &$0.80 \pm 0.61$ &$0.49 \pm 0.01$ &$4.67 \pm 0.36$ &$0.04 \pm 0.01$ &$0.88 \pm 0.10$ &$0.13 \pm 0.01$ \\
NCC & $-4.9 \pm 0.4$ &$-2.7 \pm 0.5$ &$-0.51 \pm 0.17$ &$0.70 \pm 0.48$ &$0.39 \pm 0.04$ &$2.60 \pm 0.27$ &$0.10 \pm 0.01$ &$0.38 \pm 0.02$ &$0.16 \pm 0.01$ \\
\end{tabular}

{\bf Pressure}: Eq.~\eqref{eq:nagai}

\begin{tabular}{ c | c c c c c | c c c }
Data & $P_0$ & $c_{500}$ &  $\gamma$ & $\alpha$ & $\beta$ & $\sigma_1$ & $x_0$ & $\sigma_0$ \\ 
\hline
Priors & \emph{U}(0, 14) & \emph{U}(0, 5) & \emph{U}(0, 0.8) & fix & \emph{U}(2, 8) & \emph{U}(0, 0.5) & \emph{U}(0, 2) & \emph{U}(0, 0.5) \\

ALL & $5.68 \pm 1.77$ & $1.49 \pm 0.30$ & $0.43 \pm 0.10$ & $1.33$ & $4.40 \pm 0.41$ & $0.02 \pm 0.01$ & $1.63 \pm 0.36$ & $0.25 \pm 0.02$ \\
CC & $6.03 \pm 1.61$ & $1.68 \pm 0.23$ & $0.51 \pm 0.10$ & $1.33$ & $4.37 \pm 0.26$ & $0.03 \pm 0.01$ & $1.52 \pm 0.33$ & $0.00 \pm 0.00$ \\
NCC & $7.96 \pm 2.54$ & $1.79 \pm 0.38$ & $0.29 \pm 0.11$ & $1.33$ & $4.05 \pm 0.41$ & $0.01 \pm 0.01$ & $1.28 \pm 0.52$ & $0.30 \pm 0.03$ \\
\end{tabular}

{\bf Temperature}: Eq.~\eqref{eq:vikh_temperature}

\begin{tabular}{ c | c c c c c c | c c c }
Data & $T_0$ & $\log(r_{cool}$) & $r_t$ & $\frac{T_{min}}{T_0}$ & $a_{cool}$ & $c/2$  & $\sigma_1$ & $x_0$ & $\sigma_0$ \\ 
\hline
Priors & \emph{U}(0, 2) & \emph{U}(-7, 0) & \emph{U}(0, 1) & \emph{U}(0, 1.5)  & \emph{U}(0, 3) & \emph{U}(0, 1) & \emph{U}(0, 0.5) & \emph{U}(0, 3) & \emph{U}(0, 0.5) \\

ALL & $1.21 \pm 0.23$ &$-2.8 \pm 1.1$ &$0.34 \pm 0.10$ &$0.50 \pm 0.24$ &$1.03 \pm 0.78$ &$0.27 \pm 0.04$ &$0.01 \pm 0.01$ &$2.13 \pm 0.67$ &$0.09 \pm 0.01$ \\
CC & $1.32 \pm 0.25$ &$-2.8 \pm 0.7$ &$0.40 \pm 0.10$ &$0.22 \pm 0.17$ &$0.74 \pm 0.30$ &$0.33 \pm 0.06$ &$0.01 \pm 0.01$ &$0.08 \pm 0.04$ &$0.03 \pm 0.01$ \\
NCC & $1.09 \pm 0.10$ &$-4.4 \pm 1.8$ &$0.45 \pm 0.14$ &$0.66 \pm 0.32$ &$1.33 \pm 0.89$ &$0.30 \pm 0.07$ &$0.01 \pm 0.01$ &$2.22 \pm 0.63$ &$0.11 \pm 0.01$ \\
\end{tabular}

{\bf Entropy}: Eq.~\eqref{eq:powlc_entropy}

\begin{tabular}{ c | c c c | c c c }
Data & $\log(K_0)$ & $K_1$ & $\alpha$ & $\sigma_1$ & $x_0$ & $\sigma_0$ \\ 
\hline
Priors & \emph{U}(-7, 0) & \emph{U}(1, 2) & \emph{U}(0, 2) & \emph{U}(0, 0.5) & \emph{U}(0, 1) & \emph{U}(0, 0.5) \\

ALL & $-3.98 \pm 1.22$ & $1.21 \pm 0.03$ & $0.83 \pm 0.04$ & $0.06 \pm 0.02$ & $0.56 \pm 0.11$ & $0.14 \pm 0.01$ \\
CC & $-5.50 \pm 1.10$ & $1.35 \pm 0.06$ & $0.93 \pm 0.04$ & $0.03 \pm 0.02$ & $0.60 \pm 0.22$ & $0.17 \pm 0.03$ \\
NCC & $-2.77 \pm 0.55$ & $1.14 \pm 0.03$ & $0.84 \pm 0.07$ & $0.05 \pm 0.02$ & $0.57 \pm 0.16$ & $0.14 \pm 0.01$ \\
\end{tabular}

\end{center}
\end{table*}

\subsection{CC versus NCC}
\label{sec:CC_NCC}
We divide our cluster sample into two populations based on the central entropy value, shown as the last column in Table~\ref{table:data}. We use as an indicator of dynamical state the central entropy of our clusters as measured by \textit{Chandra} \citep{cavagnolo+09}, which has a better spatial resolution than \textit{XMM-Newton}, and therefore is able to trace more accurately the behavior of the entropy profiles in the inner regions.
Using this indicator we identify four clusters as cool-core (CC) and eight as non-cool-core (NCC) using the value of $K_0 = 30 \ \rm{keV cm}^2$ as the cutoff value.

%The classification of clusters into CC and NCC populations is sometimes associated with the relaxation state of a cluster, although in general it indicates the behaviour of the thermodynamic properties in the inner regions ($<0.1R_{500}$). Since the gas in the inner regions of CC clusters has a short cooling time ($<1$ Gyr), these systems exhibit a central temperature drop in the core. The loss of thermal energy through radiative cooling leads to lower entropy in the centre, and to a contraction of the gas towards the bottom of the potential well, and thus an increase in the central density with consequent higher pressure. On the contrary, NCC clusters have flat temperature profiles, low density and high entropy in their central regions, which is usually interpreted as signs of mixing induced by merging activity.

In Fig.~\ref{fig:CCvsNCC} we show the data split into the CC and NCC populations, together with the fit using the functional forms used above, Eq.~\eqref{eq:vikh_density} for density, Eq.~\eqref{eq:nagai} for pressure, Eq.~\eqref{eq:vikh_temperature} for temperature, and Eq.~\eqref{eq:powlc_entropy} for entropy. The best fitting functional forms for the CC and NCC classes separately are provided in Table \ref{tab:forms}, and the results of piecewise power law fits to the two populations individually are given in Table~\ref{tab:piece_CCNCC} and Fig~\ref{fig:CCvsNCC_piece}.
We note that in the core, the CC and NCC systems separate out.
However, we do not observe any significant differences between CC and NCC systems outside  the core: the properties of our SZ selected clusters beyond 0.3$R_{500}$ are not influenced by the properties of the core. We remark that in the case of the temperature there is a slight difference between the two best fits, with NCC having steeper temperature profiles, however well within the 1$\sigma$ error envelope.
The only marginally significant difference is found in the entropy profiles, which appear slightly flatter in the outskirts of NCC clusters. As shown in Table \ref{tab:forms}, we measure an outer slope $\alpha_{\rm CC}=0.95\pm0.03$ for the CC populations, as opposed to $\alpha_{\rm NCC}=0.85\pm0.07$. However, we note that this difference can be an artifact of the poor fit to the data obtained with a simple power law with an entropy floor (Eq.~\eqref{eq:powlc_entropy}). Indeed, similar to the case of the fit to the overall population, we find a steeper slope at large radii when fitting the data points for the two populations with a piecewise power law ($\alpha_{\rm CC}=1.23\pm0.14$, $\alpha_{\rm NCC}=0.94\pm0.14$, see Table \ref{tab:piece_CCNCC}), which is consistent with the self-similar slope of 1.1 within $1\sigma$. Thus, the evidence for a flatter entropy slope beyond $R_{500}$ in the NCC population is marginal.

\section{Discussion}
\label{sec:disc}

\subsection{Systematic uncertainties}

In this section we describe the potential systematic errors affecting our analysis. 

\begin{itemize}
\item \textbf{Gas density:} As described in Sect. \ref{sec:red_ima}, we paid special attention to the minimization of the systematics in the subtraction of the \emph{XMM-Newton} background. The method that we used to model the contribution of each individual background component was calibrated using a large set of $\sim500$ blank-sky pointings and leads to residual systematics on the order of 3\% on the subtraction of the local background \citep[see Appendix A in][]{ghirardini18}. For the present work, we conservatively increased the level of systematics to 5\% to include potential uncertainties associated with the application of the method to a cluster field instead of a blank field. A systematic error of 5\% of the background value was thus added in quadrature to all our surface brightness measurements. We note that the systematic uncertainty becomes comparable to the statistical errors only beyond $\sim2\times R_{500}$. At $R_{200}$ the systematic uncertainty is typically 20\% or less of the measured signal. Further improvements in the modeling of the \xmm\ background could allow us in the future to provide information beyond the current limiting radii since in many cases our SZ pressure profiles extend beyond $2\times R_{500}$.

\item \textbf{Pressure profiles:} A possible source of systematics on the reconstruction of SZ pressure profiles is the relativistic corrections to the SZ effect \citep{itoh98}, which reduce the amplitude of the SZ increment in the high-frequency part of the CMB spectrum. Several recent works claimed a detection of the relativistic SZ corrections on stacked \emph{Planck} data \citep{hurier16,erler18}. In particular, \citet{erler18} noted that the relativistic corrections could lead to an underestimate of the integrated SZ signal up to 15\% for the hottest clusters, which could thus affect our pressure profiles too. However, we note that the gas temperature decreases by a factor of $2-2.5$ from the core to the outskirts, such that the impact of SZ corrections should be limited to the central regions, where spectroscopic X-ray measurements are preferred because of their higher signal-to-noise ratio and resolution. For typical temperatures of $\sim5$ keV at $R_{500}$ and beyond the expected effect is less than 5\% \citep{erler18}. For more discussion on the impact of systematic uncertainties we refer to \citet{planck+13}. 

\item \textbf{Spectroscopic temperatures:} Although our modeling of the \xmm\ spectra is fairly sophisticated (see Sect. \ref{sec:spec}), uncertainties in the subtraction of the \xmm\ background can lead to systematics in our spectral measurements in the outermost regions considered. Following \citet{lm08} we do not attempt to perform spectral measurements in the regions where our signal is less than 60\% of the background intensity to avoid introducing biases. Another potential source of systematics is the calibration of the telescope's effective area. \citet{schellenberger+15} reported systematic differences at the level of 15\% between \xmm\ and \emph{Chandra} temperature measurements for the same regions, \emph{Chandra} returning systematically higher temperatures than \xmm. As shown in Fig. \ref{fig:pressure} and demonstrated in Appendix~\ref{app:XvsSZ}, we observe a very good agreement between \xmm\ and \planck\ pressure profiles; the only exception is ZwCl 1215, for which the pressure measured by \xmm\ actually {exceeds} the SZ pressure by $\sim20\%$, which could be explained by orientation effects since the X-ray and SZ signals have different line-of-sight dependencies. Since our X-ray and SZ pressure profiles are obtained in an independent way from different instruments and different techniques, we conclude that our spectral measurements are robust.

\item \textbf{Self-similar scaling: }Given that the scaling quantities depend on the measured mass, and that we use as our reference mass model the \textit{backward NFW} mass model \citep{ettori+10,ghirardini18}, uncertainties on the mass measurements should be taken into account. In Ettori et al. (2018) we discuss the accuracy of our mass models by testing our mass measurements using various methods (forward fitting, Gaussian processes, and several functional forms for the mass model). We find that all the methods agree with the NFW mass reconstruction, with the mass profiles scattered by less than 5\% at a fixed radius of 1.5 Mpc. The uncertainty in our scaling is therefore less than 3\% on $P_{500}$ and $K_{500}$, and less than 2\% on $R_{500}$. In Eckert et al. (2018) we also assess the level of non-thermal pressure support by comparing the X-COP gas fraction profiles with the expected universal gas fraction. We find that the bias in our mass measurements at $R_{500}$ is just 6\% on average, again resulting in very small corrections in the self-similar quantities. 

%\item \textbf{Error propagation: } From a theoretical point of view our derived thermodynamic quantities, like $P_X = n_X T_X$ , are not the product of the measured quantities, like $n_X$ and $T_X$, but are the product of their distribution, meaning that the product of two random variables distributed as Gaussians does not simply produce a third Gaussian distribution with the product of the means as the new mean value, but there is an extra term caused by the covariance between the two initial gaussian distributions. However, given the small percentage errors we measure on temperature and pressure, we can neglect this second order effect.

\end{itemize}

\subsection{Regular outskirts}

The wide radial range accessible with the X-COP data allows us to study the properties of the gas at $R_{500}$ and beyond and to constrain the shape of the universal thermodynamic profiles throughout the entire cluster volume for the first time. Compared to previous works addressing the state of the gas in cluster outskirts (e.g., with \emph{Suzaku} data) the study presented here constitutes a substantial improvement in several ways: \emph{i)} the ability of our azimuthal median method to excise overdense regions down to scales of 10--20 kpc depending on the cluster redshift \citep{eckert15}, which allows us to measure gas density profiles that are free of the effects of gas clumping on the scales we are able to resolve (typically  30 kpc); \emph{ii)} a nearly uniform azimuthal coverage for all our clusters out to $R_{200}$, which guarantees that our measurements are representative of the global behavior and were not obtained along preferential directions; and \emph{iii)} an exquisite control of systematic uncertainties even in the faint cluster outskirts regime (see above).

As described in Sect. \ref{sec:thermo}, our reconstruction of clumping-free thermodynamic quantities leads to results that differ substantially from the typical results obtained with \emph{Suzaku}. We recall that several studies found relatively flat density profiles, steep temperature profiles, and entropy profiles that fall below the prediction of gravitational collapse and sometimes even roll over \citep[e.g.,][]{kawa10,walker12a,walker12b,urban14,simi17}. Conversely, our clumping-corrected reconstruction yields density and pressure profiles that steepen steadily with radius (see Figs. \ref{fig:density} and \ref{fig:pressure}), temperature profile decreasing with a mild slope of $-0.3$ that is consistent with the slopes observed inside $R_{500}$ by \xmm\ and \emph{Chandra} \citep{lm08,pratt07,vikhlini+06}, and entropy profiles rising with a slope that is consistent with the self-similar slope of 1.1 beyond $0.6 R_{500}$ and all the way out to the largest radii considered ($2\times R_{500}$). 

All the results presented here point to gas clumping as the primary origin for the deviations from the predictions reported so far by \emph{Suzaku}, in agreement with the results presented in \citet{tchernin16} for the case of Abell 2142. The low resolution of \emph{Suzaku} ($\sim2$ arcmin) indeed prevented the authors from excising cool, overdense structures that would bias at the same time the gas density toward high values and the spectroscopic temperature toward low values, resulting in underestimated values for the entropy that are not representative of the bulk of the ICM. If the gas in such infalling structures is in pressure equilibrium with its environment, as usually predicted \citep[e.g.,][]{roncarelli+13,planelles17}, pressure profiles reconstructed from the SZ effect are mildly affected by such inhomogeneities and the combination of SZ pressure and clumping-free gas density is representative of the state of the ICM well beyond $R_{500}$. 

The only exception to this scenario is the case of Abell 2319 \citep{ghirardini18}, which deviates systematically from the measured universal profiles even when the profiles are corrected for clumping. In \citet{ghirardini18} we showed that the deviations from self-similarity cannot be explained by azimuthal variations, but rather that the ongoing merging activity causes a high level of non-thermal pressure support. This conclusion is supported by the high hydrostatic gas fraction of this system and a clear deficit of entropy beyond $R_{500}$, even after excising clumps. Abell 2319 is the only system within the X-COP sample that exhibits such a behavior (see also Eckert et al. 2018), which suggests that this system is currently experiencing a transient phase of high non-thermal pressure induced by a violent merger with a mass ratio of 3 to 1 \citep{oegerle+95}.

Overall, the results presented here establish that in the majority of cases, the bulk of the ICM is virialized and follows the predictions of gravitational collapse out to $2\times R_{500}\approx R_{100}$. Accretion shocks that are expected to raise the entropy level of the smooth infalling gas should be located approximately at $3-4\times R_{500}$ \citep{lau15}, and we would expect the entropy of the ICM to turn over around this radius. These radii should also correspond to the approximate location of the splashback radius \citep{diemer14,diemer17}, which represents a natural boundary of dark matter halos. Future X-ray and SZ facilities such as \emph{ATHENA} \citep{nandra13} and CMB-S4 \citep{cmbs4} will attempt to detect the ICM at the cluster boundary to constrain the location of accretion shocks and the accretion rate. The results presented here highlight the need for relatively high angular resolution experiments with a low and highly reproducible background to reach these goals.

\subsection{Self-similarity of the profiles}

Our analysis shows that the thermodynamic profiles exhibit a high level of similarity once the profiles are rescaled according to the self-similar model \citep{kaiser86}. The level of self-similarity is particularly remarkable beyond the core ($R>0.3R_{500}$) and it reaches a maximum in the radial range $[0.2-0.8]R_{500}$. As already discussed in Sect. \ref{sec:scatter}, the region of minimum scatter observed in this study corresponds to the region where the gas is highly virialized and baryonic effects are negligible. In the central regions ($R<0.3R_{500}$) baryonic effects (cooling, AGN feedback) lead to a substantial scatter within the cluster population. Beyond $\sim R_{500}$, we again observe an increase in the measured scatter, which might be explained by different accretion rates from one system to another. Importantly, our study shows that the properties of the X-COP cluster population beyond $0.3R_{500}$ are not correlated with the core state (CC or NCC). While the core state probably retains memory of past major mergers, it does not trace the accretion rate on large scales at the present epoch. This result agrees with the predictions of \citet{planelles17}, which did not find differences in the accretion rate of simulated CC and NCC systems. For instance, the case of A2029 is striking. This cluster hosts a strong cool-core and it is one of the most regular in our sample. However, our large-scale mosaic reveals that it is located within a chain of at least three X-ray detected structures (see Fig. \ref{fig:XCOP_all}) with overlapping $R_{200}$, and the optical information shows that this system is part of a larger filamentary structure extending over $\sim20$ Mpc \citep{smith12}. 

Another important result of our study is that beyond the central regions pressure is the most scattered thermodynamic quantity (see Fig. \ref{fig:scatter_all}). The scatter in $P_{e}=T_X\times n_e$ is about 50\% larger than the scatter in either $T_X$ or $n_e$, which is expected when fluctuations in temperature and density are uncorrelated. This result is opposite to the widely accepted view that temperature and density variations are anti-correlated, which has lead people to postulate that the quantity $Y_X=M_{\rm gas}\times T_{X}$ has the lowest scatter at fixed mass \citep{kravtsov06}. Our results disagree with this conclusion and imply that the scatter in $M_{\rm gas}$ is less than the scatter in $Y_X$ at fixed mass. These results are consistent with the recent predictions of \citet{truong18}, which found that in the simulation runs including gas cooling and subgrid thermal AGN feedback, temperature and density are essentially uncorrelated (see their Fig. 10), implying that the scatter in $M_{\rm gas}$ and $T_X$ is less than that in $Y_X$. Beyond the core, X-ray observables appear to behave self-similarly to a high level of precision. In the case a selection based on the integrated gas mass or the core-excised X-ray luminosity can be achieved, future X-ray surveys such as \emph{eROSITA} \citep{erosita} will yield large cluster catalogues and low-scatter mass proxies, even in comparison to SZ surveys \citep{mantz18}. 
\iffalse
%{\bf
%We can ask: what is the role of AGN feedback? We have analysed 3D high-resolution ($\sim$ 100 pc) simulations with self-regulated mechanical AGN feedback (Gaspari et al. 2012; CC cluster akin to A1795) which can consistently prevent the cooling flow catastrophe and preserve the cool-core for several Gyr (e.g., McDonald et al. 2017). The AGN feeding is triggered via chaotic cold accretion of cold/warm clouds, while the AGN feedback occurs via massive ultra-fast outflows (mechanical efficiency $\sim 10^{-3} -- 10^{-2}$). The self-regulated AGN feedback cycle can only affect the region within $0.1 R_{500}$, where it drives a scatter in T/n/K of $\approx$\,0.1, 0.35, 0.25, respectively, which is similar to our X-COP CC clusters (Fig. A.2). The simulated scatter in pressure is similar to that of entropy, suggesting that part of the scatter in pressure for the X-COP clusters resides in additional processes (e.g., sloshing). The rapid drop in all the variables scatter beyond the $0.1 R_{500}$ region (in both X-COP and hydrodynamic simulations) is a key signal that AGN feedback has lost influence, while beyond $R_{500}$ group/galaxy infall, merger, and filamentary accretion start to elevate the scatter again.}
\fi

\section{Conclusion}

In this paper, we presented the universal thermodynamic properties of the intracluster medium for 12 SZ selected galaxy clusters observed with \xmm\ and \planck. Our observational strategy allowed us to construct radial profiles of gas density, pressure, temperature, and entropy over an unprecedentedly wide radial range from $0.01R_{500}$ to $2\times R_{500}$, i.e., covering the entire cluster volume. We fitted our self-similar scaled profiles with universal functional forms and provided estimates of the radial dependence of the slope and intrinsic scatter. Our findings can be summarized as follows:

\begin{itemize}
\item  Our gas density and pressure profiles are in excellent agreement with previous determinations of the universal density \citep{eckert12} and pressure profiles \citep{planck+13}. The typical uncertainties in the gas density and pressure at $R_{200}$ are at the level of $10\%$, allowing us to perform a detailed analysis of the shape and intrinsic scatter.
\item The logarithmic slope of the density and pressure profiles steepens steadily with radius, reaching a value of $-2.5$ and $-3.0$ at $R_{200}$ for density and pressure, respectively. These results are consistent with the expectations for an ideal gas in hydrostatic equilibrium within a NFW potential well.
\item Beyond $\sim0.3R_{500}$ the temperature profiles decrease steadily with radius with a logarithmic slope of $-0.3$, which is somewhat shallower than the slope of $\sim-1.0$ observed in the outer regions of several systems with \emph{Suzaku} \citep{reiprich13}. 
\item With the exception of one system, beyond $\sim0.5R_{500}$ all clusters follow the gravitational collapse prediction for entropy generation in galaxy clusters \citep{voit+05} out to the largest radii considered ($2\times R_{500}$). This result is at odds with the conclusions usually reached from \emph{Suzaku} observation, which often show a deficit of entropy beyond $R_{500}$. The difference is explained by the steep slope of the \emph{Suzaku} temperature profiles compared to ours and by our treatment of gas clumping. We postulate that the impossibility of properly excising clumps in low-resolution \emph{Suzaku} data is responsible for biasing the observed temperatures low and gas densities high.
\item The outer regions of galaxy clusters exhibit a high level of self-similarity. Beyond $\sim0.3R_{500}$ we find no significant difference between the cool-core and non-cool-core cluster populations in any of the quantities of interest. This result implies that the core properties are determined by the merging history of a system but do not trace efficiently the current accretion rate, which determines the state of the gas in the outskirts.
\item We determined for the first time the scatter of each thermodynamical quantity within the cluster population as a function of radius. The scatter of all quantities is maximum in the core and reaches a minimum in the radial range $[0.2-0.8]R_{500}$ (see Table~\ref{tab:piecewise} and Fig.~\ref{fig:scatter_all}). We find that the gas temperature is the least scattered quantity at nearly all radii. 
%Interestingly, pressure appears to be substantially more scattered {\bf where? } than temperature and density, implying that the latter two quantities are positively correlated. Thus, we expect that the integrated gas mass and the average spectroscopic temperature exhibit a lower scatter at fixed mass than the integrated thermal energy $Y$.
\end{itemize}

A recently accepted \xmm\ program will extend the X-COP sample to objects that were initially excluded (A754, A3667, and A3827), which will allow us to perform a similar analysis on a statistically complete SZ-selected sample. Furthermore, since pressure profiles extend beyond $2 \times R_{500}$, a further reduction of the systematics on the surface brightness profile would be useful to extend the thermodynamic profiles beyond the current limits, provided that observations with higher statistical quality can be performed. 

\begin{acknowledgements} 
X-COP data products are available for download at \href{https://www.astro.unige.ch/xcop}{https://www.astro.unige.ch/xcop}. We thank Alexis Finoguenov, Kaustuv Basu, Jens Erler, and Gus Evrard for useful discussions. The research leading to these results has received funding from the European Union’s Horizon 2020 Programme under the AHEAD project (grant agreement n. 654215). Based on observations obtained with XMM-Newton, an ESA science mission with instruments and contributions directly funded by ESA Member States and NASA. S.E. acknowledges financial contribution from the contracts NARO15 ASI-INAF I/037/12/0, ASI 2015-046-R.0 and ASI-INAF n.2017-14-H.0. F.V. acknowledges financial support from the ERC Starting Grant ``MAGCOW'', no.714196. M.G. is supported by NASA through Einstein Postdoctoral Fellowship Award Number PF5-160137 issued by the Chandra X-ray Observatory Center, which is operated by the SAO for and on behalf of NASA under contract NAS8-03060. Support for this work was also provided by Chandra grant GO7-18121X. E.R. acknowledges the ExaNeSt and Euro Exa projects, funded by the European Union’s Horizon 2020 research and innovation program under grant agreement No. 671553 and No. 754337 and financial contribution from the agreement ASI-INAF n.2017-14-H.0. H.B. and P.M. acknowledge financial support by ASI Grant 2016-24-H.0.
\end{acknowledgements}

\bibliographystyle{aa} 
\bibliography{XCOP_thermo} 

\begin{appendix}

\begin{flushright}
•
\end{flushright}\section{Consistency between X-ray and SZ pressure measurements}
\label{app:XvsSZ}
\begin{figure}[t]
\centering
\includegraphics[width=0.5\textwidth]{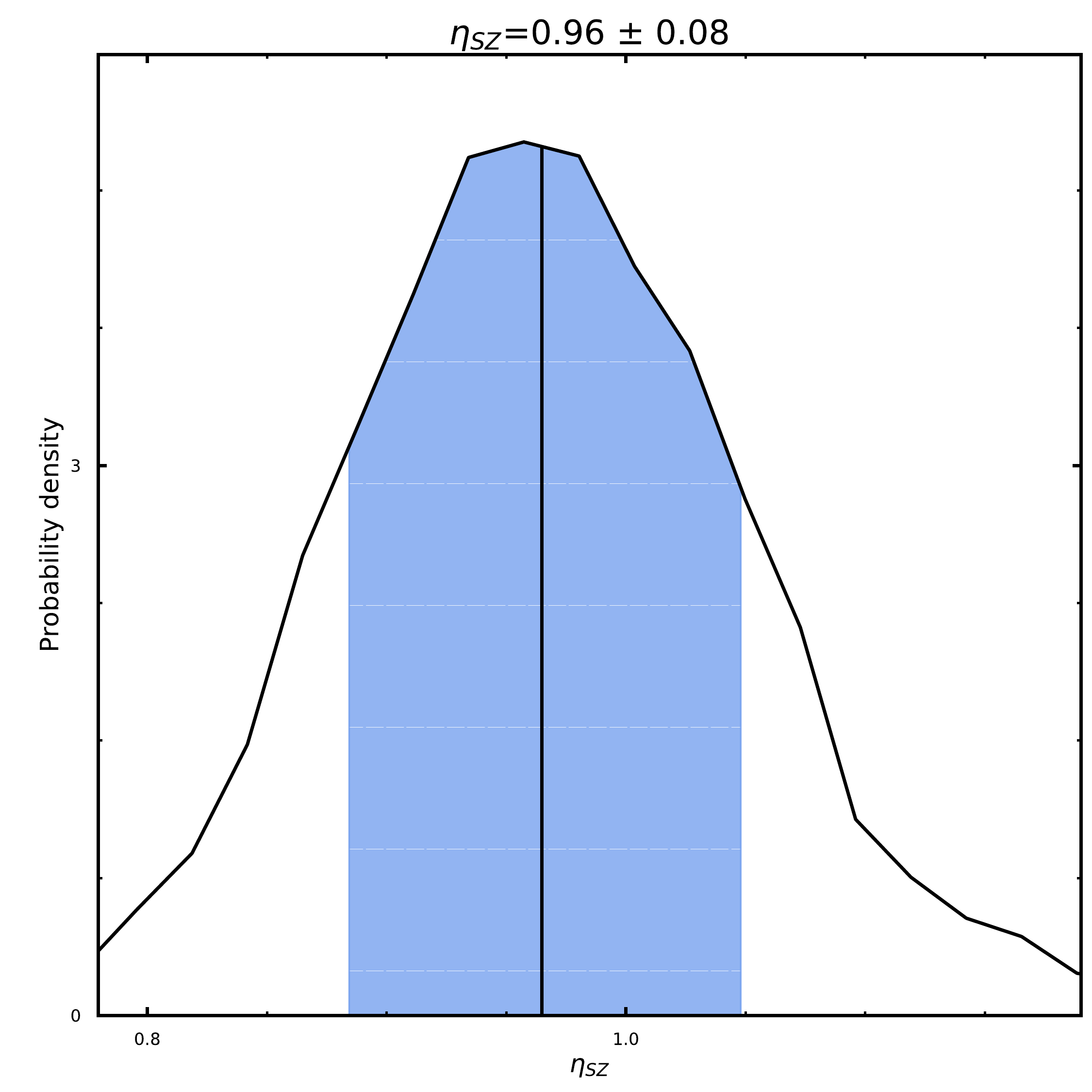}
\caption{ Posterior distribution of the parameter $\eta_{SZ} \approx P_{SZ}/P_{X}$. The shaded blue area indicates the region containing 68\% of the posterior distribution. }
\label{fig:eta_sz}
\end{figure}

We checked the consistency between X-ray and SZ pressure profiles to test how our results are affected by discrepancies between the two measurements.
To perform this check we introduced a parameter $\eta_{SZ}$, which is the ratio between the SZ and X-ray pressure profile, and proceeded with a joint fit, allowing the scatter on X-ray and SZ data to be independent.
Mathematically we can write the following system of equations:
\begin{equation}
\begin{cases} 
P_{SZ} = \eta_{SZ} P_{model} \cdot \exp [ \pm \sigma_{int,SZ} ] \\ 
P_{X} = P_{model} \cdot \exp [ \pm \sigma_{int,X} ] \\ 
\end{cases}
,\end{equation}
where $\eta_{SZ}$, $\sigma_{int,SZ}$, and $\sigma_{int,X}$ are free parameters.
For $P_{model}$ we checked both the piecewise powerlaw fit case in the radial range where we have both X-ray and SZ measurements, and the global functional form on the entire radial range.
In both cases the measured scatters are in good agreement, and are compatible with the scatter shown in Fig.~\ref{fig:scatter_all}. More importantly, the parameter $\eta_{SZ}$ shows a distribution  consistent with unity (see Fig.~\ref{fig:eta_sz}), indicating a very good general agreement between X-ray and SZ pressure measurements.

\section{Comparison between MILCA and forward-modeling pressure profiles}
\label{app:bourdin}
To test the robustness of our pressure profile measurements, we compared the MILCA results with the pressure measured using an alternative technique \citep[see][]{bourdin+17}. Following this technique, \emph{Planck}-HFI frequency maps are first wavelet cleaned for CMB and thermal dust anisotropies, a parametric pressure template \citep{nagai+07} is subsequently projected onto the sky plane, convolved with the frequency dependent HFI beams and fitted to the CMB and dust cleaned maps. Being fully parametric, this technique allow us to take advantage of the frequency dependent angular resolution of each HFI channel during the template fitting. 
This angular resolution is about 9.7 and 7.3 arcmin at 100 and 143 GHz, respectively, but reaches about 5 arcmin in the energy range [217, 857] GHz \citep{planck16}.

\begin{figure}[t]
\centering
\includegraphics[width=0.5\textwidth]{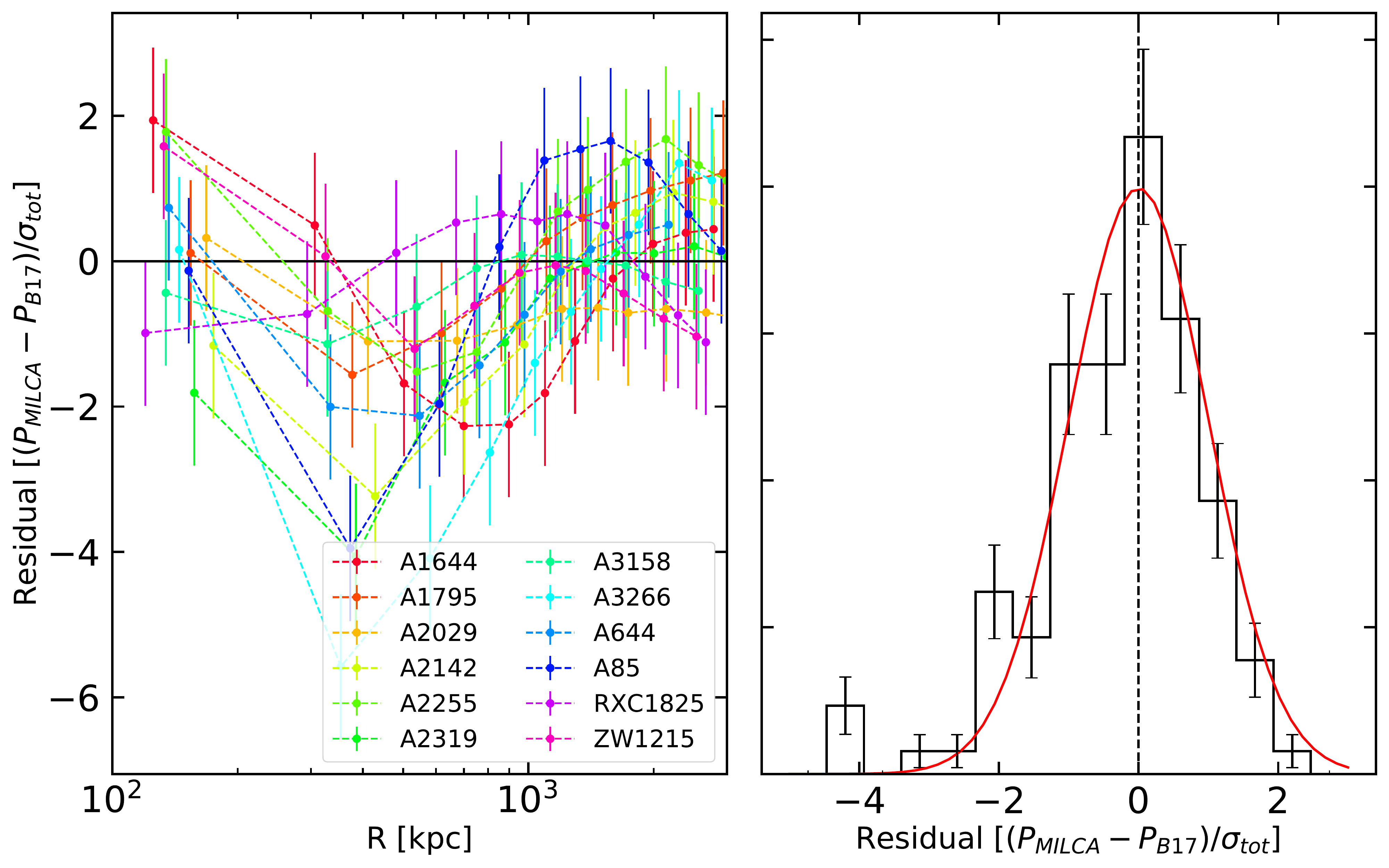}
\caption{ \emph{Left}: Residuals of the comparison of the two different methods we used to estimate the SZ pressure profile. A remarkable good agreement within the statistical uncertainties is reached at all radii, especially excluding the first 3 MILCA points which are the most affected by the \emph{Planck} PSF.
\emph{Right}: Distribution of the residuals compared with the statistical prediction of a set of residuals: a gaussian centered in zero and width one (red line). }
\label{fig:MILCAvsB17}
\end{figure}

In Fig.~\ref{fig:MILCAvsB17} we compare the resulting best fitting pressure profile from the above procedure with the MILCA maps (see Sect.~\ref{s:plck}). The residuals are shown in the left panel and  show the nice agreement, within the statistical uncertainty, of the two different methods applied. In the right panel all the residuals are grouped together to create a distribution, which is compared with a Gaussian centered at 0 and width 1, showing that the residuals follow this distribution very closely, indicating that statistically the two pressure profiles are in very good agreement.

\section{Mass-induced scatter in thermodynamic profiles}
\label{app:sigma_mass}
Since the scaling of our thermodynamic quantities depends on the cluster mass both through the scale radius $R_{500}$ and the self-similar quantities $Q_{500}$ (see Sect. \ref{sec:thermo}), the measured scatter profiles presented in Fig. \ref{fig:scatter_all} depend on the accuracy of the adopted masses. Both statistical and systematic fluctuations of the measured mass around the true mass will induce fluctuations of the scaling quantities, thus introducing an irreducible source of scatter originating from the limited accuracy of our mass calibration. 

To take this effect into account, we estimated numerically the scatter in each quantity that is induced by uncertainties in our mass scaling. We started from scatter-free universal profiles following the measurements provided in Table \ref{tab:forms}, and perturbed the mass scaling according to the known statistical uncertainties and biases in our mass scaling. Namely, for each X-COP cluster we randomly drew new values of the observed mass $M_{\rm obs}$ as
\begin{equation}
M_{\rm obs}=\mbox{Gauss}((1-b)M_{\rm true},\Delta M)
\end{equation}
with $M_{\rm true}$ the assumed true mass, $b$ the hydrostatic mass bias, and $\Delta M$ the statistical uncertainty in our hydrostatic masses (see Ettori et al. 2018 for details). For the hydrostatic mass bias, we used the distribution of non-thermal pressure values determined in Eckert et al. (2018) from the measured gas fraction. We then scaled the scatter-free profiles for each quantity $Q$ by the perturbed values of $R_{500}$ and $Q_{500}$. 

\begin{figure}
\centering
\includegraphics[width=0.5\textwidth]{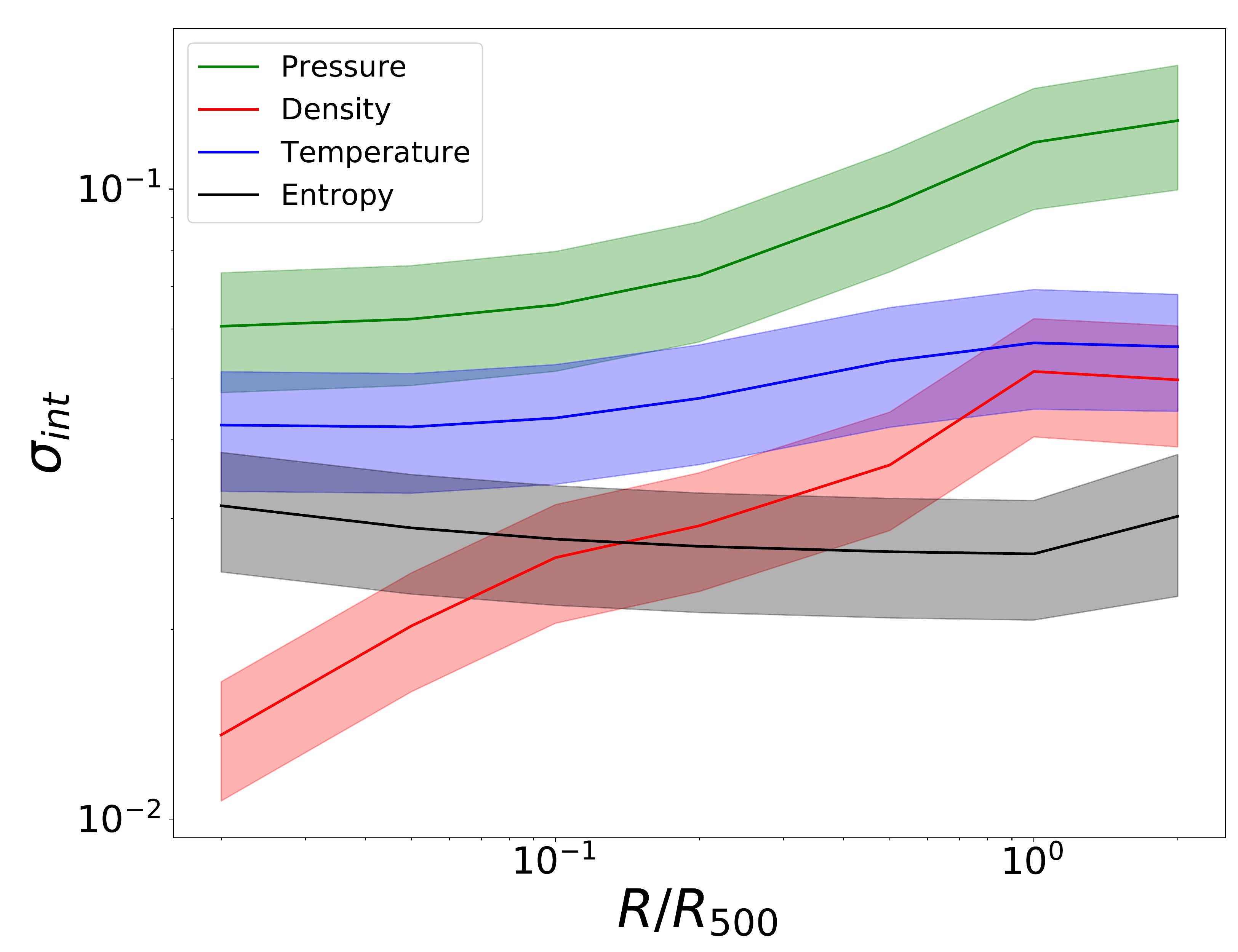}
\caption{Scatter in the various thermodynamic profiles induced by uncertainties in the mass calibration. The shaded areas show the range of intrinsic scatter obtained from 1,000 simulations.}
\label{fig:sigma_mass}
\end{figure}

We applied this procedure to each X-COP cluster and computed the resulting scatter as a function of radius. We repeated this procedure 1,000 times  to get an idea of the uncertainty  introduced by sample variance. In Fig. \ref{fig:sigma_mass} we show the resulting mass-induced scatter for the scaled pressure, density, temperature, and entropy. We can see that the effect of the mass scaling is largest on the pressure and ranges between 6\% and 12\% as a function of radius. Conversely, the effect on the entropy is minimal (2\%--3\%). The scatter in temperature and density induced by uncertainties in the mass scaling lies somewhere in between.
{ Pressure is more affected than the other thermodynamic quantities simply because its slope is the most steep among the quantities.}

\section{Piecewise power law fits for CC and NCC clusters separately}

\begin{figure*}
\includegraphics[width=0.5\textwidth]{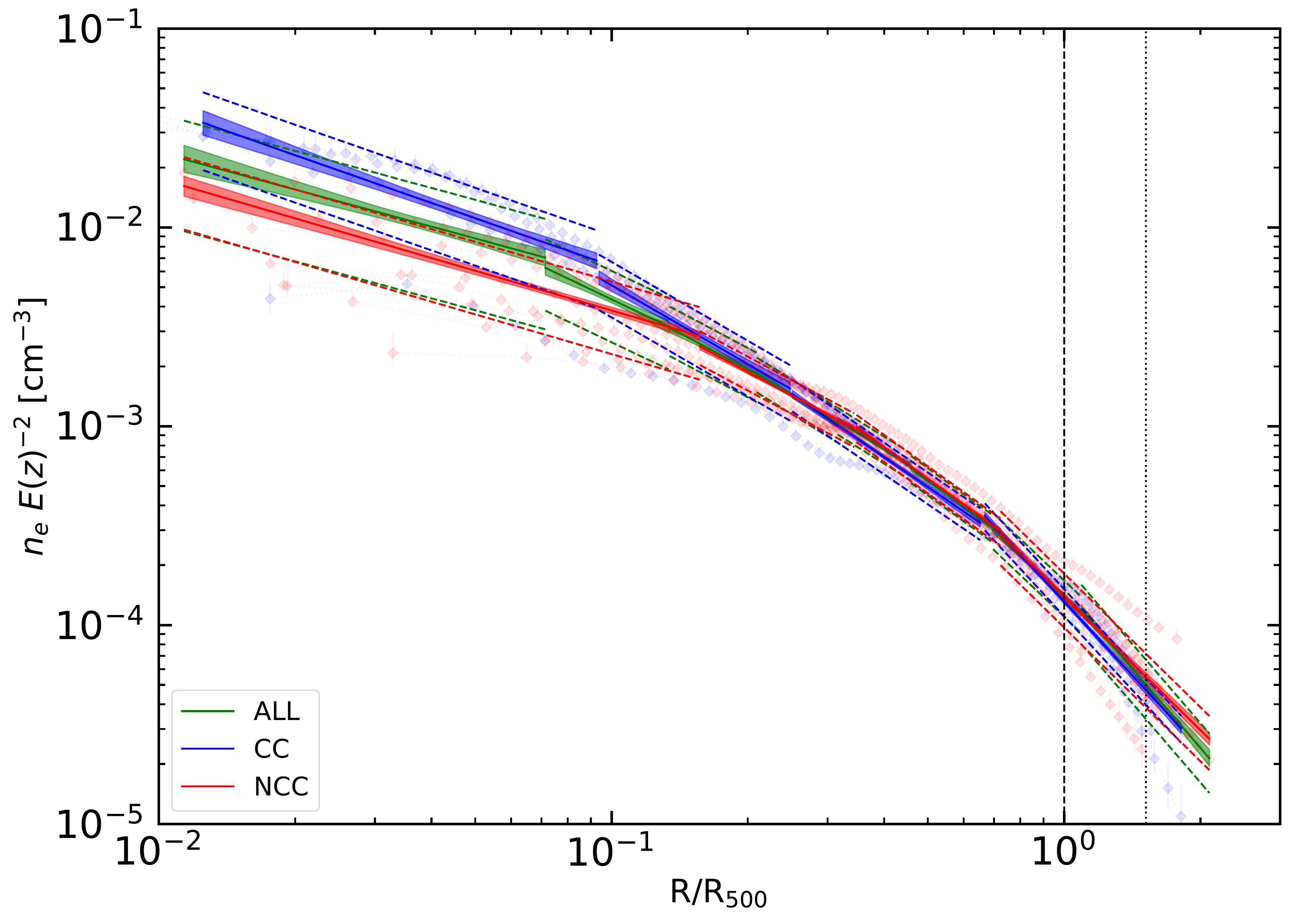}~
\includegraphics[width=0.5\textwidth]{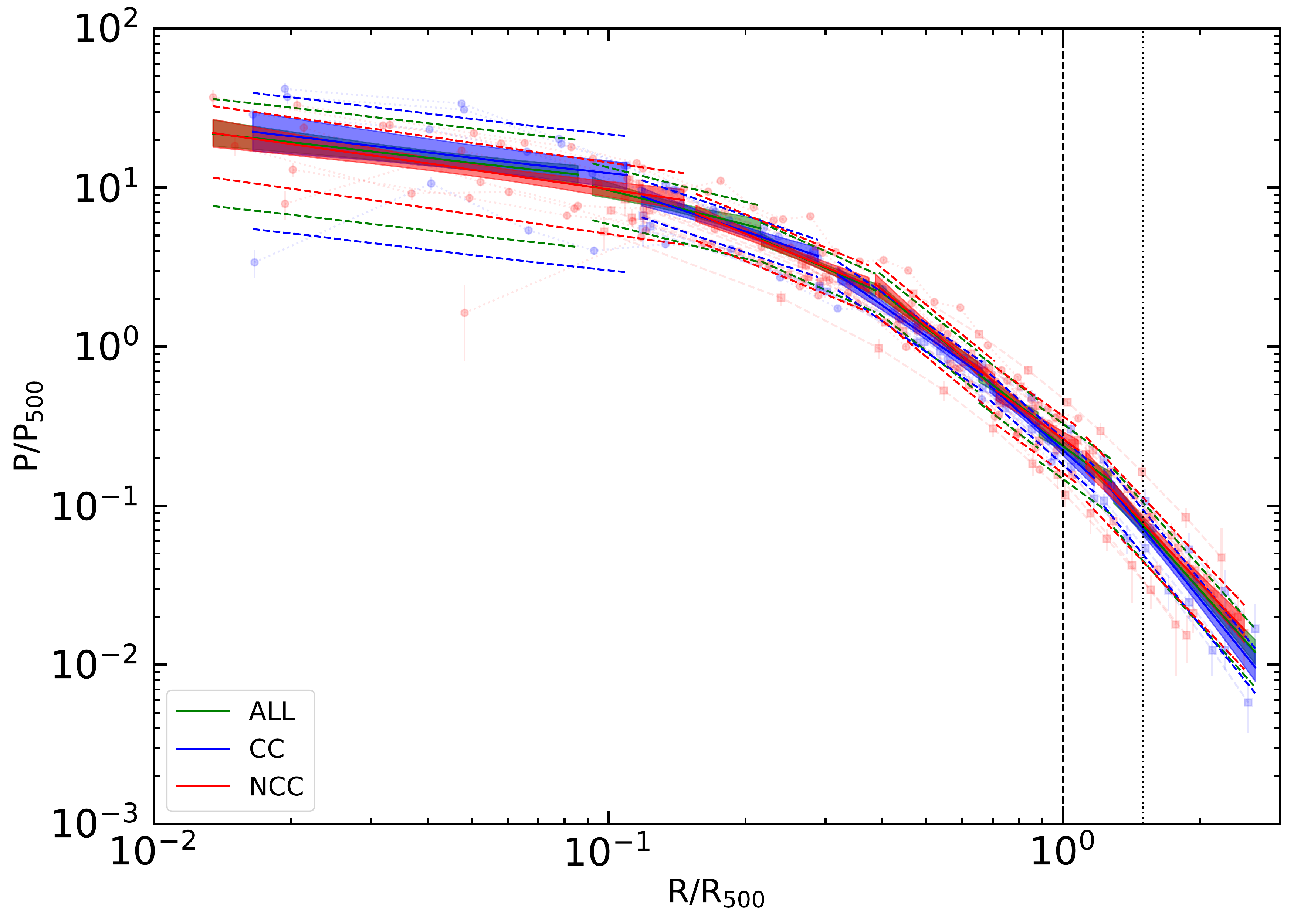}

\includegraphics[width=0.5\textwidth]{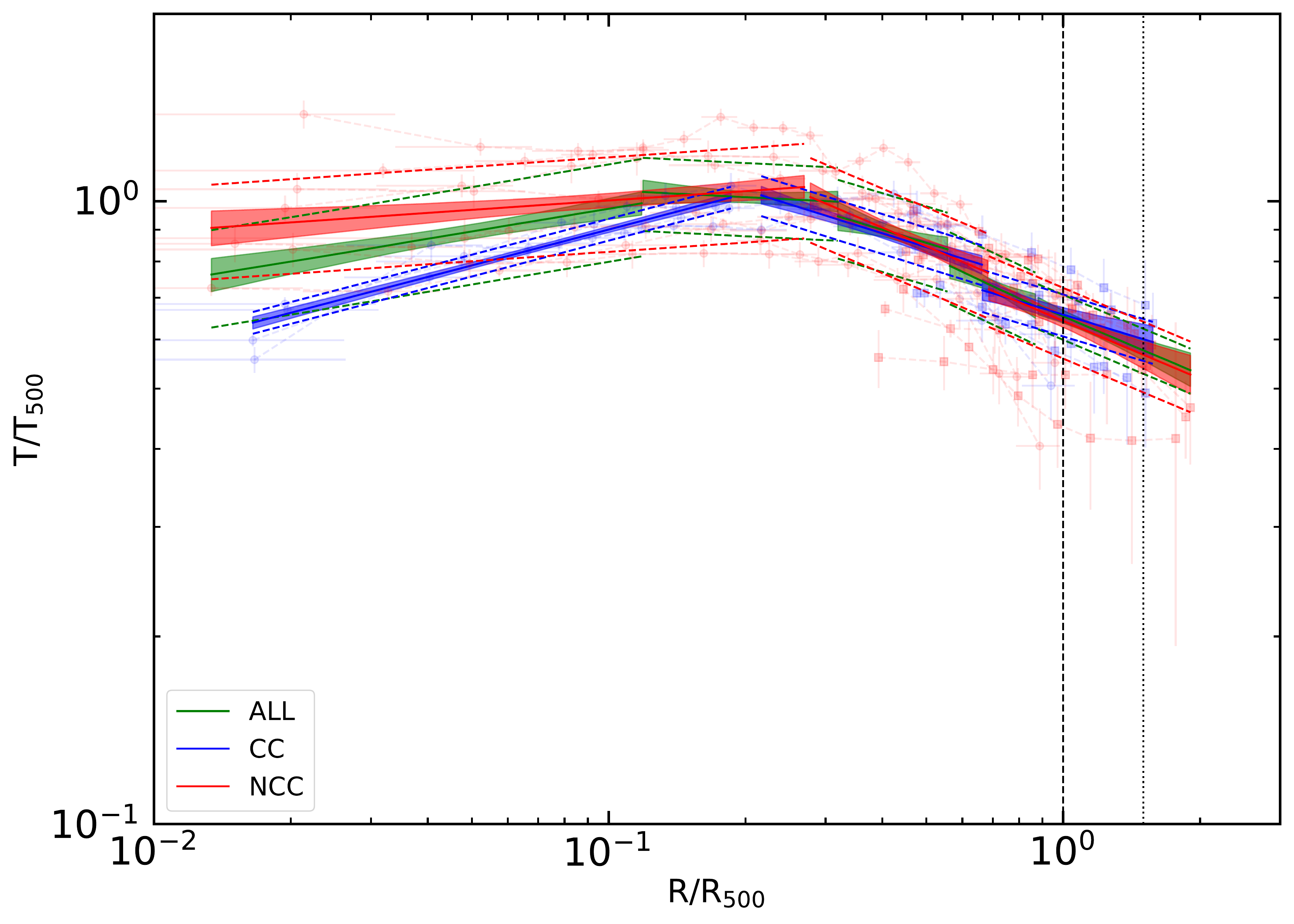}~
\includegraphics[width=0.5\textwidth]{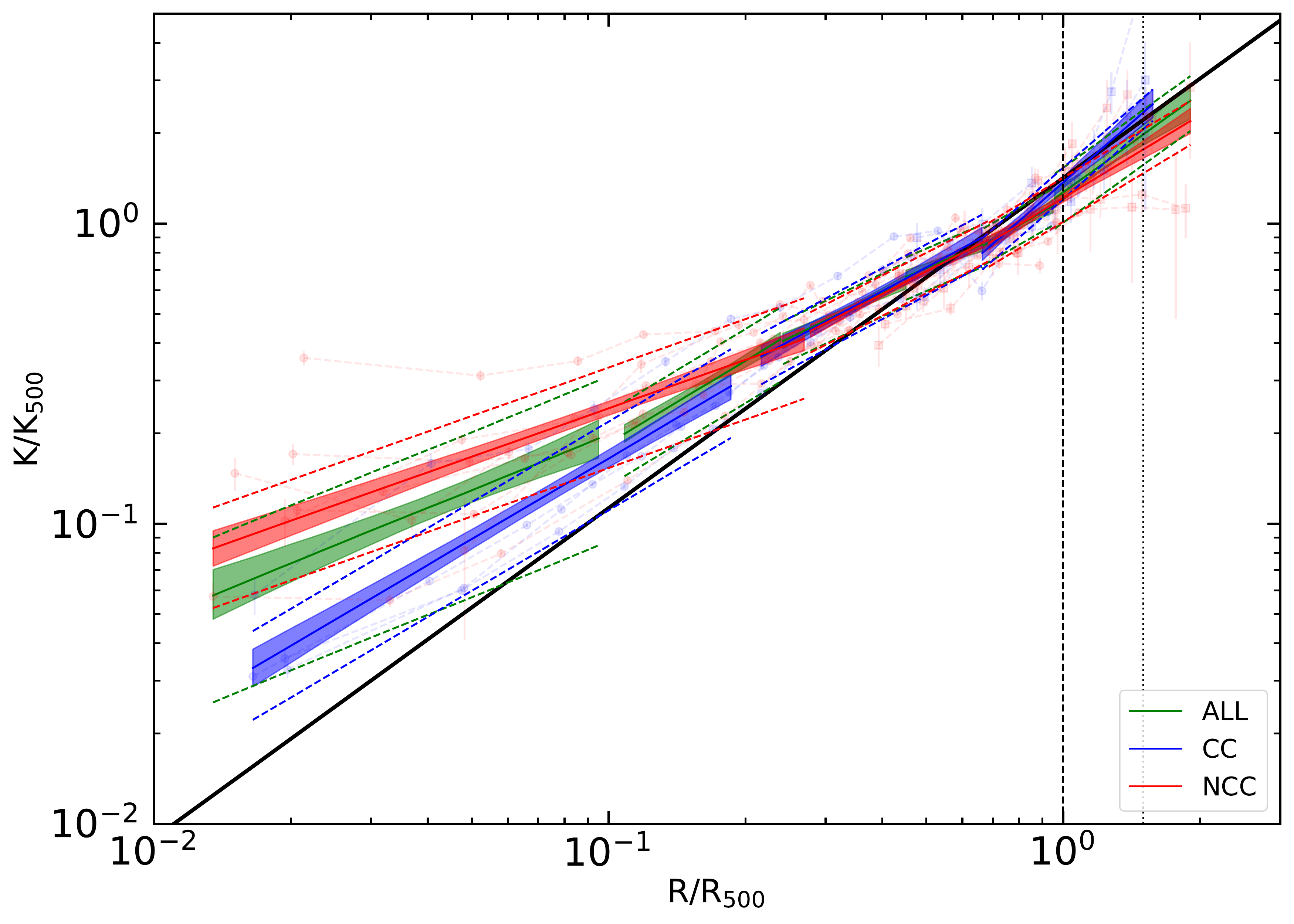}

\caption{Same as Fig.~\ref{fig:CCvsNCC}, but using the piecewise fit instead of global functional form. 
The cluster population of X-COP is divided into cool-core (CC, in blue) and non-cool-core clusters (NCC, in red), compared with the entire population (ALL, in green). }
\label{fig:CCvsNCC_piece}
\end{figure*}

\begin{figure*}
\centering
\captionsetup{justification=centering}
\includegraphics[width=0.7\textwidth]{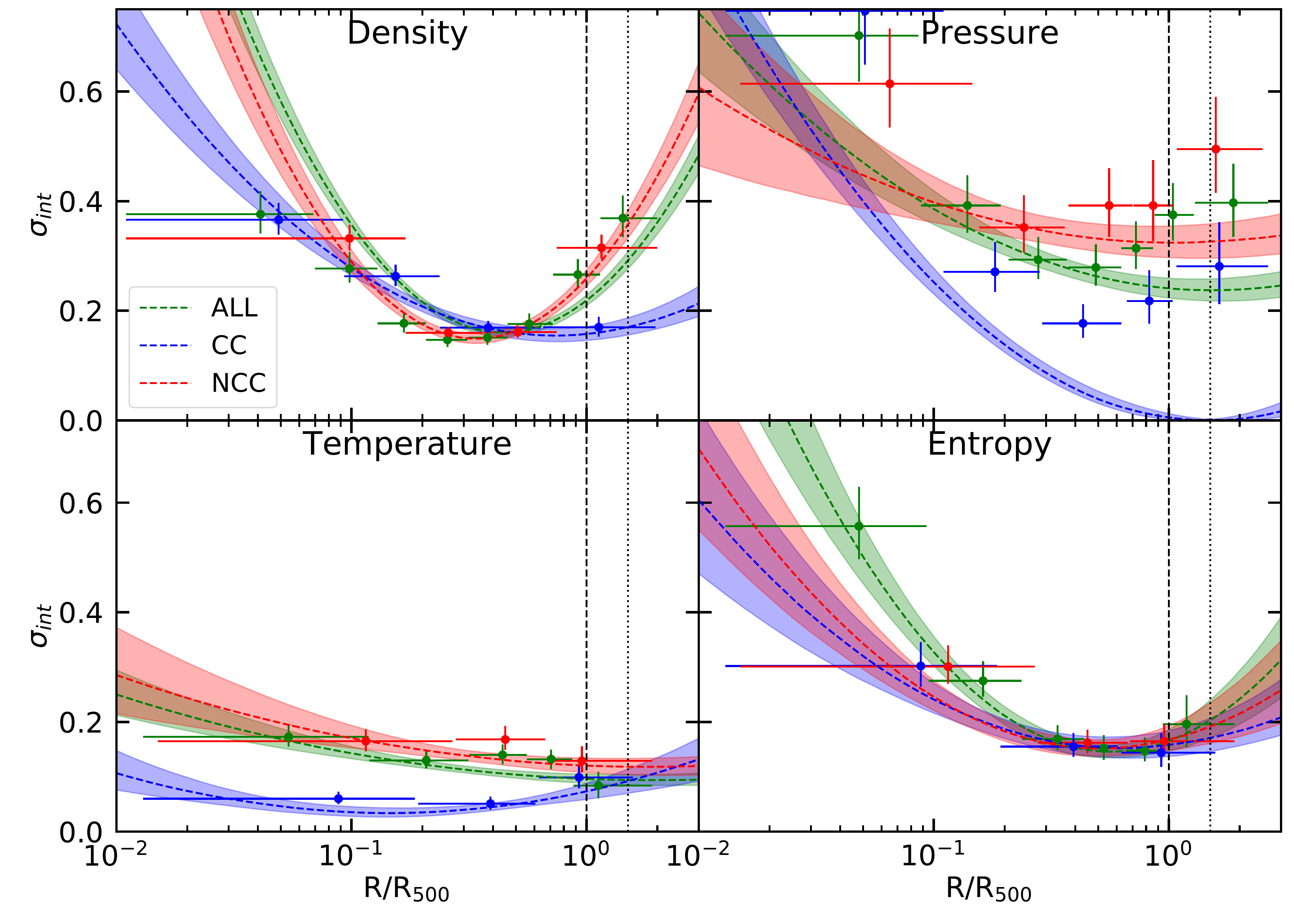}
\caption{Scatter of our thermodynamic quantities dividing our cluster sample in CC and NCC}
\label{fig:scatter_CCvsNCC}
\end{figure*}

In Fig. \ref{fig:CCvsNCC_piece} and Table \ref{tab:piece_CCNCC} we show the results of piecewise power law fits to the X-COP clusters divided into the CC and NCC populations. In Fig. \ref{fig:scatter_CCvsNCC} we also show the best fit scatter of the populations as a function of radius, split into CC and NCC clusters and compared with the full population. In this case, we caution that the number of systems in each category is small (four CC and eight NCC systems) and the measurements of the scatter may be unreliable.

\begin{table*}
\begin{center}
\caption{\label{tab:piece_CCNCC}Same as Table~\ref{tab:piecewise}, but discriminating between cool-core and non-cool-core clusters.}
{\bf Density}

\begin{tabular}{ c c c c c | c c c c c }
\multicolumn{5}{c |}{cool-cores (CC)} & \multicolumn{5}{c}{Non-cool-cores (NCC)} \\
\hline 
$x_{in}$ & $x_{out}$ & A ($10^{-4}$ cm$^{-3}$) & B (slope) & $\sigma_{int}$ & $x_{in}$ & $x_{out}$ & A & B (slope) & $\sigma_{int}$\\
\hline
0.01 & 0.09 & $10.16 \pm 3.26$ & $-0.80 \pm 0.10$ & $0.42 \pm 0.04$ & 0.01 &  0.16 & $8.33 \pm 1.27$ & $-0.66 \pm 0.06$ & $0.40 \pm 0.03$\\
0.09 & 0.25 & $2.46 \pm 0.64$ & $-1.32 \pm 0.14$ & $0.31 \pm 0.03$ & 0.16 &  0.34 & $2.66 \pm 0.29$ & $-1.21 \pm 0.08$ & $0.19 \pm 0.01$\\
0.25 & 0.65 & $1.68 \pm 0.13$ & $-1.56 \pm 0.08$ & $0.18 \pm 0.02$ & 0.35 &  0.72 & $1.75 \pm 0.08$ & $-1.64 \pm 0.07$ & $0.16 \pm 0.01$\\
0.67 & 1.81 & $1.31 \pm 0.03$ & $-2.47 \pm 0.09$ & $0.15 \pm 0.02$ & 0.72 &  2.09 & $1.39 \pm 0.04$ & $-2.23 \pm 0.10$ & $0.30 \pm 0.02$\\
\end{tabular}

{\bf Pressure}

\begin{tabular}{ c c c c c | c c c c c }
\multicolumn{5}{c |}{cool-cores (CC)} & \multicolumn{5}{c}{Non-cool-cores (NCC)} \\
\hline 
$x_{in}$ & $x_{out}$ & A & B (slope) & $\sigma_{int}$ & $x_{in}$ & $x_{out}$ & A & B (slope) & $\sigma_{int}$\\
\hline
0.02 & 0.11 & $6.23 \pm 2.63$ & $-0.31 \pm 0.15$ & $0.76 \pm 0.13$ & 0.01 &  0.15 & $3.85 \pm 1.51$ & $-0.40 \pm 0.13$ & $0.48 \pm 0.07$\\
0.12 & 0.29 & $1.12 \pm 0.48$ & $-0.97 \pm 0.25$ & $0.26 \pm 0.06$ & 0.16 &  0.37 & $0.77 \pm 0.23$ & $-1.18 \pm 0.21$ & $0.33 \pm 0.04$\\
0.32 & 0.66 & $0.30 \pm 0.05$ & $-1.98 \pm 0.23$ & $0.20 \pm 0.05$ & 0.39 &  0.71 & $0.26 \pm 0.06$ & $-2.36 \pm 0.37$ & $0.36 \pm 0.05$\\
0.69 & 1.17 & $0.22 \pm 0.02$ & $-2.52 \pm 0.35$ & $0.19 \pm 0.05$ & 0.71 &  1.08 & $0.26 \pm 0.03$ & $-2.07 \pm 0.58$ & $0.39 \pm 0.06$\\
1.23 & 2.65 & $0.30 \pm 0.07$ & $-3.54 \pm 0.39$ & $0.31 \pm 0.10$ & 1.12 &  2.50 & $0.27 \pm 0.05$ & $-3.03 \pm 0.36$ & $0.43 \pm 0.08$\\
\end{tabular}

{\bf Temperature}

\begin{tabular}{ c c c c c | c c c c c }
\multicolumn{5}{c |}{cool-cores (CC)} & \multicolumn{5}{c}{Non-cool-cores (NCC)} \\
\hline 
$x_{in}$ & $x_{out}$ & A & B (slope) & $\sigma_{int}$ & $x_{in}$ & $x_{out}$ & A & B (slope) & $\sigma_{int}$\\
\hline
0.02 & 0.19 & $1.40 \pm 0.05$ & $0.19 \pm 0.01$ & $0.04 \pm 0.01$ & 0.01 &  0.27 & $1.13 \pm 0.09$ & $0.05 \pm 0.03$ & $0.17 \pm 0.02$\\
0.22 & 0.66 & $0.72 \pm 0.04$ & $-0.23 \pm 0.05$ & $0.07 \pm 0.02$ & 0.28 &  0.68 & $0.68 \pm 0.05$ & $-0.31 \pm 0.09$ & $0.15 \pm 0.02$\\
0.66 & 1.57 & $0.66 \pm 0.02$ & $-0.22 \pm 0.10$ & $0.08 \pm 0.03$ & 0.69 &  1.90 & $0.64 \pm 0.02$ & $-0.31 \pm 0.10$ & $0.13 \pm 0.02$\\
\end{tabular}

{\bf Entropy}

\begin{tabular}{ c c c c c | c c c c c }
\multicolumn{5}{c |}{cool-cores (CC)} & \multicolumn{5}{c}{Non-cool-cores (NCC)} \\
\hline 
$x_{in}$ & $x_{out}$ & A & B (slope) & $\sigma_{int}$ & $x_{in}$ & $x_{out}$ & A & B (slope) & $\sigma_{int}$\\
\hline
0.02 & 0.19 & $1.30 \pm 0.29$ & $0.90 \pm 0.08$ & $0.33 \pm 0.05$ & 0.01 &  0.27 & $0.83 \pm 0.14$ & $0.54 \pm 0.06$ & $0.37 \pm 0.04$\\
0.22 & 0.66 & $1.26 \pm 0.16$ & $0.81 \pm 0.12$ & $0.19 \pm 0.03$ & 0.28 &  0.68 & $1.19 \pm 0.09$ & $0.78 \pm 0.09$ & $0.15 \pm 0.02$\\
0.66 & 1.57 & $1.37 \pm 0.06$ & $1.32 \pm 0.17$ & $0.12 \pm 0.04$ & 0.69 &  1.90 & $1.22 \pm 0.04$ & $0.92 \pm 0.13$ & $0.17 \pm 0.03$\\
\end{tabular}

\end{center}
\end{table*}

\begin{table*}
\begin{center}
\caption{\label{tab:stack}Stacked thermodynamic profiles. $N_{X}$ and $N_{SZ}$ indicates  the number or objects reaching the indicated radius in X.ray and SZ data respectively. {We indicate the median of the 12 cluster profiles, and with the subscripts $_{low}$ and $_{high}$ we indicate the values that contain 68\% of the objects.}}

\begin{tabular}{ c c c c c }
\multicolumn{5}{c}{\bf Density} \\
$\frac{R}{R_{500}}$ & $n_e E(z)^{-2}$ & $(n_e E(z)^{-2})_{low}$  & $(n_e E(z)^{-2})_{high}$ & $N_X$  \\
- & cm$^{-3}$ & cm$^{-3}$  & cm$^{-3}$ & -  \\
\hline
1.058e-02 & 8.814e-03 & 6.543e-03 & 1.503e-02 & 12 \\
1.179e-02 & 8.751e-03 & 6.224e-03 & 1.416e-02 & 12 \\
1.313e-02 & 8.684e-03 & 6.018e-03 & 1.457e-02 & 12 \\
1.463e-02 & 8.712e-03 & 5.929e-03 & 1.421e-02 & 12 \\
1.630e-02 & 8.409e-03 & 5.839e-03 & 1.324e-02 & 12 \\
1.816e-02 & 8.110e-03 & 5.666e-03 & 1.269e-02 & 12 \\
2.024e-02 & 8.007e-03 & 5.560e-03 & 1.314e-02 & 12 \\
2.255e-02 & 7.720e-03 & 5.432e-03 & 1.100e-02 & 12 \\
2.513e-02 & 7.627e-03 & 5.440e-03 & 1.167e-02 & 12 \\
2.799e-02 & 7.593e-03 & 5.445e-03 & 1.165e-02 & 12 \\
3.119e-02 & 7.513e-03 & 5.503e-03 & 1.133e-02 & 12 \\
3.475e-02 & 7.334e-03 & 5.534e-03 & 1.089e-02 & 12 \\
3.872e-02 & 7.086e-03 & 5.271e-03 & 1.058e-02 & 12 \\
4.314e-02 & 6.889e-03 & 5.086e-03 & 1.000e-02 & 12 \\
4.807e-02 & 6.675e-03 & 4.839e-03 & 9.519e-03 & 12 \\
5.356e-02 & 6.408e-03 & 4.531e-03 & 8.860e-03 & 12 \\
5.967e-02 & 6.161e-03 & 4.246e-03 & 8.219e-03 & 12 \\
6.649e-02 & 5.628e-03 & 3.817e-03 & 7.509e-03 & 12 \\
7.408e-02 & 5.255e-03 & 3.550e-03 & 6.841e-03 & 12 \\
8.254e-02 & 4.875e-03 & 3.359e-03 & 6.049e-03 & 12 \\
9.197e-02 & 4.473e-03 & 3.196e-03 & 5.461e-03 & 12 \\
1.025e-01 & 4.141e-03 & 2.992e-03 & 4.983e-03 & 12 \\
1.142e-01 & 3.813e-03 & 2.728e-03 & 4.495e-03 & 12 \\
1.272e-01 & 3.429e-03 & 2.527e-03 & 4.023e-03 & 12 \\
1.417e-01 & 3.011e-03 & 2.327e-03 & 3.565e-03 & 12 \\
1.579e-01 & 2.678e-03 & 2.123e-03 & 3.089e-03 & 12 \\
1.759e-01 & 2.385e-03 & 1.854e-03 & 2.676e-03 & 12 \\
1.960e-01 & 2.142e-03 & 1.650e-03 & 2.296e-03 & 12 \\
2.184e-01 & 1.892e-03 & 1.478e-03 & 1.992e-03 & 12 \\
2.434e-01 & 1.675e-03 & 1.328e-03 & 1.723e-03 & 12 \\
2.712e-01 & 1.444e-03 & 1.159e-03 & 1.475e-03 & 12 \\
3.021e-01 & 1.193e-03 & 1.038e-03 & 1.238e-03 & 12 \\
3.366e-01 & 1.013e-03 & 9.677e-04 & 1.039e-03 & 12 \\
3.751e-01 & 8.583e-04 & 8.375e-04 & 8.741e-04 & 12 \\
4.179e-01 & 7.145e-04 & 6.885e-04 & 7.349e-04 & 12 \\
4.656e-01 & 5.997e-04 & 5.689e-04 & 6.211e-04 & 12 \\
5.188e-01 & 4.926e-04 & 4.631e-04 & 5.208e-04 & 12 \\
5.780e-01 & 4.043e-04 & 3.744e-04 & 4.390e-04 & 12 \\
6.440e-01 & 3.391e-04 & 3.190e-04 & 3.530e-04 & 12 \\
7.176e-01 & 2.742e-04 & 2.657e-04 & 2.832e-04 & 12 \\
7.995e-01 & 2.199e-04 & 2.106e-04 & 2.277e-04 & 12 \\
8.908e-01 & 1.783e-04 & 1.691e-04 & 1.843e-04 & 12 \\
9.925e-01 & 1.423e-04 & 1.312e-04 & 1.493e-04 & 12 \\
1.106e+00 & 1.142e-04 & 1.051e-04 & 1.213e-04 & 11 \\
1.232e+00 & 8.635e-05 & 7.469e-05 & 9.571e-05 & 11 \\
1.373e+00 & 6.274e-05 & 5.512e-05 & 6.934e-05 & 11 \\
1.530e+00 & 4.767e-05 & 4.269e-05 & 5.336e-05 & 10 \\
1.704e+00 & 3.720e-05 & 3.132e-05 & 4.402e-05 & 5 \\
1.899e+00 & 2.769e-05 & 2.355e-05 & 3.228e-05 & 2 \\
\hline \multicolumn{5}{r}{{Continued on next page}} \\ \hline
\end{tabular}
\end{center}
\end{table*}
\addtocounter{table}{-1}

\begin{table*}
\begin{center}
\caption{continued. }
\renewcommand{\arraystretch}{0.95}% Tighter

\begin{tabular}{ c c c c c c c c c }
\multicolumn{9}{c}{\bf Pressure} \\
$\frac{R}{R_{500}}$ & $\left( \frac{P}{P_{500}} \right)_{X}$ & $\left( \frac{P}{P_{500}} \right)_{X,low}$ & $\left( \frac{P}{P_{500}} \right)_{X,high}$ & $N_{X}$ & $\left( \frac{P}{P_{500}} \right)_{SZ}$ & $\left( \frac{P}{P_{500}} \right)_{SZ,low}$ & $\left( \frac{P}{P_{500}} \right)_{SZ,high}$ & $N_{SZ}$ \\
\hline
2.283e-02 & 2.396e+01 & 1.380e+01 & 2.829e+01 & 10 & - & - & - & - \\
2.909e-02 & 2.110e+01 & 1.243e+01 & 2.603e+01 & 10 & - & - & - & - \\
3.706e-02 & 2.198e+01 & 1.420e+01 & 2.376e+01 & 11 & - & - & - & - \\
4.723e-02 & 1.995e+01 & 1.447e+01 & 2.123e+01 & 11 & - & - & - & - \\
6.018e-02 & 1.569e+01 & 1.025e+01 & 1.847e+01 & 12 & - & - & - & - \\
7.669e-02 & 1.318e+01 & 8.647e+00 & 1.557e+01 & 12 & - & - & - & - \\
9.772e-02 & 1.086e+01 & 7.665e+00 & 1.283e+01 & 12 & - & - & - & - \\
1.245e-01 & 8.800e+00 & 7.024e+00 & 1.022e+01 & 12 & - & - & - & - \\
1.587e-01 & 6.847e+00 & 5.763e+00 & 7.680e+00 & 12 & - & - & - & - \\
2.022e-01 & 5.144e+00 & 4.492e+00 & 5.694e+00 & 12 & - & - & - & - \\
2.577e-01 & 3.896e+00 & 3.390e+00 & 4.181e+00 & 12 & - & - & - & - \\
3.283e-01 & 2.702e+00 & 2.537e+00 & 2.837e+00 & 12 & - & - & - & - \\
4.184e-01 & 1.834e+00 & 1.748e+00 & 1.921e+00 & 12 & - & - & - & - \\
5.331e-01 & 1.093e+00 & 1.059e+00 & 1.152e+00 & 12 & 9.885e-01 & 9.060e-01 & 1.077e+00 & 11 \\
6.794e-01 & 6.140e-01 & 5.636e-01 & 6.594e-01 & 12 & 6.075e-01 & 5.461e-01 & 6.765e-01 & 12 \\
8.657e-01 & 3.260e-01 & 2.643e-01 & 3.803e-01 & 8 & 3.571e-01 & 3.031e-01 & 4.056e-01 & 12 \\
1.103e+00 & - & - & - & - & 1.939e-01 & 1.562e-01 & 2.242e-01 & 12 \\
1.406e+00 & - & - & - & - & 9.760e-02 & 7.288e-02 & 1.132e-01 & 12 \\
1.791e+00 & - & - & - & - & 4.649e-02 & 3.710e-02 & 5.343e-02 & 11 \\
\hline
\hline
\end{tabular}

%\end{center}
%\end{table*}

%\begin{table*}
%\begin{center}
%\caption{\label{tab:stack}Stacked thermodynamic profiles. $N_{X}$ and $N_{SZ}$ indicates respectively the number or objects reaching the indicated radius. }

\begin{tabular}{ c c c c c c c c c }
\multicolumn{9}{c}{\bf Temperature} \\
$\frac{R}{R_{500}}$ & $\left( \frac{T}{T_{500}} \right)_{X}$ & $\left( \frac{T}{T_{500}} \right)_{X,low}$ & $\left( \frac{T}{T_{500}} \right)_{X,high}$ & $N_{X}$ & $\left( \frac{T}{T_{500}} \right)_{SZ}$ & $\left( \frac{T}{T_{500}} \right)_{SZ,low}$ & $\left( \frac{T}{T_{500}} \right)_{SZ,high}$ & $N_{SZ}$ \\
\hline
2.283e-02 & 7.606e-01 & 7.049e-01 & 8.741e-01 & 10 & - & - & - & - \\
2.909e-02 & 7.776e-01 & 7.338e-01 & 8.823e-01 & 10 & - & - & - & - \\
3.706e-02 & 8.259e-01 & 7.814e-01 & 9.534e-01 & 11 & - & - & - & - \\
4.723e-02 & 8.485e-01 & 8.086e-01 & 9.598e-01 & 11 & - & - & - & - \\
6.018e-02 & 8.759e-01 & 8.434e-01 & 9.510e-01 & 12 & - & - & - & - \\
7.669e-02 & 9.102e-01 & 8.705e-01 & 9.650e-01 & 12 & - & - & - & - \\
9.772e-02 & 9.460e-01 & 9.028e-01 & 9.913e-01 & 12 & - & - & - & - \\
1.245e-01 & 9.669e-01 & 9.314e-01 & 1.011e+00 & 12 & - & - & - & - \\
1.587e-01 & 9.717e-01 & 9.425e-01 & 1.021e+00 & 12 & - & - & - & - \\
2.022e-01 & 9.816e-01 & 9.354e-01 & 1.024e+00 & 12 & - & - & - & - \\
2.577e-01 & 9.751e-01 & 9.518e-01 & 1.005e+00 & 12 & - & - & - & - \\
3.283e-01 & 9.540e-01 & 9.347e-01 & 9.794e-01 & 12 & - & - & - & - \\
4.184e-01 & 9.317e-01 & 8.946e-01 & 9.662e-01 & 12 & - & - & - & - \\
5.331e-01 & 8.714e-01 & 8.409e-01 & 8.970e-01 & 12 & 7.616e-01 & 7.053e-01 & 8.134e-01 & 11 \\
6.794e-01 & 7.529e-01 & 7.093e-01 & 7.937e-01 & 12 & 7.253e-01 & 6.779e-01 & 7.748e-01 & 12 \\
8.657e-01 & 6.540e-01 & 6.061e-01 & 6.917e-01 & 8 & 6.872e-01 & 6.461e-01 & 7.280e-01 & 12 \\
1.103e+00 & - & - & - & - & 6.381e-01 & 6.001e-01 & 6.673e-01 & 12 \\
1.406e+00 & - & - & - & - & 5.664e-01 & 5.317e-01 & 6.012e-01 & 11 \\
1.791e+00 & - & - & - & - & 4.852e-01 & 4.380e-01 & 5.406e-01 & 6 \\
\hline
\hline
\end{tabular}

%\end{center}
%\end{table*}

%\begin{table*}
%\begin{center}
%\caption{\label{tab:stack}Stacked thermodynamic profiles. $N_{X}$ and $N_{SZ}$ indicates respectively the number or objects reaching the indicated radius. }

\begin{tabular}{ c c c c c c c c c }
\multicolumn{9}{c}{\bf Entropy} \\
$\frac{R}{R_{500}}$ & $\left( \frac{K}{K_{500}} \right)_{X}$ & $\left( \frac{K}{K_{500}} \right)_{X,low}$ & $\left( \frac{K}{K_{500}} \right)_{X,high}$ & $N_{X}$ & $\left( \frac{K}{K_{500}} \right)_{SZ}$ & $\left( \frac{K}{K_{500}} \right)_{SZ,low}$ & $\left( \frac{K}{K_{500}} \right)_{SZ,high}$ & $N_{SZ}$ \\
\hline
2.283e-02 & 9.245e-02 & 5.696e-02 & 1.159e-01 & 10 & - & - & - & - \\
2.909e-02 & 1.058e-01 & 5.736e-02 & 1.192e-01 & 10 & - & - & - & - \\
3.706e-02 & 1.116e-01 & 7.998e-02 & 1.377e-01 & 11 & - & - & - & - \\
4.723e-02 & 1.303e-01 & 8.800e-02 & 1.516e-01 & 11 & - & - & - & - \\
6.018e-02 & 1.455e-01 & 1.059e-01 & 1.643e-01 & 12 & - & - & - & - \\
7.669e-02 & 1.701e-01 & 1.403e-01 & 1.863e-01 & 12 & - & - & - & - \\
9.772e-02 & 2.024e-01 & 1.745e-01 & 2.263e-01 & 12 & - & - & - & - \\
1.245e-01 & 2.367e-01 & 2.107e-01 & 2.637e-01 & 12 & - & - & - & - \\
1.587e-01 & 2.879e-01 & 2.584e-01 & 3.121e-01 & 12 & - & - & - & - \\
2.022e-01 & 3.412e-01 & 3.060e-01 & 3.960e-01 & 12 & - & - & - & - \\
2.577e-01 & 4.147e-01 & 3.895e-01 & 4.615e-01 & 12 & - & - & - & - \\
3.283e-01 & 4.881e-01 & 4.737e-01 & 5.074e-01 & 12 & - & - & - & - \\
4.184e-01 & 6.452e-01 & 5.985e-01 & 6.999e-01 & 12 & - & - & - & - \\
5.331e-01 & 7.886e-01 & 7.595e-01 & 8.156e-01 & 12 & 6.631e-01 & 6.304e-01 & 6.973e-01 & 11 \\
6.794e-01 & 8.852e-01 & 8.467e-01 & 9.227e-01 & 12 & 8.607e-01 & 8.225e-01 & 8.982e-01 & 12 \\
8.657e-01 & 1.027e+00 & 9.461e-01 & 1.104e+00 & 8 & 1.107e+00 & 1.046e+00 & 1.168e+00 & 12 \\
1.103e+00 & - & - & - & - & 1.460e+00 & 1.360e+00 & 1.569e+00 & 12 \\
1.406e+00 & - & - & - & - & 2.060e+00 & 1.779e+00 & 2.389e+00 & 11 \\
1.791e+00 & - & - & - & - & 2.774e+00 & 1.658e+00 & 4.491e+00 & 6 \\
\hline
\end{tabular}

\end{center}
\end{table*}

\section{Marginalized posterior likelihood}

\begin{figure*}
\includegraphics[width=\textwidth]{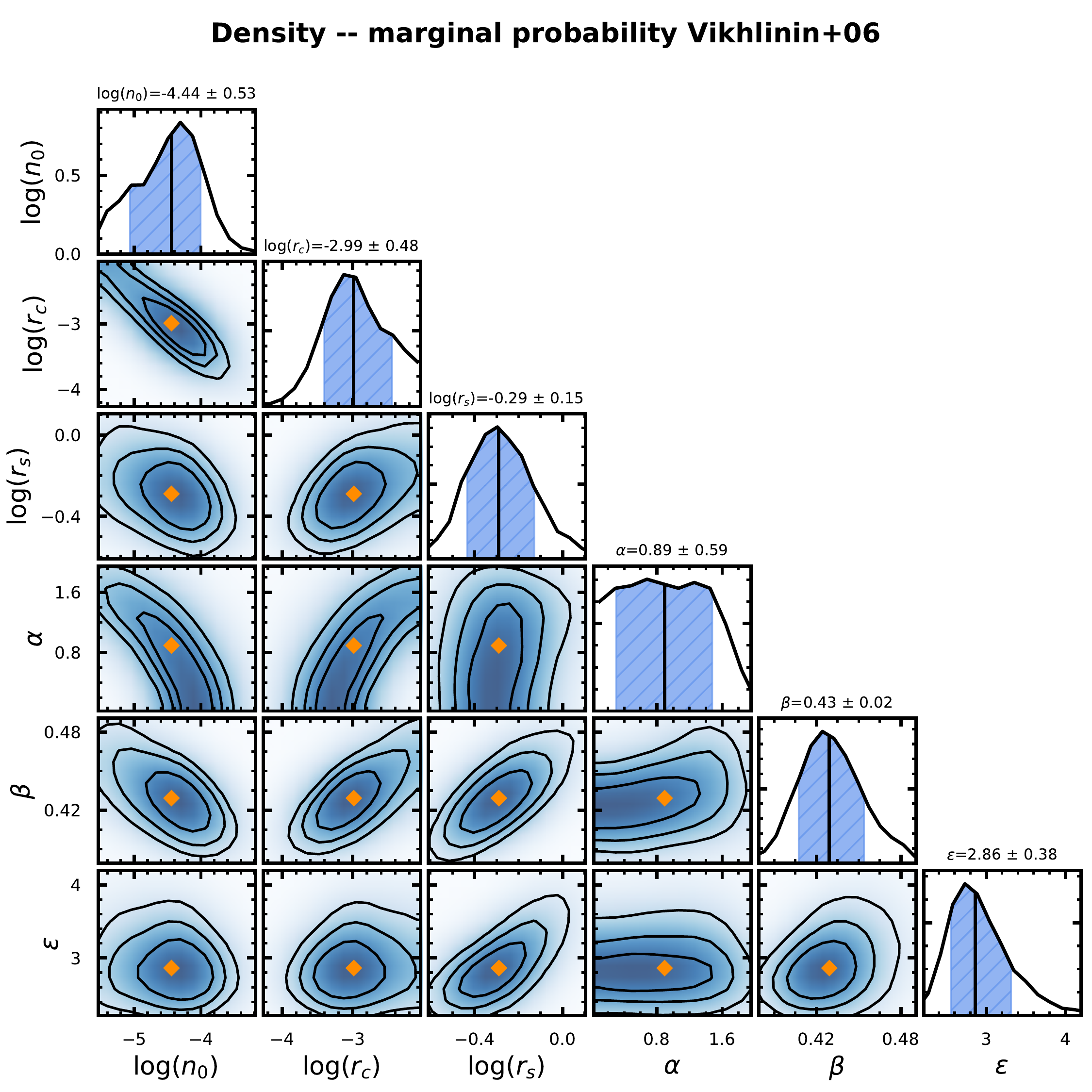}
\caption{Parameter distribution for the best fit on the density of all clusters using the \cite{vikhlini+06} functional form, Eq.~\eqref{eq:vikh_density}. The priors on the parameters are shown in Table~\ref{tab:forms}.}
\label{fig:posterior_density_vikh}
\end{figure*}

\begin{figure*}
\includegraphics[width=\textwidth]{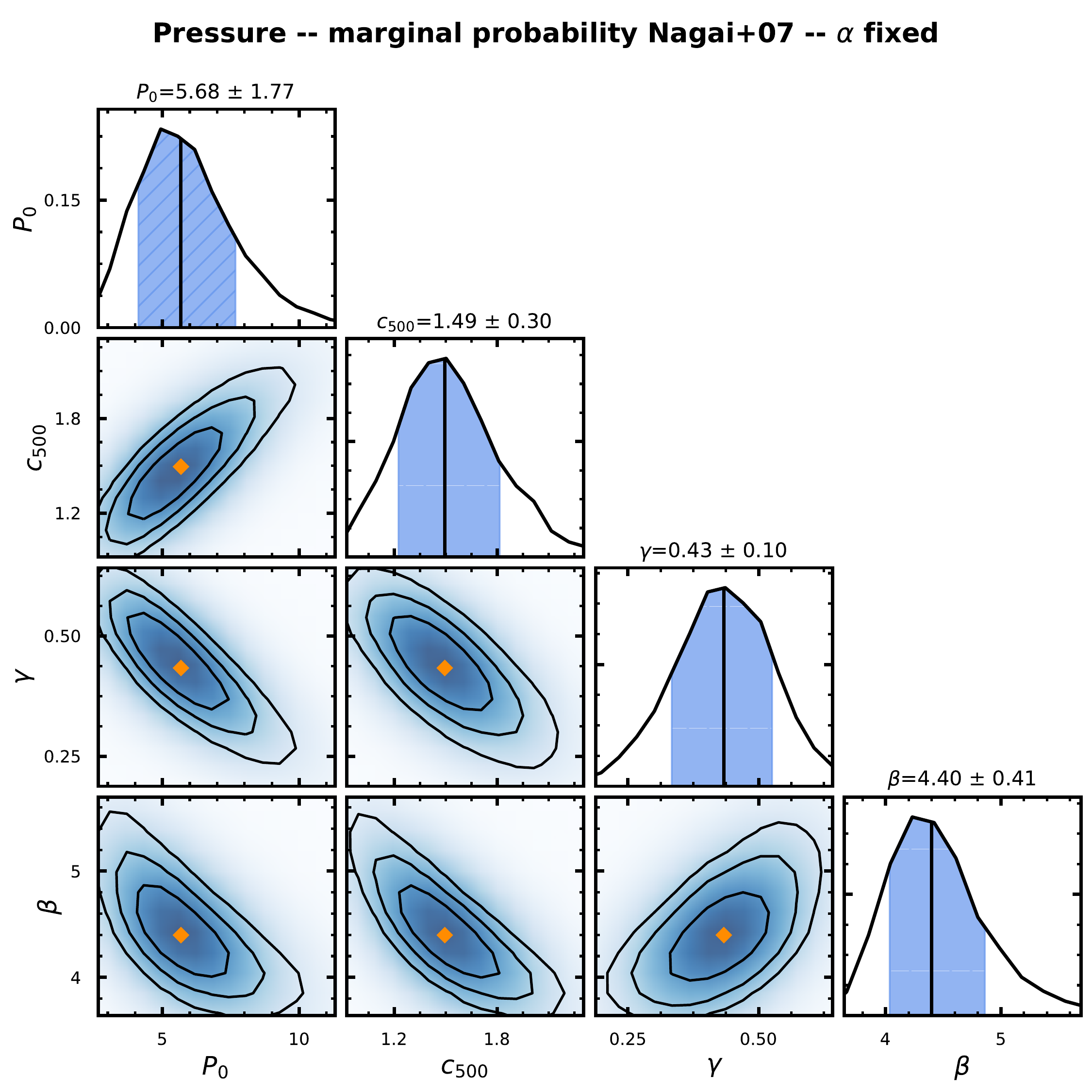}
\caption{Parameter distribution for the best fit on the pressure of all clusters using the \cite{nagai+07} gNFW functional form, Eq.~\eqref{eq:nagai}, fixing the intermediate slope $\alpha$. The priors on the parameters are shown in Table~\ref{tab:forms}.}
\label{fig:posterior_pressure_nagai_fix1}
\end{figure*}

\begin{figure*}
\includegraphics[width=\textwidth]{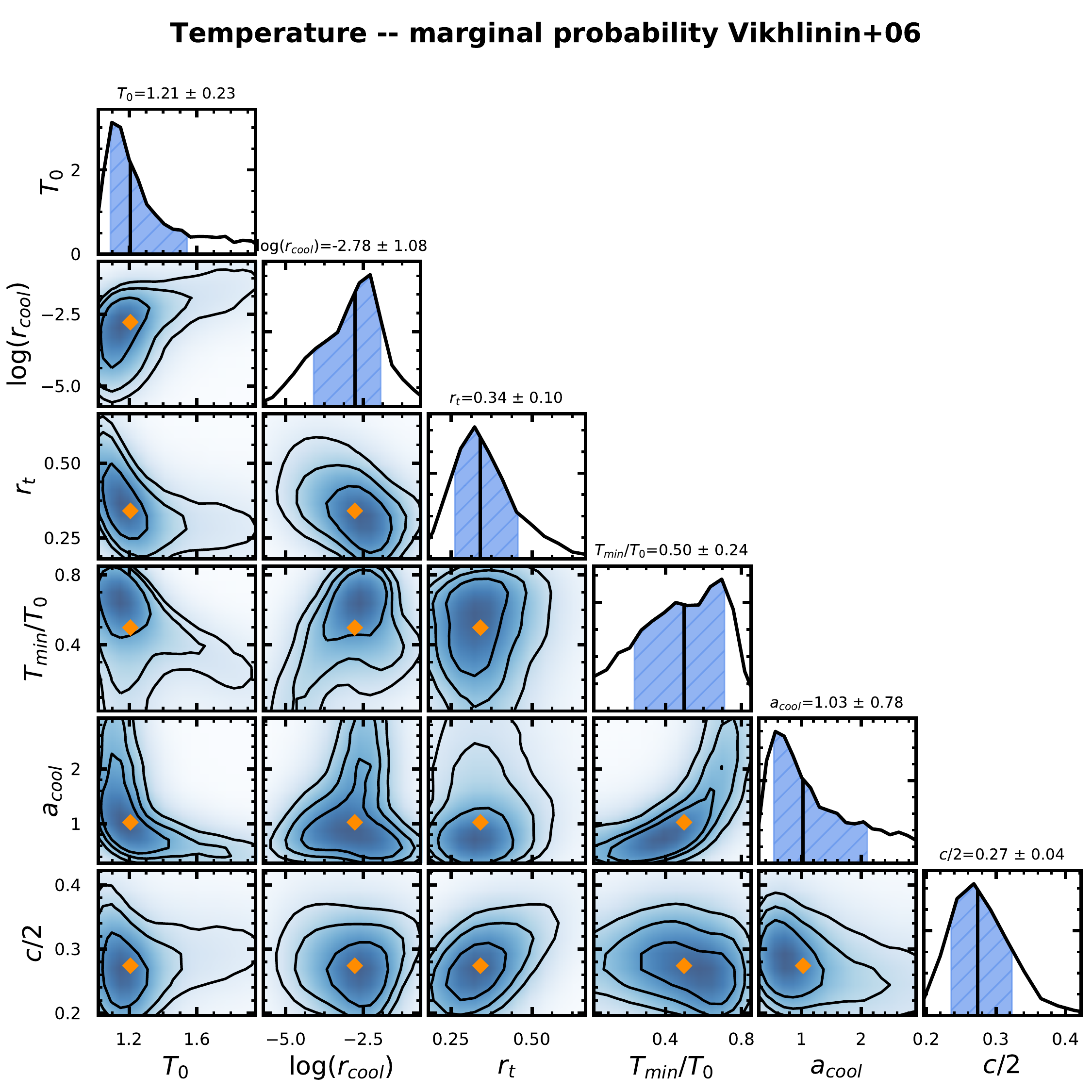}
\caption{Parameter distribution for the best fit on the temperature of all clusters using the \cite{vikhlini+06} functional form, Eq.~\eqref{eq:vikh_temperature}. The priors on the parameters are shown in Table~\ref{tab:forms}.        }
\label{fig:posterior_temperature_vikh}
\end{figure*}

\begin{figure*}
\includegraphics[width=\textwidth]{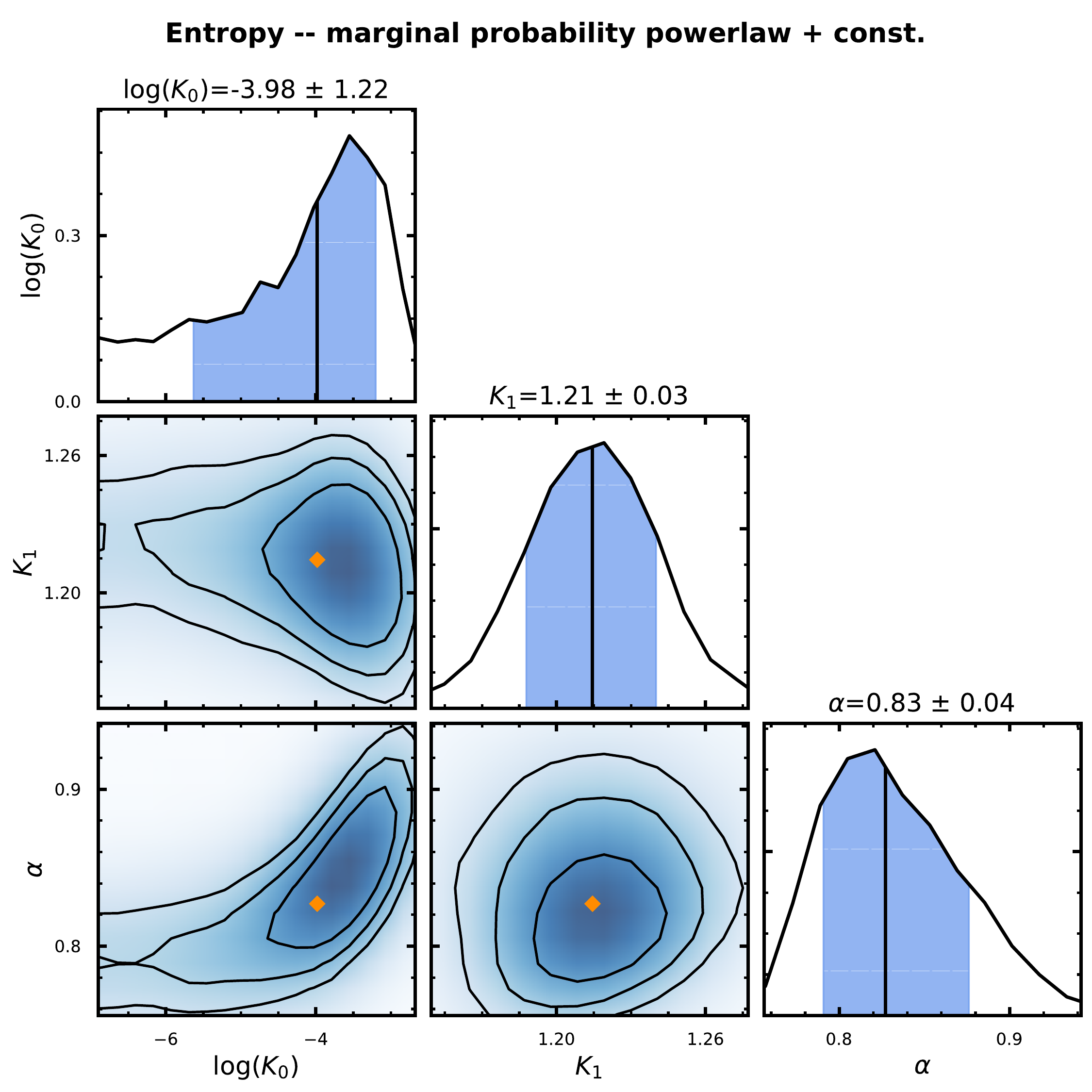}
\caption{Parameter distribution for the best fit on the entropy of all clusters using a power law plus constant functional form, Eq.~\eqref{eq:powlc_entropy}.     The priors on the parameters are shown in Table~\ref{tab:forms}.}
\label{fig:posterior_entropy_powlc}
\end{figure*}

\section{Log of scientific observations}

\onecolumn
{\setstretch{1.3}
%\begin{longtab}
\begin{longtable}{lcccccc}
\caption{\label{tab:log}Log of X-COP observations.}\\
\hline\hline
Target & Obs.Id. & Obs.Date & $N_H$ & $t_{M1}$ &  $t_{M2}$ & $t_{pn}$\\
\, & \, & [yr/mm/dd] & [$10^{20} \rm cm^{-2}$] & [ks] & [ks] & [ks] \\
\hline
\endfirsthead
\caption{continued.}\\
\hline\hline
Target & Obs.Id. & Obs.Date & $N_H$ & $t_{M1}$ &  $t_{M2}$ & $t_{pn}$\\
\, & \, & [yr/mm/dd] & [$10^{20} \rm cm^{-2}$] & [ks] & [ks] & [ks] \\
\hline
\endhead
\hline
\endfoot
\input tablea1.tex
\end{longtable}
\noindent \textbf{Columns:} 1. Target name; 2. Observation identifier; 3. Observation date; 4. Equivalent hydrogen column density as estimated from 21 cm maps \citep{kalberla05}; 4. Exposure time for MOS1 detector after flare removal; 5. Exposure time for MOS2 detector after flare removal; 6. Exposure time for pn detector after flare removal. 
%\end{longtab}
}
\twocolumn

\end{appendix}

\end{document}

%% file: tablea1.tex
A1644 Center & 0010420201 & 2001-01-08 & 4.2 & 13.5 & 13.6 & 11.9 \\
A1644 E & 0744413001 & 2015-01-01 & 4.3 & 15.5 & 17.5 & 10.3 \\
A1644 N & 0744412701 & 2014-12-29 & 3.9 & 27.5 & 28.4 & 14.8 \\
A1644 S & 0744412901 & 2014-12-31 & 4.4 & 10.2 & 11.3 & 4.7 \\
A1644 W & 0744412801 & 2015-06-30 & 3.8 & 29.2 & 28.7 & 21.7 \\
A1795 Center & 0097820101 & 2000-06-26 & 1.2 & 36.1 & 36.5 & 26.0 \\
A1795 E & 0744412101 & 2015-06-16 & 1.2 & 19.3 & 25.1 & 10.4 \\
A1795 N & 0744412001 & 2015-01-05 & 1.2 & 15.3 & 16.3 & 11.2 \\
A1795 NW & 0205190201 & 2004-01-25 & 1.2 & 22.5 & 22.9 & 21.1 \\
A1795 S & 0109070201 & 2003-01-13 & 1.2 & 52.6 & 53.5 & 52.1 \\
A1795 W & 0205190101 & 2004-01-25 & 1.1 & 29.2 & 28.9 & 27.0 \\
A2029 Center 1 & 0551780201 & 2008-07-17 & 3.2 & 33.1 & 34.0 & 15.1 \\
A2029 Center 2 & 0551780301 & 2008-07-19 & 3.2 & 39.0 & 40.9 & 27.4 \\
A2029 E & 0744411201 & 2015-01-31 & 3.2 & 26.0 & 26.7 & 20.6 \\
A2029 N & 0744410901 & 2015-02-08 & 3.0 & 17.5 & 22.3 & 6.6 \\
A2029 S & 0744411101 & 2015-02-22 & 3.4 & 11.0 & 16.6 & 4.9 \\
A2029 W & 0744411001 & 2015-07-27 & 3.2 & 42.6 & 43.3 & 39.2 \\
A2142 Center & 0674560201 & 2011-07-13  & 3.8 & 52.3 & 53.8 & 48.8 \\
A2142 NE & 0694440201 & 2012-07-14 & 3.8 & 33.2 & 33.1 & 29.8 \\
A2142 NW & 0694440101 & 2012-07-14 & 3.7 & 19.6 & 18.5 & 12.5 \\
A2142 SE & 0694440501 & 2012-07-16 & 4.0 & 33.2 & 32.5 & 29.8 \\
A2142 SW & 0694440601 & 2012-07-18 & 3.9 & 30.3 & 31.5 & 24.1 \\
A2255 E & 0744410801 & 2014-03-30 & 2.5 & 12.1 & 17.8 & 9.4 \\
A2255 N & 0744410501 & 2014-03-14 & 2.5 & 14.0 & 19.9 & 7.5 \\
A2255 S & 0744410701 & 2014-03-28 & 2.5 & 23.4 & 23.8 & 12.4 \\
A2255 W & 0744410601 & 2014-04-27 & 2.4 & 28.6 & 30.2 & 22.1 \\
A2255 Center & 0112260801 & 2002-12-07 & 2.5 & 7.7 & 8.2 & 2.3 \\
A2319 E & 0744410401 & 2014-04-08 & 9.1 & 14.5 & 15.9 & 10.9 \\
A2319 N & 0744410101 & 2014-03-15 & 8.8 & 23.9 & 24.2 & 21.8 \\
A2319 S & 0744410301 & 2014-04-04 & 8.2 & 13.8 & 14.3 & 8.2 \\
A2319 W & 0744410201 & 2014-04-09 & 7.7 & 23.8 & 25.4 & 11.1 \\
A2319 Center 1 & 0302150101 & 2005-10-10 & 8.1 & 15.7 & 15.6 & 10.6 \\
A2319 Center 2 & 0302150201 & 2005-11-14 & 8.1 & 16.0 & 15.5 & 12.4 \\
A3158 Center & 0300210201 & 2005-11-22 & 1.4 & 19.8 & 19.7 & 11.2 \\
A3158 E & 0744411601 & 2015-08-29 & 1.4 & 31.2 & 31.7 & 26.2 \\
A3158 N & 0744411301 & 2014-11-12 & 1.3 & 28.1 & 28.6 & 21.5 \\
A3158 S & 0744411501 & 2015-05-31 & 1.3 & 28.0 & 28.6 & 19.3 \\
A3158 W & 0744411401 & 2015-03-01 & 1.3 & 20.8 & 20.2 & 16.4 \\
A3266 f1 & 0105260701 & 2000-10-01 & 1.5 & 19.1 & 19.5 & 15.5 \\
A3266 f2 & 0105260801 & 2000-10-11 & 1.5 & 19.6 & 19.6 & 15.5 \\
A3266 f3 & 0105260901 & 2000-10-09 & 1.6 & 23.4 & 23.1 & 17.9 \\
A3266 f4 & 0105262201 & 2000-09-27 & 1.5 & 3.0 & 2.9 & 3.2 \\
A3266 f5 & 0105262101 & 2000-09-25 & 1.8 & 5.8 & 6.1 & 4.1 \\
A3266 f5b & 0105261101 & 2000-09-25 & 1.8 & 11.3 & 12.3 & 7.1 \\
A3266 f6 & 0105262001 & 2000-09-23 & 1.7 & 6.2 & 5.6 & 2.5 \\
A3266 f6c & 0105262501 & 2003-03-15 & 1.7 & 6.3 & 6.8 & 3.1 \\
A644 Center & 0744412201 & 2014-04-07 & 7.5 & 19.7 & 25.4 & 11.8 \\
A644 E & 0744412601 & 2014-05-18 & 6.7 & 20.5 & 24.3 & 8.6 \\
A644 N & 0744412301 & 2014-10-22 & 7.4 & 35.1 & 35.8 & 30.3 \\
A644 S & 0744412501 & 2014-10-24 & 7.0 & 32.3 & 32.2 & 26.2 \\
A644 W & 0744412401 & 2015-04-08 & 7.5 & 30.6 & 30.4 & 18.9 \\
A85 Center & 0723802201 & 2013-06-18 & 2.8 & 95.8 & 97.9 & 85.2 \\
A85 E & 0744411901 & 2014-12-11 & 2.8 & 28.2 & 29.0 & 19.5 \\
A85 N & 0744411701 & 2015-01-12 & 2.9 & 22.0 & 23.4 & 14.8 \\
A85 S & 0065140201 & 2002-01-07 & 2.7 & 12.2 & 12.1 & 9.4 \\
A85 W & 0744411801 & 2015-06-06 & 2.8 & 28.4 & 28.2 & 21.1 \\
A85 NW & 0744930301 & 2014-06-23 & 2.8 & 29.2 & 30.4 & 21.0 \\
%Hydra A Center & 0504260101 & 2007-05-11 & 4.7 & 65.4 & 70.6 & 50.2 \\ 
%Hydra A NE & 0694440401 & 2012-06-02 & 4.7 & 29.1 & 29.9 & 23.5 \\ 
%Hydra A SE1 & 0694440301 & 2012-06-07 & 4.5 & 4.4 & 4.4 & 2.8 \\ 
%Hydra A SE2 & 0725240101 & 2013-11-02 & 4.5 & 17.9 & 21.1 & 8.6 \\ 
%Hydra A SW & 0694440801 & 2012-05-13 & 4.7 & 20.4 & 21.9 & 14.9 \\ 
%Hydra A NW & 0694440701 & 2012-05-19 & 4.7 & 18.7 & 21.1 & 11.2 \\ 
RXCJ1825 Center & 0744413501 & 2014-04-11 & 9.4 & 48.1 & 48.2 & 39.1 \\
RXCJ1825 E & 0744413901 & 2014-04-13 & 10.0 & 31.7 & 32.4 & 16.2 \\
RXCJ1825 N & 0744413601 & 2014-04-12 & 8.9 & 20.0 & 20.2 & 14.3 \\
RXCJ1825 S & 0744413801 & 2014-04-14 & 9.5 & 25.5 & 29.5 & 12.4 \\
RXCJ1825 W & 0744413701 & 2014-10-02 & 9.2 & 41.4 & 40.8 & 36.8 \\
ZwCl1215 Center & 0300211401 & 2006-06-24 & 1.7 & 23.3 & 24.0 & 16.3 \\
ZwCl1215 E & 0744413401 & 2015-12-17 & 1.7 & 26.0 & 26.0 & 21.1 \\
ZwCl1215 N & 0744413101 & 2014-12-06 & 1.7 & 18.2 & 18.5 & 10.1 \\
ZwCl1215 S & 0744413301 & 2015-06-04 & 1.8 & 28.9 & 28.7 & 24.4 \\
ZwCl1215 W & 0744413201 & 2015-06-11 & 1.8 & 21.4 & 24.2 & 12.9 \\

%% file: XCOP_thermo.bbl
\begin{thebibliography}{114}
\expandafter\ifx\csname natexlab\endcsname\relax\def\natexlab#1{#1}\fi

\bibitem[{{Abazajian} {et~al.}(2016){Abazajian}, {Adshead}, {Ahmed}, {Allen},
  {Alonso}, {Arnold}, {Baccigalupi}, {Bartlett}, {Battaglia}, {Benson},
  {Bischoff}, {Borrill}, {Buza}, {Calabrese}, {Caldwell}, {Carlstrom}, {Chang},
  {Crawford}, {Cyr-Racine}, {De Bernardis}, {de Haan}, {di Serego Alighieri},
  {Dunkley}, {Dvorkin}, {Errard}, {Fabbian}, {Feeney}, {Ferraro}, {Filippini},
  {Flauger}, {Fuller}, {Gluscevic}, {Green}, {Grin}, {Grohs}, {Henning},
  {Hill}, {Hlozek}, {Holder}, {Holzapfel}, {Hu}, {Huffenberger}, {Keskitalo},
  {Knox}, {Kosowsky}, {Kovac}, {Kovetz}, {Kuo}, {Kusaka}, {Le Jeune}, {Lee},
  {Lilley}, {Loverde}, {Madhavacheril}, {Mantz}, {Marsh}, {McMahon},
  {Meerburg}, {Meyers}, {Miller}, {Munoz}, {Nguyen}, {Niemack}, {Peloso},
  {Peloton}, {Pogosian}, {Pryke}, {Raveri}, {Reichardt}, {Rocha}, {Rotti},
  {Schaan}, {Schmittfull}, {Scott}, {Sehgal}, {Shandera}, {Sherwin}, {Smith},
  {Sorbo}, {Starkman}, {Story}, {van Engelen}, {Vieira}, {Watson}, {Whitehorn},
  \& {Kimmy Wu}}]{cmbs4}
{Abazajian}, K.~N., {Adshead}, P., {Ahmed}, Z., {et~al.} 2016, ArXiv e-prints
  [\eprint[arXiv]{1610.02743}]

\bibitem[{{Adam} {et~al.}(2017){Adam}, {Arnaud}, {Bartalucci}, {Ade},
  {Andr{\'e}}, {Beelen}, {Beno{\^\i}t}, {Bideaud}, {Billot}, {Bourdin},
  {Bourrion}, {Calvo}, {Catalano}, {Coiffard}, {Comis}, {D'Addabbo},
  {D{\'e}sert}, {Doyle}, {Ferrari}, {Goupy}, {Kramer}, {Lagache}, {Leclercq},
  {Mac{\'{\i}}as-P{\'e}rez}, {Maurogordato}, {Mauskopf}, {Mayet}, {Monfardini},
  {Pajot}, {Pascale}, {Perotto}, {Pisano}, {Pointecouteau}, {Ponthieu},
  {Pratt}, {Rev{\'e}ret}, {Ritacco}, {Rodriguez}, {Romero}, {Ruppin},
  {Schuster}, {Sievers}, {Triqueneaux}, {Tucker}, \& {Zylka}}]{adam17}
{Adam}, R., {Arnaud}, M., {Bartalucci}, I., {et~al.} 2017, \aap, 606, A64

\bibitem[{{Adam} {et~al.}(2015){Adam}, {Comis}, {Mac{\'{\i}}as-P{\'e}rez},
  {Adane}, {Ade}, {Andr{\'e}}, {Beelen}, {Belier}, {Beno{\^i}t}, {Bideaud},
  {Billot}, {Blanquer}, {Bourrion}, {Calvo}, {Catalano}, {Coiffard},
  {Cruciani}, {D'Addabbo}, {D{\'e}sert}, {Doyle}, {Goupy}, {Kramer},
  {Leclercq}, {Martino}, {Mauskopf}, {Mayet}, {Monfardini}, {Pajot}, {Pascale},
  {Perotto}, {Pointecouteau}, {Ponthieu}, {Rev{\'e}ret}, {Ritacco},
  {Rodriguez}, {Savini}, {Schuster}, {Sievers}, {Tucker}, \& {Zylka}}]{adam15}
{Adam}, R., {Comis}, B., {Mac{\'{\i}}as-P{\'e}rez}, J.-F., {et~al.} 2015, \aap,
  576, A12

\bibitem[{{Akamatsu} {et~al.}(2011){Akamatsu}, {Hoshino}, {Ishisaki}, {Ohashi},
  {Sato}, {Takei}, \& {Ota}}]{akamatsu11}
{Akamatsu}, H., {Hoshino}, A., {Ishisaki}, Y., {et~al.} 2011, \pasj, 63, 1019

\bibitem[{{Ameglio} {et~al.}(2007){Ameglio}, {Borgani}, {Pierpaoli}, \&
  {Dolag}}]{ameglio+07}
{Ameglio}, S., {Borgani}, S., {Pierpaoli}, E., \& {Dolag}, K. 2007, \mnras,
  382, 397

\bibitem[{{Anders} \& {Grevesse}(1989)}]{ag89}
{Anders}, E. \& {Grevesse}, N. 1989, \gca, 53, 197

\bibitem[{{Andrade-Santos} {et~al.}(2017){Andrade-Santos}, {Jones}, {Forman},
  {Lovisari}, {Vikhlinin}, {van Weeren}, {Murray}, {Arnaud}, {Pratt},
  {D{\'e}mocl{\`e}s}, {Kraft}, {Mazzotta}, {B{\"o}hringer}, {Chon},
  {Giacintucci}, {Clarke}, {Borgani}, {David}, {Douspis}, {Pointecouteau},
  {Dahle}, {Brown}, {Aghanim}, \& {Rasia}}]{andrade-santos17}
{Andrade-Santos}, F., {Jones}, C., {Forman}, W.~R., {et~al.} 2017, \apj, 843,
  76

\bibitem[{{Arnaud} {et~al.}(2010){Arnaud}, {Pratt}, {Piffaretti},
  {B{\"o}hringer}, {Croston}, \& {Pointecouteau}}]{arnaud+10}
{Arnaud}, M., {Pratt}, G.~W., {Piffaretti}, R., {et~al.} 2010, \aap, 517, A92

\bibitem[{{Avestruz} {et~al.}(2015){Avestruz}, {Nagai}, {Lau}, \&
  {Nelson}}]{avestruz15}
{Avestruz}, C., {Nagai}, D., {Lau}, E.~T., \& {Nelson}, K. 2015, \apj, 808, 176

\bibitem[{{Barnes} {et~al.}(2017){Barnes}, {Vogelsberger}, {Kannan},
  {Marinacci}, {Weinberger}, {Springel}, {Torrey}, {Pillepich}, {Nelson},
  {Pakmor}, {Naiman}, {Hernquist}, \& {McDonald}}]{barnes17}
{Barnes}, D.~J., {Vogelsberger}, M., {Kannan}, R., {et~al.} 2017, ArXiv
  e-prints [\eprint[arXiv]{1710.08420}]

\bibitem[{{Basu} {et~al.}(2010){Basu}, {Zhang}, {Sommer}, {Bender}, {Bertoldi},
  {Dobbs}, {Eckmiller}, {Halverson}, {Holzapfel}, {Horellou}, {Jaritz},
  {Johansson}, {Johnson}, {Kennedy}, {Kneissl}, {Lanting}, {Lee}, {Mehl},
  {Menten}, {Navarrete}, {Pacaud}, {Reichardt}, {Reiprich}, {Richards},
  {Schwan}, \& {Westbrook}}]{basu10}
{Basu}, K., {Zhang}, Y.-Y., {Sommer}, M.~W., {et~al.} 2010, \aap, 519, A29

\bibitem[{{Borgani} {et~al.}(2005){Borgani}, {Finoguenov}, {Kay}, {Ponman},
  {Springel}, {Tozzi}, \& {Voit}}]{borgani05}
{Borgani}, S., {Finoguenov}, A., {Kay}, S.~T., {et~al.} 2005, \mnras, 361, 233

\bibitem[{{Borgani} {et~al.}(2004){Borgani}, {Murante}, {Springel}, {Diaferio},
  {Dolag}, {Moscardini}, {Tormen}, {Tornatore}, \& {Tozzi}}]{borgani04}
{Borgani}, S., {Murante}, G., {Springel}, V., {et~al.} 2004, \mnras, 348, 1078

\bibitem[{{Bourdin} {et~al.}(2017){Bourdin}, {Mazzotta}, {Kozmanyan}, {Jones},
  \& {Vikhlinin}}]{bourdin+17}
{Bourdin}, H., {Mazzotta}, P., {Kozmanyan}, A., {Jones}, C., \& {Vikhlinin}, A.
  2017, \apj, 843, 72

\bibitem[{{Bryan} \& {Norman}(1998)}]{bryan98}
{Bryan}, G.~L. \& {Norman}, M.~L. 1998, \apj, 495, 80

\bibitem[{{Burns} {et~al.}(2010){Burns}, {Skillman}, \& {O'Shea}}]{burns10}
{Burns}, J.~O., {Skillman}, S.~W., \& {O'Shea}, B.~W. 2010, \apj, 721, 1105

\bibitem[{{Cappellari} \& {Copin}(2003)}]{cappellari03}
{Cappellari}, M. \& {Copin}, Y. 2003, \mnras, 342, 345

\bibitem[{{Cash}(1979)}]{cash79}
{Cash}, W. 1979, \apj, 228, 939

\bibitem[{{Cavagnolo} {et~al.}(2009){Cavagnolo}, {Donahue}, {Voit}, \&
  {Sun}}]{cavagnolo+09}
{Cavagnolo}, K.~W., {Donahue}, M., {Voit}, G.~M., \& {Sun}, M. 2009, \apjs,
  182, 12

\bibitem[{{Croston} {et~al.}(2006){Croston}, {Arnaud}, {Pointecouteau}, \&
  {Pratt}}]{croston+06}
{Croston}, J.~H., {Arnaud}, M., {Pointecouteau}, E., \& {Pratt}, G.~W. 2006,
  \aap, 459, 1007

\bibitem[{{De Grandi} \& {Molendi}(2002)}]{degrandi2002}
{De Grandi}, S. \& {Molendi}, S. 2002, \apj, 567, 163

\bibitem[{{De Luca} \& {Molendi}(2004)}]{deluca04}
{De Luca}, A. \& {Molendi}, S. 2004, \aap, 419, 837

\bibitem[{{Diaz-Rodriguez} {et~al.}(2017){Diaz-Rodriguez}, {Eckert},
  {Monajemi}, {Paltani}, \& {Sardy}}]{diaz17}
{Diaz-Rodriguez}, J., {Eckert}, D., {Monajemi}, H., {Paltani}, M., \& {Sardy},
  S. 2017, Submitted to Annals of Applied Statistics
  [\eprint[arXiv]{1703.00654}]

\bibitem[{{Diemer} \& {Kravtsov}(2014)}]{diemer14}
{Diemer}, B. \& {Kravtsov}, A.~V. 2014, \apj, 789, 1

\bibitem[{{Diemer} {et~al.}(2017){Diemer}, {Mansfield}, {Kravtsov}, \&
  {More}}]{diemer17}
{Diemer}, B., {Mansfield}, P., {Kravtsov}, A.~V., \& {More}, S. 2017, \apj,
  843, 140

\bibitem[{{Dolag} {et~al.}(1999){Dolag}, {Bartelmann}, \& {Lesch}}]{dolag+99}
{Dolag}, K., {Bartelmann}, M., \& {Lesch}, H. 1999, \aap, 348, 351

\bibitem[{{Dolag} {et~al.}(2005){Dolag}, {Vazza}, {Brunetti}, \&
  {Tormen}}]{dolag05}
{Dolag}, K., {Vazza}, F., {Brunetti}, G., \& {Tormen}, G. 2005, \mnras, 364,
  753

\bibitem[{{Ebeling} {et~al.}(2006){Ebeling}, {White}, \&
  {Rangarajan}}]{ebeling06}
{Ebeling}, H., {White}, D.~A., \& {Rangarajan}, F.~V.~N. 2006, \mnras, 368, 65

\bibitem[{{Eckert} {et~al.}(2016){Eckert}, {Ettori}, {Coupon}, {Gastaldello},
  {Pierre}, {Melin}, {Le Brun}, {McCarthy}, {Adami}, {Chiappetti}, {Faccioli},
  {Giles}, {Lavoie}, {Lef{\`e}vre}, {Lieu}, {Mantz}, {Maughan}, {McGee},
  {Pacaud}, {Paltani}, {Sadibekova}, {Smith}, \& {Ziparo}}]{eckert16}
{Eckert}, D., {Ettori}, S., {Coupon}, J., {et~al.} 2016, \aap, 592, A12

\bibitem[{{Eckert} {et~al.}(2017){Eckert}, {Ettori}, {Pointecouteau},
  {Molendi}, {Paltani}, \& {Tchernin}}]{xcop}
{Eckert}, D., {Ettori}, S., {Pointecouteau}, E., {et~al.} 2017, Astronomische
  Nachrichten, 338, 293

\bibitem[{{Eckert} {et~al.}(2014){Eckert}, {Molendi}, {Owers}, {Gaspari},
  {Venturi}, {Rudnick}, {Ettori}, {Paltani}, {Gastaldello}, \&
  {Rossetti}}]{eckert14}
{Eckert}, D., {Molendi}, S., {Owers}, M., {et~al.} 2014, \aap, 570, A119

\bibitem[{{Eckert} {et~al.}(2013){Eckert}, {Molendi}, {Vazza}, {Ettori}, \&
  {Paltani}}]{eckert13a}
{Eckert}, D., {Molendi}, S., {Vazza}, F., {Ettori}, S., \& {Paltani}, S. 2013,
  \aap, 551, A22

\bibitem[{{Eckert} {et~al.}(2015){Eckert}, {Roncarelli}, {Ettori}, {Molendi},
  {Vazza}, {Gastaldello}, \& {Rossetti}}]{eckert15}
{Eckert}, D., {Roncarelli}, M., {Ettori}, S., {et~al.} 2015, \mnras, 447, 2198

\bibitem[{{Eckert} {et~al.}(2012){Eckert}, {Vazza}, {Ettori}, {Molendi},
  {Nagai}, {Lau}, {Roncarelli}, {Rossetti}, {Snowden}, \&
  {Gastaldello}}]{eckert12}
{Eckert}, D., {Vazza}, F., {Ettori}, S., {et~al.} 2012, \aap, 541, A57

\bibitem[{{Erler} {et~al.}(2018){Erler}, {Basu}, {Chluba}, \&
  {Bertoldi}}]{erler18}
{Erler}, J., {Basu}, K., {Chluba}, J., \& {Bertoldi}, F. 2018, \mnras
  [\eprint[arXiv]{1709.01187}]

\bibitem[{{Ettori} {et~al.}(2002){Ettori}, {De Grandi}, \&
  {Molendi}}]{ettori+02}
{Ettori}, S., {De Grandi}, S., \& {Molendi}, S. 2002, \aap, 391, 841

\bibitem[{{Ettori} {et~al.}(2010){Ettori}, {Gastaldello}, {Leccardi},
  {Molendi}, {Rossetti}, {Buote}, \& {Meneghetti}}]{ettori+10}
{Ettori}, S., {Gastaldello}, F., {Leccardi}, A., {et~al.} 2010, \aap, 524, A68

\bibitem[{{Feroz} {et~al.}(2009){Feroz}, {Hobson}, \& {Bridges}}]{multinest}
{Feroz}, F., {Hobson}, M.~P., \& {Bridges}, M. 2009, \mnras, 398, 1601

\bibitem[{{Foreman-Mackey} {et~al.}(2013){Foreman-Mackey}, {Hogg}, {Lang}, \&
  {Goodman}}]{foreman-mackey13}
{Foreman-Mackey}, D., {Hogg}, D.~W., {Lang}, D., \& {Goodman}, J. 2013, \pasp,
  125, 306

\bibitem[{{Frenk} {et~al.}(1999){Frenk}, {White}, {Bode}, {Bond}, {Bryan},
  {Cen}, {Couchman}, {Evrard}, {Gnedin}, {Jenkins}, {Khokhlov}, {Klypin},
  {Navarro}, {Norman}, {Ostriker}, {Owen}, {Pearce}, {Pen}, {Steinmetz},
  {Thomas}, {Villumsen}, {Wadsley}, {Warren}, {Xu}, \& {Yepes}}]{frenk99}
{Frenk}, C.~S., {White}, S.~D.~M., {Bode}, P., {et~al.} 1999, \apj, 525, 554

\bibitem[{{Gaspari} {et~al.}(2014){Gaspari}, {Brighenti}, {Temi}, \&
  {Ettori}}]{gaspari14}
{Gaspari}, M., {Brighenti}, F., {Temi}, P., \& {Ettori}, S. 2014, \apjl, 783,
  L10

\bibitem[{{Gaspari} {et~al.}(2018){Gaspari}, {McDonald}, {Hamer}, {Brighenti},
  {Temi}, {Gendron-Marsolais}, {Hlavacek-Larrondo}, {Edge}, {Werner}, {Tozzi},
  {Sun}, {Stone}, {Tremblay}, {Hogan}, {Eckert}, {Ettori}, {Yu}, {Biffi}, \&
  {Planelles}}]{gaspari18}
{Gaspari}, M., {McDonald}, M., {Hamer}, S.~L., {et~al.} 2018, \apj, 854, 167

\bibitem[{{Gaspari} {et~al.}(2012){Gaspari}, {Ruszkowski}, \&
  {Sharma}}]{gaspari12}
{Gaspari}, M., {Ruszkowski}, M., \& {Sharma}, P. 2012, \apj, 746, 94

\bibitem[{{Ghirardini} {et~al.}(2017){Ghirardini}, {Ettori}, {Eckert},
  {Molendi}, {Gastaldello}, {Pointecouteau}, {Hurier}, \&
  {Bourdin}}]{ghirardini18}
{Ghirardini}, V., {Ettori}, S., {Eckert}, D., {et~al.} 2017, ArXiv e-prints
  [\eprint[arXiv]{1708.02954}]

\bibitem[{{Hahn} {et~al.}(2017){Hahn}, {Martizzi}, {Wu}, {Evrard}, {Teyssier},
  \& {Wechsler}}]{hahn17}
{Hahn}, O., {Martizzi}, D., {Wu}, H.-Y., {et~al.} 2017, \mnras, 470, 166

\bibitem[{{Hoshino} {et~al.}(2010){Hoshino}, {Patrick Henry}, {Sato},
  {Akamatsu}, {Yokota}, {Sasaki}, {Ishisaki}, {Ohashi}, {Bautz}, {Fukazawa},
  {Kawano}, {Furuzawa}, {Hayashida}, {Tawa}, {Hughes}, {Kokubun}, \&
  {Tamura}}]{hoshino10}
{Hoshino}, A., {Patrick Henry}, J., {Sato}, K., {et~al.} 2010, \pasj, 62, 371

\bibitem[{{Hurier}(2016)}]{hurier16}
{Hurier}, G. 2016, \aap, 596, A61

\bibitem[{{Hurier} {et~al.}(2013){Hurier}, {Mac{\'{\i}}as-P{\'e}rez}, \&
  {Hildebrandt}}]{hurier+13}
{Hurier}, G., {Mac{\'{\i}}as-P{\'e}rez}, J.~F., \& {Hildebrandt}, S. 2013,
  \aap, 558, A118

\bibitem[{{Itoh} {et~al.}(1998){Itoh}, {Kohyama}, \& {Nozawa}}]{itoh98}
{Itoh}, N., {Kohyama}, Y., \& {Nozawa}, S. 1998, \apj, 502, 7

\bibitem[{{Jones} \& {Forman}(1984)}]{Jones+84}
{Jones}, C. \& {Forman}, W. 1984, \apj, 276, 38

\bibitem[{{Kaiser}(1986)}]{kaiser86}
{Kaiser}, N. 1986, \mnras, 222, 323

\bibitem[{{Kalberla} {et~al.}(2005){Kalberla}, {Burton}, {Hartmann}, {Arnal},
  {Bajaja}, {Morras}, \& {P{\"o}ppel}}]{kalberla05}
{Kalberla}, P.~M.~W., {Burton}, W.~B., {Hartmann}, D., {et~al.} 2005, \aap,
  440, 775

\bibitem[{{Kawaharada} {et~al.}(2010){Kawaharada}, {Okabe}, {Umetsu},
  {Takizawa}, {Matsushita}, {Fukazawa}, {Hamana}, {Miyazaki}, {Nakazawa}, \&
  {Ohashi}}]{kawa10}
{Kawaharada}, M., {Okabe}, N., {Umetsu}, K., {et~al.} 2010, \apj, 714, 423

\bibitem[{{Kay} {et~al.}(2002){Kay}, {Pearce}, {Frenk}, \& {Jenkins}}]{kay02}
{Kay}, S.~T., {Pearce}, F.~R., {Frenk}, C.~S., \& {Jenkins}, A. 2002, \mnras,
  330, 113

\bibitem[{{Khatri} \& {Gaspari}(2016)}]{khatri16}
{Khatri}, R. \& {Gaspari}, M. 2016, \mnras, 463, 655

\bibitem[{{Khedekar} {et~al.}(2013){Khedekar}, {Churazov}, {Kravtsov},
  {Zhuravleva}, {Lau}, {Nagai}, \& {Sunyaev}}]{khedekar13}
{Khedekar}, S., {Churazov}, E., {Kravtsov}, A., {et~al.} 2013, \mnras, 431, 954

\bibitem[{{Kravtsov} \& {Borgani}(2012)}]{kravtsov12}
{Kravtsov}, A.~V. \& {Borgani}, S. 2012, \araa, 50, 353

\bibitem[{{Kravtsov} {et~al.}(2005){Kravtsov}, {Nagai}, \&
  {Vikhlinin}}]{kravtsov05}
{Kravtsov}, A.~V., {Nagai}, D., \& {Vikhlinin}, A.~A. 2005, \apj, 625, 588

\bibitem[{{Kravtsov} {et~al.}(2006){Kravtsov}, {Vikhlinin}, \&
  {Nagai}}]{kravtsov06}
{Kravtsov}, A.~V., {Vikhlinin}, A., \& {Nagai}, D. 2006, \apj, 650, 128

\bibitem[{{Kriss} {et~al.}(1983){Kriss}, {Cioffi}, \& {Canizares}}]{kriss+83}
{Kriss}, G.~A., {Cioffi}, D.~F., \& {Canizares}, C.~R. 1983, \apj, 272, 439

\bibitem[{{Lamarre} {et~al.}(2010){Lamarre}, {Puget}, {Ade}, {Bouchet},
  {Guyot}, {Lange}, {Pajot}, {Arondel}, {Benabed}, {Beney}, {Beno{\^i}t},
  {Bernard}, {Bhatia}, {Blanc}, {Bock}, {Br{\'e}elle}, {Bradshaw}, {Camus},
  {Catalano}, {Charra}, {Charra}, {Church}, {Couchot}, {Coulais}, {Crill},
  {Crook}, {Dassas}, {de Bernardis}, {Delabrouille}, {de Marcillac}, {Delouis},
  {D{\'e}sert}, {Dumesnil}, {Dupac}, {Efstathiou}, {Eng}, {Evesque},
  {Fourmond}, {Ganga}, {Giard}, {Gispert}, {Guglielmi}, {Haissinski},
  {Henrot-Versill{\'e}}, {Hivon}, {Holmes}, {Jones}, {Koch}, {Lagard{\`e}re},
  {Lami}, {Land{\'e}}, {Leriche}, {Leroy}, {Longval},
  {Mac{\'{\i}}as-P{\'e}rez}, {Maciaszek}, {Maffei}, {Mansoux}, {Marty}, {Masi},
  {Mercier}, {Miville-Desch{\^e}nes}, {Moneti}, {Montier}, {Murphy},
  {Narbonne}, {Nexon}, {Paine}, {Pahn}, {Perdereau}, {Piacentini}, {Piat},
  {Plaszczynski}, {Pointecouteau}, {Pons}, {Ponthieu}, {Prunet}, {Rambaud},
  {Recouvreur}, {Renault}, {Ristorcelli}, {Rosset}, {Santos}, {Savini},
  {Serra}, {Stassi}, {Sudiwala}, {Sygnet}, {Tauber}, {Torre}, {Tristram},
  {Vibert}, {Woodcraft}, {Yurchenko}, \& {Yvon}}]{lamarre+10}
{Lamarre}, J.-M., {Puget}, J.-L., {Ade}, P.~A.~R., {et~al.} 2010, \aap, 520, A9

\bibitem[{{Lapi} {et~al.}(2010){Lapi}, {Fusco-Femiano}, \&
  {Cavaliere}}]{lapi10}
{Lapi}, A., {Fusco-Femiano}, R., \& {Cavaliere}, A. 2010, \aap, 516, A34+

\bibitem[{{Lau} {et~al.}(2015){Lau}, {Nagai}, {Avestruz}, {Nelson}, \&
  {Vikhlinin}}]{lau15}
{Lau}, E.~T., {Nagai}, D., {Avestruz}, C., {Nelson}, K., \& {Vikhlinin}, A.
  2015, \apj, 806, 68

\bibitem[{{Le Brun} {et~al.}(2014){Le Brun}, {McCarthy}, {Schaye}, \&
  {Ponman}}]{lebrun14}
{Le Brun}, A.~M.~C., {McCarthy}, I.~G., {Schaye}, J., \& {Ponman}, T.~J. 2014,
  \mnras, 441, 1270

\bibitem[{{Leccardi} \& {Molendi}(2008)}]{lm08}
{Leccardi}, A. \& {Molendi}, S. 2008, \aap, 486, 359

\bibitem[{{Lovisari} {et~al.}(2017){Lovisari}, {Forman}, {Jones}, {Ettori},
  {Andrade-Santos}, {Arnaud}, {D{\'e}mocl{\`e}s}, {Pratt}, {Randall}, \&
  {Kraft}}]{lovisari17}
{Lovisari}, L., {Forman}, W.~R., {Jones}, C., {et~al.} 2017, \apj, 846, 51

\bibitem[{{Mantz} {et~al.}(2018){Mantz}, {Allen}, {Morris}, \& {von der
  Linden}}]{mantz18}
{Mantz}, A.~B., {Allen}, S.~W., {Morris}, R.~G., \& {von der Linden}, A. 2018,
  \mnras, 473, 3072

\bibitem[{{Mazzotta} {et~al.}(2004){Mazzotta}, {Rasia}, {Moscardini}, \&
  {Tormen}}]{mazzotta+04}
{Mazzotta}, P., {Rasia}, E., {Moscardini}, L., \& {Tormen}, G. 2004, \mnras,
  354, 10

\bibitem[{{McCammon} {et~al.}(2002){McCammon}, {Almy}, {Apodaca}, {Bergmann
  Tiest}, {Cui}, {Deiker}, {Galeazzi}, {Juda}, {Lesser}, {Mihara},
  {Morgenthaler}, {Sanders}, {Zhang}, {Figueroa-Feliciano}, {Kelley},
  {Moseley}, {Mushotzky}, {Porter}, {Stahle}, \& {Szymkowiak}}]{mccammon02}
{McCammon}, D., {Almy}, R., {Apodaca}, E., {et~al.} 2002, \apj, 576, 188

\bibitem[{{Morandi} {et~al.}(2015){Morandi}, {Sun}, {Forman}, \&
  {Jones}}]{morandi15}
{Morandi}, A., {Sun}, M., {Forman}, W., \& {Jones}, C. 2015, \mnras, 450, 2261

\bibitem[{{Moretti} {et~al.}(2003){Moretti}, {Campana}, {Lazzati}, \&
  {Tagliaferri}}]{moretti03}
{Moretti}, A., {Campana}, S., {Lazzati}, D., \& {Tagliaferri}, G. 2003, \apj,
  588, 696

\bibitem[{{Nagai} {et~al.}(2007){Nagai}, {Kravtsov}, \& {Vikhlinin}}]{nagai+07}
{Nagai}, D., {Kravtsov}, A.~V., \& {Vikhlinin}, A. 2007, \apj, 668, 1

\bibitem[{{Nagai} \& {Lau}(2011)}]{nagai11}
{Nagai}, D. \& {Lau}, E.~T. 2011, \apjl, 731, L10

\bibitem[{{Nandra} {et~al.}(2013){Nandra}, {Barret}, {Barcons}, {Fabian}, {den
  Herder}, {Piro}, {Watson}, {Adami}, {Aird}, {Afonso}, \& et~al.}]{nandra13}
{Nandra}, K., {Barret}, D., {Barcons}, X., {et~al.} 2013, ArXiv e-prints
  [\eprint[arXiv]{1306.2307}]

\bibitem[{{Navarro} {et~al.}(1996){Navarro}, {Frenk}, \& {White}}]{nfw96}
{Navarro}, J.~F., {Frenk}, C.~S., \& {White}, S.~D.~M. 1996, \apj, 462, 563

\bibitem[{{Oegerle} {et~al.}(1995){Oegerle}, {Hill}, \&
  {Fitchett}}]{oegerle+95}
{Oegerle}, W.~R., {Hill}, J.~M., \& {Fitchett}, M.~J. 1995, \aj, 110, 32

\bibitem[{{Okabe} {et~al.}(2014){Okabe}, {Umetsu}, {Tamura}, {Fujita},
  {Takizawa}, {Zhang}, {Matsushita}, {Hamana}, {Fukazawa}, {Futamase},
  {Kawaharada}, {Miyazaki}, {Mochizuki}, {Nakazawa}, {Ohashi}, {Ota}, {Sasaki},
  {Sato}, \& {Tam}}]{okabe14}
{Okabe}, N., {Umetsu}, K., {Tamura}, T., {et~al.} 2014, \pasj, 66, 99

\bibitem[{{Pfrommer} {et~al.}(2007){Pfrommer}, {Springel}, {Jubelgas}, \&
  {Ensslin}}]{pfrommer+07}
{Pfrommer}, C., {Springel}, V., {Jubelgas}, M., \& {Ensslin}, T.~A. 2007, in
  Astronomical Society of the Pacific Conference Series, Vol. 379, Cosmic
  Frontiers, ed. N.~{Metcalfe} \& T.~{Shanks}, 221

\bibitem[{{Planck Collaboration} {et~al.}(2016{\natexlab{a}}){Planck
  Collaboration}, {Adam}, {Ade}, {Aghanim}, {Akrami}, {Alves}, {Arg{\"u}eso},
  {Arnaud}, {Arroja}, {Ashdown}, \& et~al.}]{planckdr2015}
{Planck Collaboration}, {Adam}, R., {Ade}, P.~A.~R., {et~al.}
  2016{\natexlab{a}}, \aap, 594, A1

\bibitem[{{Planck Collaboration} {et~al.}(2016{\natexlab{b}}){Planck
  Collaboration}, {Adam}, {Ade}, {Aghanim}, {Arnaud}, {Ashdown}, {Aumont},
  {Baccigalupi}, {Banday}, {Barreiro}, \& et~al.}]{planck16}
{Planck Collaboration}, {Adam}, R., {Ade}, P.~A.~R., {et~al.}
  2016{\natexlab{b}}, \aap, 594, A7

\bibitem[{{Planck Collaboration} {et~al.}(2013){Planck Collaboration}, {Ade},
  {Aghanim}, {Arnaud}, {Ashdown}, {Atrio-Barandela}, {Aumont}, {Baccigalupi},
  {Balbi}, {Banday}, \& et~al.}]{planck+13}
{Planck Collaboration}, {Ade}, P.~A.~R., {Aghanim}, N., {et~al.} 2013, \aap,
  550, A131

\bibitem[{{Planck Collaboration} {et~al.}(2011){Planck Collaboration}, {Ade},
  {Aghanim}, {Arnaud}, {Ashdown}, {Aumont}, {Baccigalupi}, {Balbi}, {Banday},
  {Barreiro}, \& et~al.}]{PESZ}
{Planck Collaboration}, {Ade}, P.~A.~R., {Aghanim}, N., {et~al.} 2011, \aap,
  536, A8

\bibitem[{{Planck Collaboration} {et~al.}(2016{\natexlab{c}}){Planck
  Collaboration}, {Ade}, {Aghanim}, {Arnaud}, {Ashdown}, {Aumont},
  {Baccigalupi}, {Banday}, {Barreiro}, {Bartlett}, \& et~al.}]{planck+16}
{Planck Collaboration}, {Ade}, P.~A.~R., {Aghanim}, N., {et~al.}
  2016{\natexlab{c}}, \aap, 594, A13

\bibitem[{{Planck Collaboration XXIX}(2014)}]{psz1}
{Planck Collaboration XXIX}. 2014, \aap, 571, A29

\bibitem[{{Planck HFI Core Team} {et~al.}(2011){Planck HFI Core Team}, {Ade},
  {Aghanim}, {Ansari}, {Arnaud}, {Ashdown}, {Aumont}, {Banday}, {Bartelmann},
  {Bartlett}, {Battaner}, {Benabed}, {Beno{\^i}t}, {Bernard}, {Bersanelli},
  {Bhatia}, {Bock}, {Bond}, {Borrill}, {Bouchet}, {Boulanger}, {Bradshaw},
  {Br{\'e}elle}, {Bucher}, {Camus}, {Cardoso}, {Catalano}, {Challinor},
  {Chamballu}, {Charra}, {Charra}, {Chary}, {Chiang}, {Church}, {Clements},
  {Colombi}, {Couchot}, {Coulais}, {Cressiot}, {Crill}, {Crook}, {de
  Bernardis}, {Delabrouille}, {Delouis}, {D{\'e}sert}, {Dolag}, {Dole},
  {Dor{\'e}}, {Douspis}, {Efstathiou}, {Eng}, {Filliard}, {Forni}, {Fosalba},
  {Fourmond}, {Ganga}, {Giard}, {Girard}, {Giraud-H{\'e}raud}, {Gispert},
  {G{\'o}rski}, {Gratton}, {Griffin}, {Guyot}, {Haissinski}, {Harrison},
  {Helou}, {Henrot-Versill{\'e}}, {Hern{\'a}ndez-Monteagudo}, {Hildebrandt},
  {Hills}, {Hivon}, {Hobson}, {Holmes}, {Huffenberger}, {Jaffe}, {Jones},
  {Kaplan}, {Kneissl}, {Knox}, {Lagache}, {Lamarre}, {Lami}, {Lange},
  {Lasenby}, {Lavabre}, {Lawrence}, {Leriche}, {Leroy}, {Longval},
  {Mac{\'{\i}}as-P{\'e}rez}, {Maciaszek}, {MacTavish}, {Maffei}, {Mandolesi},
  {Mann}, {Mansoux}, {Masi}, {Matsumura}, {McGehee}, {Melin}, {Mercier},
  {Miville-Desch{\^e}nes}, {Moneti}, {Montier}, {Mortlock}, {Murphy}, {Nati},
  {Netterfield}, {N{\o}rgaard-Nielsen}, {North}, {Noviello}, {Novikov},
  {Osborne}, {Paine}, {Pajot}, {Patanchon}, {Peacocke}, {Pearson}, {Perdereau},
  {Perotto}, {Piacentini}, {Piat}, {Plaszczynski}, {Pointecouteau}, {Pons},
  {Ponthieu}, {Pr{\'e}zeau}, {Prunet}, {Puget}, {Reach}, {Renault},
  {Ristorcelli}, {Rocha}, {Rosset}, {Roudier}, {Rowan-Robinson}, {Rusholme},
  {Santos}, {Savini}, {Schaefer}, {Shellard}, {Spencer}, {Starck}, {Stassi},
  {Stolyarov}, {Stompor}, {Sudiwala}, {Sunyaev}, {Sygnet}, {Tauber}, {Thum},
  {Torre}, {Touze}, {Tristram}, {van Leeuwen}, {Vibert}, {Vibert}, {Wade},
  {Wandelt}, {White}, {Wiesemeyer}, {Woodcraft}, {Yurchenko}, {Yvon}, \&
  {Zacchei}}]{planckHFI}
{Planck HFI Core Team}, {Ade}, P.~A.~R., {Aghanim}, N., {et~al.} 2011, \aap,
  536, A4

\bibitem[{{Planelles} {et~al.}(2017){Planelles}, {Fabjan}, {Borgani},
  {Murante}, {Rasia}, {Biffi}, {Truong}, {Ragone-Figueroa}, {Granato}, {Dolag},
  {Pierpaoli}, {Beck}, {Steinborn}, \& {Gaspari}}]{planelles17}
{Planelles}, S., {Fabjan}, D., {Borgani}, S., {et~al.} 2017, \mnras, 467, 3827

\bibitem[{{Pratt} {et~al.}(2010){Pratt}, {Arnaud}, {Piffaretti},
  {B{\"o}hringer}, {Ponman}, {Croston}, {Voit}, {Borgani}, \&
  {Bower}}]{pratt+10}
{Pratt}, G.~W., {Arnaud}, M., {Piffaretti}, R., {et~al.} 2010, \aap, 511, A85

\bibitem[{{Pratt} {et~al.}(2007){Pratt}, {B{\"o}hringer}, {Croston}, {Arnaud},
  {Borgani}, {Finoguenov}, \& {Temple}}]{pratt07}
{Pratt}, G.~W., {B{\"o}hringer}, H., {Croston}, J.~H., {et~al.} 2007, \aap,
  461, 71

\bibitem[{{Predehl} {et~al.}(2010){Predehl}, {Andritschke}, {B{\"o}hringer},
  {Bornemann}, {Br{\"a}uninger}, {Brunner}, {Brusa}, {Burkert}, {Burwitz},
  {Cappelluti}, {Churazov}, {Dennerl}, {Eder}, {Elbs}, {Freyberg}, {Friedrich},
  {F{\"u}rmetz}, {Gaida}, {H{\"a}lker}, {Hartner}, {Hasinger}, {Hermann},
  {Huber}, {Kendziorra}, {von Kienlin}, {Kink}, {Kreykenbohm}, {Lamer},
  {Lapchov}, {Lehmann}, {Meidinger}, {Mican}, {Mohr}, {M{\"u}hlegger},
  {M{\"u}ller}, {Nandra}, {Pavlinsky}, {Pfeffermann}, {Reiprich}, {Robrade},
  {Roh{\'e}}, {Santangelo}, {Sch{\"a}chner}, {Schanz}, {Schmid}, {Schmitt},
  {Schreib}, {Schrey}, {Schwope}, {Steinmetz}, {Str{\"u}der}, {Sunyaev},
  {Tenzer}, {Tiedemann}, {Vongehr}, \& {Wilms}}]{erosita}
{Predehl}, P., {Andritschke}, R., {B{\"o}hringer}, H., {et~al.} 2010, in
  \procspie, Vol. 7732, Space Telescopes and Instrumentation 2010: Ultraviolet
  to Gamma Ray, 77320U

\bibitem[{{Rasia} {et~al.}(2006){Rasia}, {Ettori}, {Moscardini}, {Mazzotta},
  {Borgani}, {Dolag}, {Tormen}, {Cheng}, \& {Diaferio}}]{rasia+06}
{Rasia}, E., {Ettori}, S., {Moscardini}, L., {et~al.} 2006, \mnras, 369, 2013

\bibitem[{{Reiprich} {et~al.}(2013){Reiprich}, {Basu}, {Ettori}, {Israel},
  {Lovisari}, {Molendi}, {Pointecouteau}, \& {Roncarelli}}]{reiprich13}
{Reiprich}, T.~H., {Basu}, K., {Ettori}, S., {et~al.} 2013, \ssr
  [\eprint[arXiv]{1303.3286}]

\bibitem[{{Roncarelli} {et~al.}(2013){Roncarelli}, {Ettori}, {Borgani},
  {Dolag}, {Fabjan}, \& {Moscardini}}]{roncarelli+13}
{Roncarelli}, M., {Ettori}, S., {Borgani}, S., {et~al.} 2013, \mnras, 432, 3030

\bibitem[{{Rossetti} {et~al.}(2017){Rossetti}, {Gastaldello}, {Eckert}, {Della
  Torre}, {Pantiri}, {Cazzoletti}, \& {Molendi}}]{rossetti17}
{Rossetti}, M., {Gastaldello}, F., {Eckert}, D., {et~al.} 2017, \mnras, 468,
  1917

\bibitem[{{Ruppin} {et~al.}(2017){Ruppin}, {Adam}, {Comis}, {Ade}, {Andr{\'e}},
  {Arnaud}, {Beelen}, {Beno{\^i}t}, {Bideaud}, {Billot}, {Bourrion}, {Calvo},
  {Catalano}, {Coiffard}, {D'Addabbo}, {De Petris}, {D{\'e}sert}, {Doyle},
  {Goupy}, {Kramer}, {Leclercq}, {Mac{\'{\i}}as-P{\'e}rez}, {Mauskopf},
  {Mayet}, {Monfardini}, {Pajot}, {Pascale}, {Perotto}, {Pisano},
  {Pointecouteau}, {Ponthieu}, {Pratt}, {Rev{\'e}ret}, {Ritacco}, {Rodriguez},
  {Romero}, {Schuster}, {Sievers}, {Triqueneaux}, {Tucker}, \&
  {Zylka}}]{ruppin17}
{Ruppin}, F., {Adam}, R., {Comis}, B., {et~al.} 2017, \aap, 597, A110

\bibitem[{{Salvetti} {et~al.}(2017){Salvetti}, {Marelli}, {Gastaldello},
  {Ghizzardi}, {Molendi}, {De Luca}, {Moretti}, {Rossetti}, \&
  {Tiengo}}]{salvetti17}
{Salvetti}, D., {Marelli}, M., {Gastaldello}, F., {et~al.} 2017, ArXiv e-prints
  [\eprint[arXiv]{1705.04172}]

\bibitem[{{Schellenberger} {et~al.}(2015){Schellenberger}, {Reiprich},
  {Lovisari}, {Nevalainen}, \& {David}}]{schellenberger+15}
{Schellenberger}, G., {Reiprich}, T.~H., {Lovisari}, L., {Nevalainen}, J., \&
  {David}, L. 2015, \aap, 575, A30

\bibitem[{{Simionescu} {et~al.}(2011){Simionescu}, {Allen}, {Mantz}, {Werner},
  {Takei}, {Morris}, {Fabian}, {Sanders}, {Nulsen}, {George}, \&
  {Taylor}}]{simi11}
{Simionescu}, A., {Allen}, S.~W., {Mantz}, A., {et~al.} 2011, Science, 331,
  1576

\bibitem[{{Simionescu} {et~al.}(2017){Simionescu}, {Werner}, {Mantz}, {Allen},
  \& {Urban}}]{simi17}
{Simionescu}, A., {Werner}, N., {Mantz}, A., {Allen}, S.~W., \& {Urban}, O.
  2017, \mnras, 469, 1476

\bibitem[{{Smith} {et~al.}(2012){Smith}, {Hopkins}, {Hunstead}, \&
  {Pimbblet}}]{smith12}
{Smith}, A.~G., {Hopkins}, A.~M., {Hunstead}, R.~W., \& {Pimbblet}, K.~A. 2012,
  \mnras, 422, 25

\bibitem[{{Snowden} {et~al.}(2008){Snowden}, {Mushotzky}, {Kuntz}, \&
  {Davis}}]{snowden08}
{Snowden}, S.~L., {Mushotzky}, R.~F., {Kuntz}, K.~D., \& {Davis}, D.~S. 2008,
  \aap, 478, 615

\bibitem[{{Sunyaev} \& {Zeldovich}(1972)}]{SZ}
{Sunyaev}, R.~A. \& {Zeldovich}, Y.~B. 1972, Comments on Astrophysics and Space
  Physics, 4, 173

\bibitem[{{Tauber} {et~al.}(2010){Tauber}, {Mandolesi}, {Puget}, {Banos},
  {Bersanelli}, {Bouchet}, {Butler}, {Charra}, {Crone}, {Dodsworth}, \&
  et~al.}]{tauber+10}
{Tauber}, J.~A., {Mandolesi}, N., {Puget}, J.-L., {et~al.} 2010, \aap, 520, A1

\bibitem[{{Tchernin} {et~al.}(2016){Tchernin}, {Eckert}, {Ettori},
  {Pointecouteau}, {Paltani}, {Molendi}, {Hurier}, {Gastaldello}, {Lau},
  {Nagai}, {Roncarelli}, \& {Rossetti}}]{tchernin16}
{Tchernin}, C., {Eckert}, D., {Ettori}, S., {et~al.} 2016, \aap, 595, A42

\bibitem[{{Tozzi} \& {Norman}(2001)}]{tozzi01}
{Tozzi}, P. \& {Norman}, C. 2001, \apj, 546, 63

\bibitem[{{Truong} {et~al.}(2018){Truong}, {Rasia}, {Mazzotta}, {Planelles},
  {Biffi}, {Fabjan}, {Beck}, {Borgani}, {Dolag}, {Gaspari}, {Granato},
  {Murante}, {Ragone-Figueroa}, \& {Steinborn}}]{truong18}
{Truong}, N., {Rasia}, E., {Mazzotta}, P., {et~al.} 2018, \mnras, 474, 4089

\bibitem[{{Urban} {et~al.}(2014){Urban}, {Simionescu}, {Werner}, {Allen},
  {Ehlert}, {Zhuravleva}, {Morris}, {Fabian}, {Mantz}, {Nulsen}, {Sanders}, \&
  {Takei}}]{urban14}
{Urban}, O., {Simionescu}, A., {Werner}, N., {et~al.} 2014, \mnras, 437, 3939

\bibitem[{{Vazza} {et~al.}(2009){Vazza}, {Brunetti}, {Kritsuk}, {Wagner},
  {Gheller}, \& {Norman}}]{vazza09}
{Vazza}, F., {Brunetti}, G., {Kritsuk}, A., {et~al.} 2009, \aap, 504, 33

\bibitem[{{Vazza} {et~al.}(2013){Vazza}, {Eckert}, {Simionescu}, {Br{\"u}ggen},
  \& {Ettori}}]{vazza13}
{Vazza}, F., {Eckert}, D., {Simionescu}, A., {Br{\"u}ggen}, M., \& {Ettori}, S.
  2013, \mnras, 429, 799

\bibitem[{{Vikhlinin} {et~al.}(2006){Vikhlinin}, {Kravtsov}, {Forman}, {Jones},
  {Markevitch}, {Murray}, \& {Van Speybroeck}}]{vikhlini+06}
{Vikhlinin}, A., {Kravtsov}, A., {Forman}, W., {et~al.} 2006, \apj, 640, 691

\bibitem[{{Voit}(2005)}]{voit05rev}
{Voit}, G.~M. 2005, Reviews of Modern Physics, 77, 207

\bibitem[{{Voit} {et~al.}(2005){Voit}, {Kay}, \& {Bryan}}]{voit+05}
{Voit}, G.~M., {Kay}, S.~T., \& {Bryan}, G.~L. 2005, \mnras, 364, 909

\bibitem[{{Walker} {et~al.}(2012{\natexlab{a}}){Walker}, {Fabian}, {Sanders},
  \& {George}}]{walker12a}
{Walker}, S.~A., {Fabian}, A.~C., {Sanders}, J.~S., \& {George}, M.~R.
  2012{\natexlab{a}}, \mnras, 424, 1826

\bibitem[{{Walker} {et~al.}(2012{\natexlab{b}}){Walker}, {Fabian}, {Sanders},
  {George}, \& {Tawara}}]{walker12b}
{Walker}, S.~A., {Fabian}, A.~C., {Sanders}, J.~S., {George}, M.~R., \&
  {Tawara}, Y. 2012{\natexlab{b}}, \mnras, 422, 3503

\bibitem[{{Zhuravleva} {et~al.}(2013){Zhuravleva}, {Churazov}, {Kravtsov},
  {Lau}, {Nagai}, \& {Sunyaev}}]{zhuravleva13}
{Zhuravleva}, I., {Churazov}, E., {Kravtsov}, A., {et~al.} 2013, \mnras, 428,
  3274

\end{thebibliography}
